\documentclass{emulateapj}
\newcommand{\msun}{\ensuremath{\rm M_\odot}}
\newcommand{\msunyr}{\ensuremath{\rm M_{\odot}\;{\rm yr}^{-1}}}
\newcommand{\Ha}{\ensuremath{\rm H\alpha}}
\newcommand{\Hb}{\ensuremath{\rm H\beta}}
\newcommand{\lya}{\ensuremath{\rm Ly\alpha}}
\newcommand{\Ntwo}{[\ion{N}{2}]}
\newcommand{\kms}{km~s\ensuremath{^{-1}\,}}
\newcommand{\ztwo}{\ensuremath{z\sim2}}
\newcommand{\zthree}{\ensuremath{z\sim3}}

\begin{document}

\title{THE STELLAR, GAS AND DYNAMICAL MASSES OF STAR-FORMING GALAXIES
  AT $z\sim2$\altaffilmark{1}} \author{\sc Dawn
  K. Erb\altaffilmark{2}, Charles C. Steidel\altaffilmark{3}, Alice
  E. Shapley\altaffilmark{4},\\ Max Pettini\altaffilmark{5}, Naveen
  A. Reddy\altaffilmark{3}, Kurt L. Adelberger\altaffilmark{6}}

\shorttitle{THE STELLAR, GAS AND DYNAMICAL MASSES OF $z\sim2$ GALAXIES}
\shortauthors{ERB ET AL.}
\slugcomment{Accepted for publication in \apj}

\altaffiltext{1}{Based on data obtained at the 
W.M. Keck Observatory, which is operated as a scientific partnership
among the California Institute of Technology, the University of
California, and NASA, and was made possible by the generous financial
support of the W.M. Keck Foundation.}  
\altaffiltext{2}{Harvard-Smithsonian Center for Astrophysics, MS 20,
  60 Garden St, Cambridge, MA 02138; derb@cfa.harvard.edu}
\altaffiltext{3}{California Institute of Technology, MS 105--24,
  Pasadena, CA 91125}  
\altaffiltext{4}{Department of Astrophysical Sciences, Princeton
  University, Peyton Hall, Ivy Lane, Princeton, NJ 08544}
\altaffiltext{5}{Institute of Astronomy, Madingley Road, Cambridge CB3
  0HA, UK}
\altaffiltext{6}{McKinsey and Company, 1420 Fifth Avenue, Suite 3100,
  Seattle, WA, 98101}

\begin{abstract}
We present analysis of the near-infrared spectra of 114 rest-frame
UV-selected star-forming galaxies at \ztwo.  By combining the
\Ha\ spectra with photometric measurements from observed 0.3--8
\micron, we assess the relationships between kinematics, dynamical
masses, inferred gas fractions, and stellar masses and ages.  The
\Ha\ line widths give a mean velocity dispersion $\sigma=101\pm5$
\kms\ and a mean dynamical mass $M_{\rm dyn}=6.9\pm0.6\times10^{10}$
\msun\ within a typical radius of $\sim6$ kpc, after excluding AGN.
The average dynamical mass is $\sim2$ times larger than the average
stellar mass, and the two are correlated at the 2.5$\sigma$ level and
agree to within a factor of several for most objects, consistent with
observational and systematic uncertainties.  However, $\sim15$\% of
the sample has $M_{\rm dyn}\gg M_{\star}$.  These objects are best fit
by young stellar populations and tend to have high \Ha\ equivalent
widths $W_{\Ha}\gtrsim200$ \AA, suggesting that they are young
starbursts with large gas masses.  Rest-frame optical luminosity and
velocity dispersion are correlated with 4$\sigma$ significance; the
correlation and its accompanying large scatter suggest that the
processes which produce the correlation between luminosity and
velocity dispersion in local galaxies are underway at \ztwo.  Fourteen
of the 114 galaxies in the sample have spatially resolved and tilted
\Ha\ emission lines indicative of velocity shear.  It is not yet clear
whether the shear indicates merging or rotation, but if the galaxies
are rotating disks and follow relations between velocity dispersion
and circular velocity similar to those seen in local galaxies, our
observations underestimate the circular velocities by an average
factor of $\sim2$ and the sample has $\langle V_c\rangle\sim190$ \kms.
Using the local empirical correlation between star formation rate per
unit area and gas surface density, we estimate the mass of the gas
associated with star formation, and find a mean gas fraction of
$\sim50$\% and a strong decrease in gas fraction with increasing
stellar mass.  The masses of gas and stars combined are considerably
better correlated with the dynamical masses than are the stellar
masses alone, and agree to within a factor of three for 85\% of the
sample.  The combination of kinematic measurements, estimates of gas
masses, and stellar population properties suggest that the factor of
$\sim500$ range in stellar mass across the sample cannot be fully
explained by intrinsic differences in the total masses of the
galaxies, which vary by a factor of $\sim40$; the remaining variation
is due to the evolution of the stellar population and the conversion
of gas into stars.
 
\end{abstract}


\section{Introduction}
A galaxy's mass is one of its most fundamental properties.  Both
popular models of galaxy formation and recent observations suggest
that mass is an important factor in determining when and how quickly
galaxies form their stars; star formation in more massive galaxies is
both expected and observed to start at earlier times
(e.g.\ \citealt{dfh+05,hpjd04,jgc+05}).  The redshift range
$1.5\lesssim z \lesssim 2.5$ has been shown to be particularly
important both for the buildup of stellar mass and for accretion onto
massive black holes.  A large fraction of the stellar mass in the
universe today likely formed at $z>1$ \citep{dpfb03,rrf+03}, and this
redshift range also sees the peak of bright QSO activity
\citep{fss+01,dcsh03}.  Effective techniques now exist for the
selection of galaxies at \ztwo; these use the galaxies' observed
optical \citep{ssp+04} or near-infrared \citep{flr+03} colors, or a
combination of the two \citep{bzk}, and can be used to select both
star-forming and passively evolving galaxies.

Even with large galaxy samples mass is difficult to measure,
especially at high redshift.  While galaxy formation models most
naturally parameterize galaxies as a function of mass, observers can
only determine mass through the measurement of luminosity or
(preferably but with more difficulty) kinematic properties.
Measurements of galaxy mass at high (and low) redshift take a variety
of approaches.  The most direct methods use kinematic tracers to
measure the depth of a galaxy's potential well.  At relatively low
redshifts, when rotation curves can be traced to large radii and
galaxies' mass distributions modeled, this technique yields relatively
robust measurements, as shown by the good agreement between results
derived from optical emission lines and 21 cm mapping of \ion{H}{1}
(e.g.\ \citealt{court97,kg00}).  At high redshift the situation is more
complicated.  While spatially resolved ``rotation curves'' can
sometimes be obtained using nebular emission lines in the near-IR
\citep{pss+01,ess+03,ess+04,fgl+06}, the limitations imposed by the
seeing mean that they do not necessarily provide a unique kinematic
model.  A more robust measurement is the velocity dispersion, which
can be measured for most high redshift galaxies and is less affected
by the seeing.  Velocity dispersions have been used to estimate the
dynamical masses of relatively small samples of galaxies at \ztwo\ and
\zthree; results typically range from $\sim10^{10}$ to $\sim10^{11}$
\msun\ \citep{pss+01,ess+03,ess+04,vff+04,ssc+04}.  A model for the
mass distribution is still required to turn a measurement of the
velocity dispersion into a mass estimate, and this is a significant
source of uncertainty in such dynamical masses.

Dynamical masses measured from velocity dispersions do not, of course,
trace the full halo masses of high redshift galaxies; detectable
nebular line emission is typically confined to sites of active star
formation within a central region of a few kpc.  Halo masses can be
estimated less directly, however, through the galaxies' clustering
properties and the predicted clustering properties of halos as a
function of mass.  \citet{asp+05} find that the correlation lengths of
the \ztwo\ rest-frame UV-selected galaxies on which this paper focuses
correspond to halos with a typical mass of $\sim10^{12}$ \msun.

Estimations of stellar mass at high redshift are increasingly common,
driven by wide-field IR detectors which allow large samples of
rest-frame optical photometry, and by the Spitzer Space Telescope
which enables detection of rest-frame near-IR light.  Stellar masses
of \ztwo--3 galaxies are determined through population synthesis
models applied to such broadband photometry
(e.g.\ \citealt{pdf01,ssa+01,fvf+04,sse+05}, among many others); the
masses obtained range from $\sim10^{9}$ to $\gtrsim 10^{11}$ \msun.
Such masses usually represent lower limits to the true stellar mass,
however, since it is possible for the light from older populations of
stars to be obscured by current star formation.

The most difficult component of a galaxy's baryonic mass to measure at
high redshift is its gas.  For the most luminous, gas-rich, or
gravitationally magnified examples, this can be estimated with
millimeter observations (e.g.\ \citealt{btg+04,gbs+05}), but for the
typical \ztwo\ galaxy such observations are still out of reach.  As
the gas probably accounts for a significant fraction of the baryonic
mass in young galaxies at high redshift, we have made use of the
empirical correlation between star formation rate density and gas
surface density to estimate the gas masses of the galaxies considered
here.

This paper is one of several making use of a large sample of
\Ha\ spectra of 114 \ztwo\ galaxies.  In this work we discuss the
galaxies' kinematic properties and their stellar and dynamical masses,
while \citet{ess+06} is devoted to star formation.  This paper is
organized as follows.  We describe the selection of our sample, the
observations, and our data reduction procedures in \S\ref{sec:sample}.
In \S\ref{sec:smass} we outline the modeling procedure by which we
determine stellar masses and other stellar population
parameters. \S\ref{sec:dynmass} is devoted to the galaxies' dynamical
masses as derived from the \Ha\ line widths; we compare them with
stellar masses in \S\ref{sec:masscomp}, and assess the relationship
between velocity dispersion and rest-frame optical luminosity in
\S\ref{sec:siglum}.  We compare the distributions of velocity
dispersions of galaxies at \ztwo\ and \zthree\ in \S\ref{sec:lbgcomp}.
In \S\ref{sec:tilt} we examine galaxies with spatially resolved and
tilted emission lines.  \S\ref{sec:gasmass} describes our estimates of
gas masses from the local correlation between star formation rate and
gas density per unit area, and comparisons of stellar, gas and
dynamical masses.  We summarize our conclusions and discuss our
results in \S\ref{sec:disc}.  Separately, we use the same sample of
\Ha\ spectra to construct composite spectra according to stellar mass
to show that there is a strong correlation between increasing oxygen
abundance as measured by the \Ntwo/\Ha\ ratio and increasing stellar
mass \citep{esp+06}.  Galactic outflows in this sample are discussed
by Steidel et al (2006, in preparation).

A cosmology with $H_0=70\;{\rm km}\;{\rm s}^{-1}\;{\rm
  Mpc}^{-1}$, $\Omega_m=0.3$, and $\Omega_{\Lambda}=0.7$ is assumed throughout.
In such a cosmology, 1\arcsec\ at $z=2.24$ (the mean redshift of the current
sample) corresponds to 8.2 kpc, and at this redshift the universe is 2.9 Gyr
old, or 21\% of its present age.  For calculations of stellar masses and star
formation rates we use a \citet{c03} initial mass function (IMF),
which results in stellar masses and SFRs 1.8 times smaller than would
be obtained with a \citet{s55} IMF.
 
\section{Sample Selection, Observations and Data Reduction}
\label{sec:sample}
The galaxies discussed herein are drawn from the rest-frame
UV-selected $z\sim2$ spectroscopic sample described by \citet{ssp+04}.
The candidate galaxies were selected by their $U_nG\cal R$ colors
(from deep optical images discussed by \citealt{ssp+04}), with
redshifts for most of the objects in the current sample then confirmed
in the rest-frame UV using the LRIS-B spectrograph on the Keck I
telescope.  The total sample of 114 \Ha\ spectra consists of 75 new
observations and 39 that have been published previously
\citep{ess+03,ess+04,sep+04}.  In total we attempted 132 galaxies with
previously known redshifts, but 23 were not detected.  We also
attempted to observe 28 galaxies without previously known redshifts
but which met the photometric selection criteria; only 5 of these
yielded secure \Ha\ redshifts.  Most of the undetected galaxies have
$K>21.5$, while most of those we detect have $K<21.5$; we are less
likely to detect \Ha\ emission for objects that are faint in $K$.
This is not surprising, given the correlation between $K$ magnitude
and star formation rate recently found for the \ztwo\ sample by
\citet{res+05}.  This correlation is confirmed and discussed by
\citet{ess+06}

The set of galaxies presented here is not necessarily representative
of the UV-selected sample as a whole, because objects were chosen for
near-IR spectroscopy for a wide variety of reasons.  Criteria for
selection included: 1) galaxies near the line of sight to a QSO, for
studies of correlations between galaxies and metal systems seen in
absorption in the QSO spectra \citep{ass+05}; 2)
morphologies---elongated in most cases, with a few more compact
objects for comparison; most of these have been previously discussed
by \citealt{ess+04}; 3) galaxies with red or bright near-IR colors or
magnitudes, or occasionally those whose photometry suggested an
unusual spectral energy distribution (SED; 13 objects selected because
they have $K<20$ are discussed by \citealt{sep+04}); 4) galaxies that
have excellent deep rest-frame UV spectra, or are bright in the
rest-frame UV; 5) in order to confirm classification as AGN; or 6)
members of close pairs with redshifts known to be favorable relative
to night sky lines.  The remaining objects are galaxies within
$\sim25$\arcsec\ of our primary targets, which were placed along the
slit and observed simultaneously with the primary targets.  A
significant number of galaxies meet more than one of these criteria.
The sightline and secondary pair objects are therefore generally
representative of the UV-selected sample as a whole, while the
addition of the IR red or bright objects means that the total
\Ha\ sample is somewhat biased toward more massive galaxies (see
\S\ref{sec:smass}).  We identify AGN as galaxies that show broad
and/or high ionization emission lines in their rest-frame UV spectra,
or as objects with broad \Ha\ lines or very high \Ntwo/\Ha\ ratios.
The AGN fraction (5/114) is similar to the fraction found for the full
spectroscopic sample in the GOODS-N field by \citet{res+05}, using
direct detections in the 2-Ms {\it Chandra} Deep Field North images.

\begin{figure*}[htbp]
\begin{center}
\includegraphics[width=3.3in]{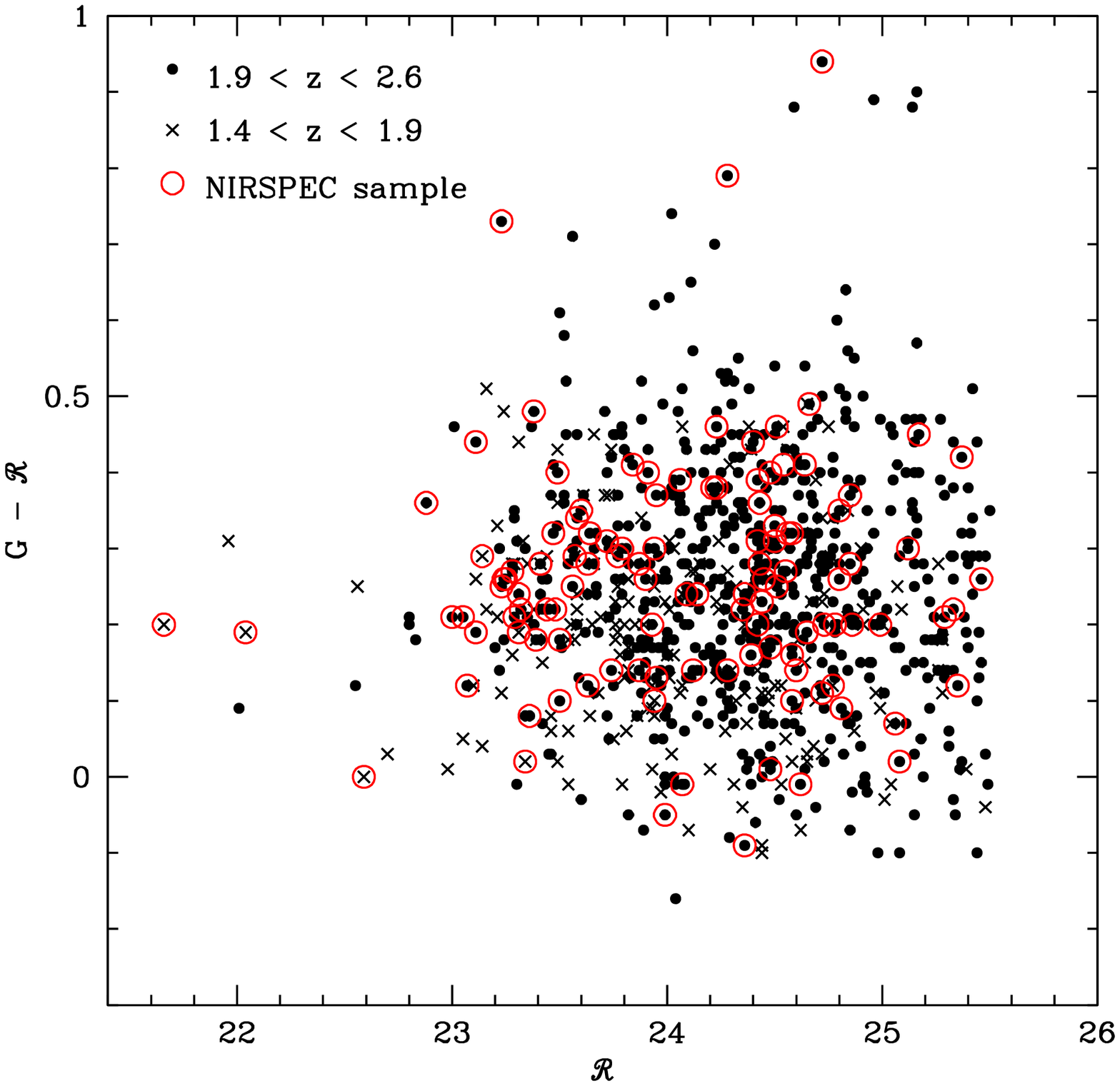}\includegraphics[width=3.3in]{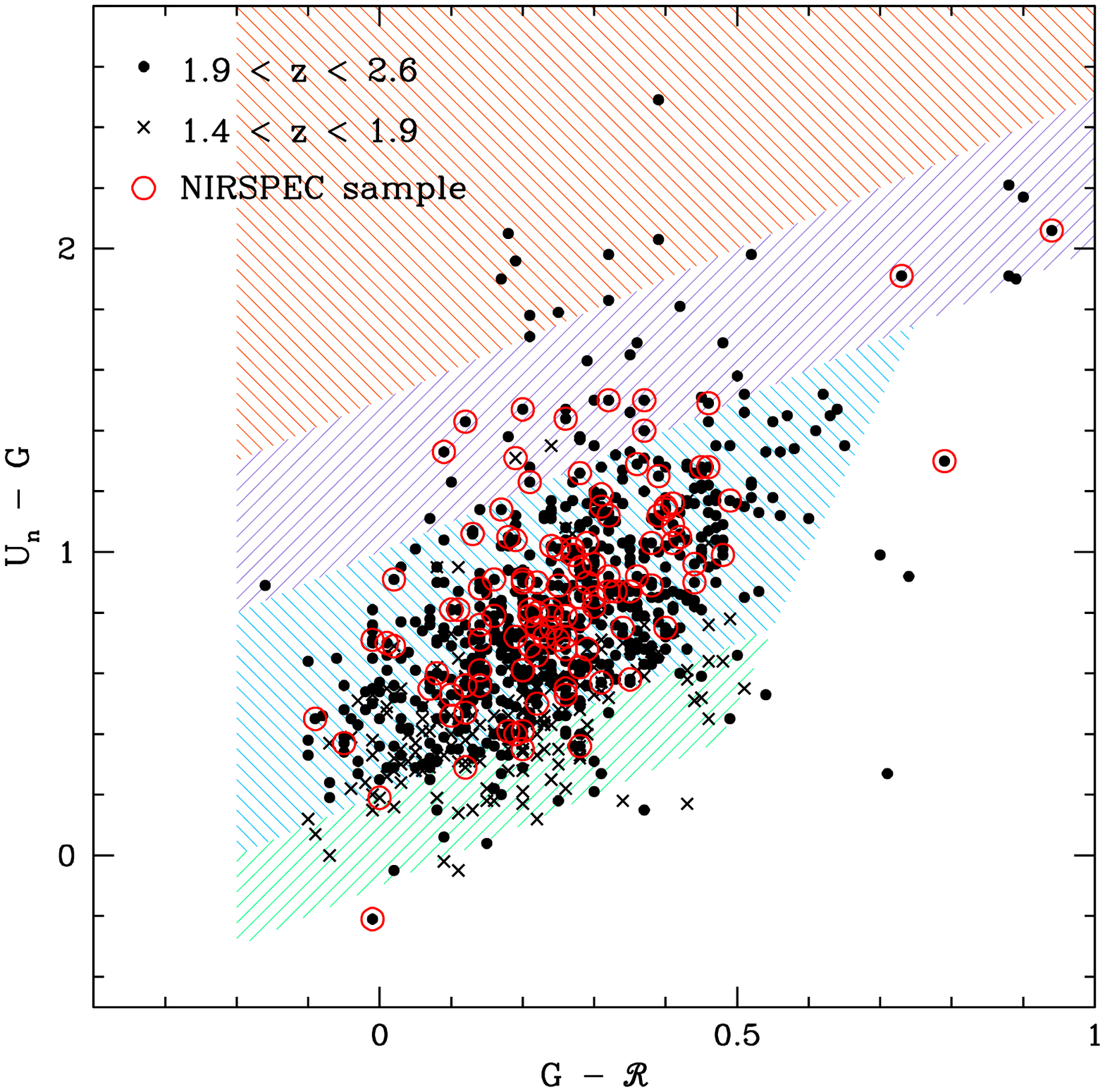}
\includegraphics[width=3.3in]{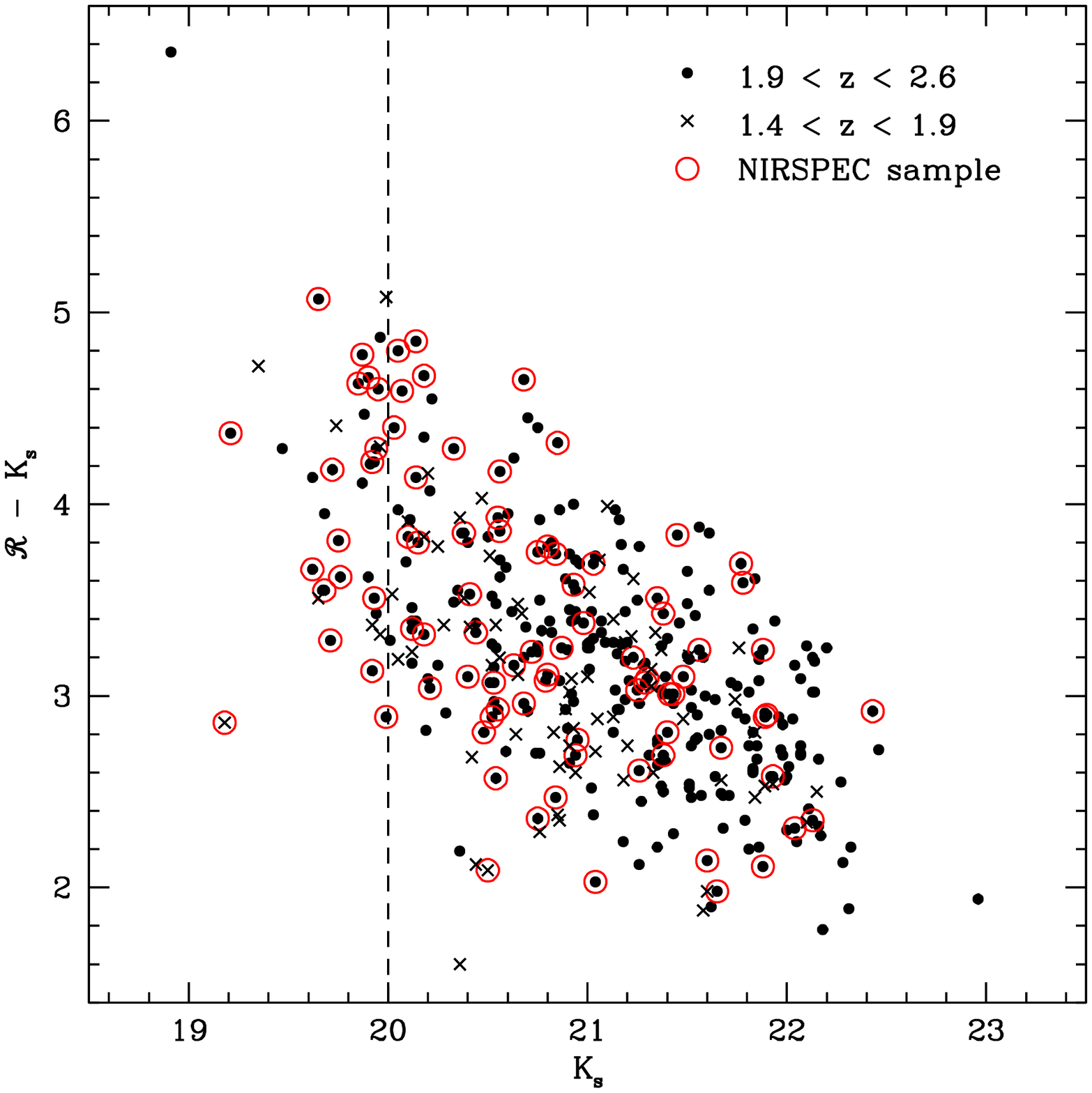}\includegraphics[width=3.3in]{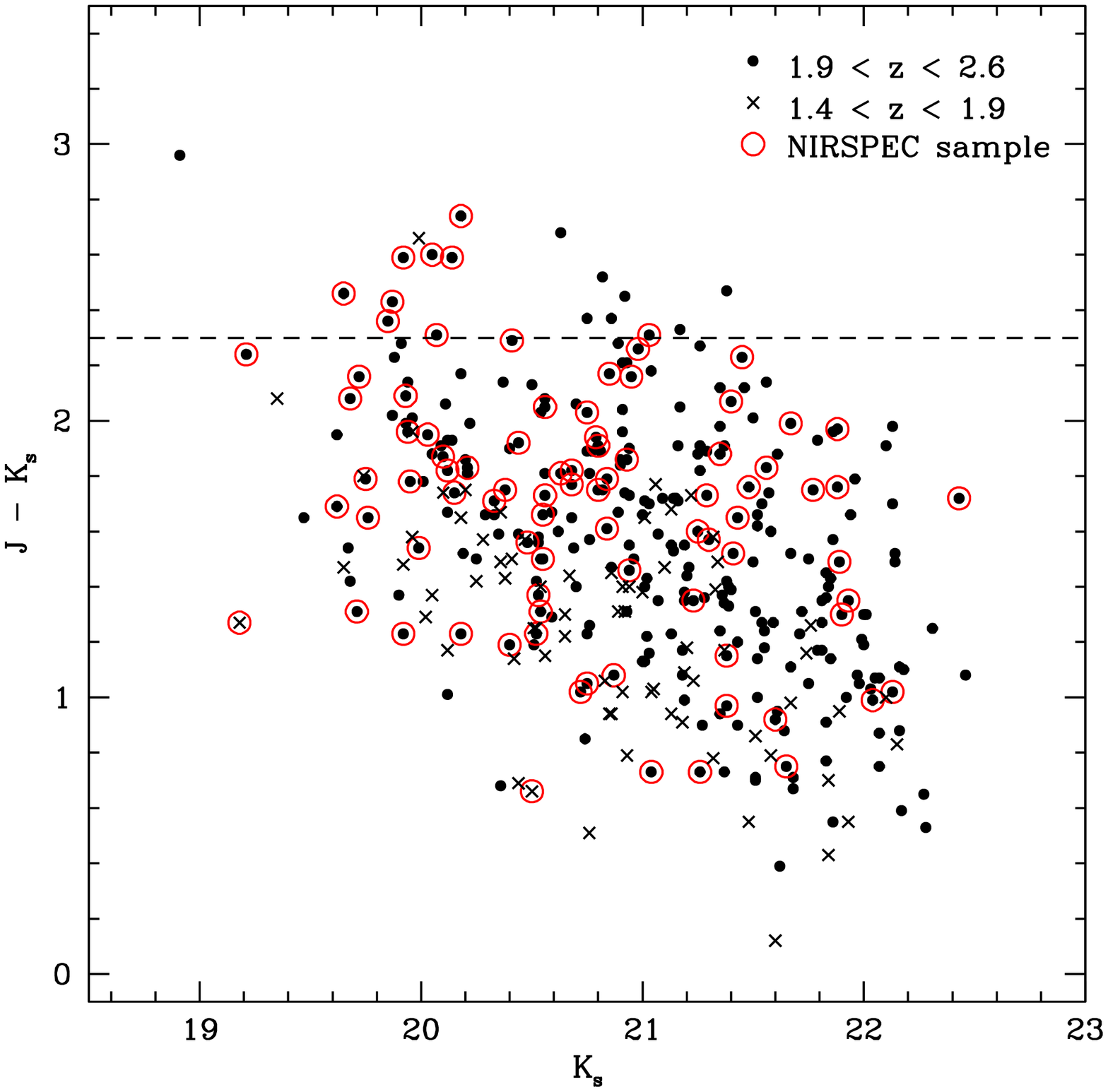}
\end{center}
\caption{Photometric comparisons of the NIRSPEC sample with all
  spectroscopically confirmed $U_nG{\cal R}$-selected galaxies in the
  same fields and redshift range.  The top panels compare the
  rest-frame UV photometry: ${\cal R}$ vs.\ $G-{\cal R}$ at upper left,
  and $G-{\cal R}$ vs.\ $U_n-G$ at upper right.  From top to bottom,
  the shaded regions show selection criteria for the \zthree\ LBGs
  (C/D and MD, \citealt{sas+03}) and for BX and BM galaxies
  \citep{ssp+04}.  Objects that fall outside the selection windows
  were either selected using earlier photometry or criteria extending
  slightly blueward in $U_n-G$.  In the bottom panels we compare
  the near-IR photometry: $K_s$ vs.\ ${\cal R}-K_s$ at lower left,
  and $K_s$  vs.\ $J-K_s$ at lower right.  The dashed lines show
  $K_s=20$ and $J-K_s=2.3$ in the left and right panels respectively.
  We include objects with $1.4 < z < 1.9$; 6 of the 114 objects in the
  NIRSPEC sample are in this redshift range.}
\label{fig:photsamples}
\end{figure*}

In order to assess the representativeness of the NIRSPEC sample, in
Figure~\ref{fig:photsamples} we compare the colors and magnitudes of
the galaxies in the NIRSPEC sample with that of all spectroscopically
confirmed UV-selected galaxies in the same fields and redshift range.
The top panels show ${\cal R}$ vs.\ $G-{\cal R}$ (upper left) and
$G-{\cal R}$ vs.\ $U_n-G$ (upper right).  The shaded regions in the
upper right panel show the selection criteria for, from top to bottom,
the \zthree\ C/D and MD galaxies \citep{sas+03} and the \ztwo\ BX and
BM galaxies \citep{ssp+04}.  Objects that fall outside the selection
windows were either selected using earlier photometry or criteria
extending slightly blueward in $U_n-G$.  The lower panels show $K_s$
vs.\ ${\cal R}-K_s$ at lower left, and $K_s$ vs.\ $J-K_s$ at lower
right.  These comparisons show that the NIRSPEC sample spans nearly
the full range in ${\cal R}$ and $K_s$ magnitudes and in ${\cal
  R}-K_s$ and $J-K_s$ colors.  We have observed $\sim30$--50\% of the
galaxies with NIRSPEC at the bright and red ends of the distributions,
and $\sim10$\% at the faint and blue ends.  Most objects selected by
the BX/BM criteria have $G-{\cal R}<0.5$, as do all but three galaxies
in the NIRSPEC sample; the reddest and bluest galaxies in $G-{\cal R}$
are somewhat underrepresented.

For the purposes of comparisons with other surveys, we note that 10 of
the 87 galaxies for which we have \Ha\ spectra and $JK_s$ photometry
have $J-K_s>2.3$ (the selection criterion for the FIRES survey,
\citealt{flr+03}); this is similar to the $\sim$12\% of UV-selected
galaxies which meet this criterion \citep{res+05}.  18 of the 93
galaxies for which we have $K$ magnitudes have $K_s<20$, the selection
criterion for the K20 survey \citep{cdm+02}; this is a higher fraction
than is found in the full UV-selected sample ($\sim10$\%), because we
intentionally targeted many $K$-bright galaxies \citep{sep+04}.  Four
of the 10 galaxies with $J-K_s>2.3$ also have $K_s<20$.

\subsection{Near-IR Spectra}
\label{sec:irspec}
All of the \Ha\ spectra (with the exceptions of CDFb-BN88, Q0201-B13,
and SSA22a-MD41, which were observed with the ISAAC spectrograph on
the VLT and previously discussed by \citealt{ess+03}) were obtained
using the near-IR spectrograph NIRSPEC \citep{mbb+98} on the Keck II
telescope.  Observing runs were in May 2002, May, July and September
2003, and June and September 2004.  The conditions were generally
photometric with good ($\sim0.5$--0.6\arcsec) seeing, with the
exceptions of the May 2003 and June 2004 runs which suffered from
occasional clouds and seeing up to $\sim1$\arcsec.  \Ha\ falls in the
$K$-band for the redshift range $2.0\lesssim z \lesssim 2.6$, which
includes the vast majority of the current sample; 6 of the 114 objects
have $1.4 \lesssim z \lesssim 1.8$, and were observed in the $H$-band.
We use the $0.76\arcsec \times42$\arcsec\ slit, which, in NIRSPEC's
low resolution mode, provides a resolution of $\sim15$ \AA\ ($\sim200$
\kms; $R\simeq1400$) in the $K$-band.

Our observing procedure is described by \citet{ess+03}.  For a typical
object, we use four 15 minute integrations, although the number of
exposure varied from 2 to 6 for total integrations of 0.5 to 1.5
hours.  We perform blind offsets from a nearby bright star, returning
to the offset star between each integration on the science target to
recenter and dither along the slit.  In most cases we attempt to
observe two galaxies with separation $<25$\arcsec\ simultaneously by
placing them both on the slit; thus the position angle is set by the
positions of the two galaxies.  

\begin{figure*}[htbp]
\plotone{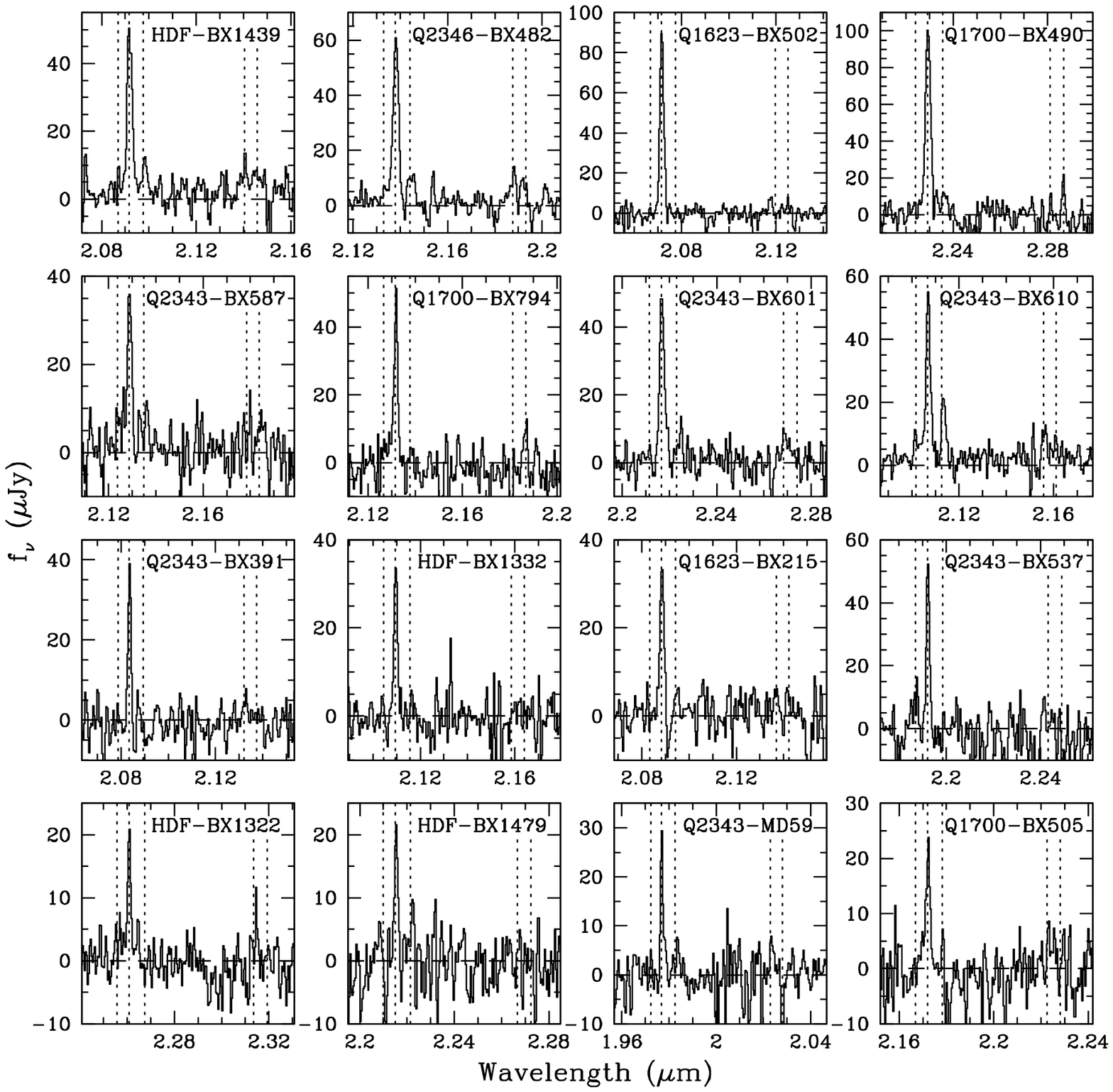}
\caption{Representative examples of the \Ha\ spectra.  Each row shows objects
  drawn from each quartile of \Ha\ flux: the top row contains galaxies
  drawn from the highest quartile in flux, the second row from the
  second highest quartile, the third row from the third highest
  quartile, and the bottom row from the lowest quartile.  The dotted
  lines in each panel show, from left to right, the locations of
  [\ion{N}{2}]$\lambda$6548, \Ha, [\ion{N}{2}]$\lambda$6583,
  [\ion{S}{2}]$\lambda$6716, and [\ion{S}{2}]$\lambda$6732.}
\label{fig:spectra}
\end{figure*}

The spectra were reduced using the standard procedures described by
\citet{ess+03}.  Uncertainties were accounted for via the creation of
a two-dimensional frame of the statistical 1$\sigma$ error appropriate
to each pixel. The last step in the reduction was to extract
one-dimensional spectra of each galaxy; this was done by summing the
pixels containing a signal along the slit. The same aperture was then
used to extract a variance spectrum from the square of the error image
described above; the square root of this is a 1$\sigma$ error
spectrum, which was used to determine the uncertainties in the line
fluxes and widths.  Line centroids, fluxes and widths were measured by
fitting a Gaussian profile to the line using the IRAF task {\it
  splot}, with the base fixed to the average value of the continuum
(we do not detect significant continuum flux for most of the objects
in the sample).  The {\it splot} task provides 1$\sigma$ statistical
uncertainties for each of the fitted quantities, which are determined
by a series of Monte Carlo simulations which perturb the spectrum
according to the 1$\sigma$ uncertainties determined from the error
spectrum described above.  The objects observed are listed in
Table~\ref{tab:obs}, and some representative examples of the spectra
are shown in Figure~\ref{fig:spectra}.  Others are presented by
\citet{ess+03} and \citet{sep+04}.

\subsection{Near-IR Imaging}
$J$-band and $K_s$-band images were obtained with the Wide-field IR Camera
(WIRC; \citealt{wirc}) on the 5-m Palomar Hale telescope, which uses a
Rockwell HgCdTe Hawaii-2 2k$\times$2k array, with a field of view
of $8.5\arcmin \times 8.5\arcmin$ and spatial sampling of
0.249 arcseconds per pixel.  Observations were conducted in June and October
2003 and April, May and August 2004.  Some images were also
taken in June 2004, courtesy of A. Blain and J. Bird.  We used
120-second integrations (four 30-second coadds in $K$, one 120-second
coadd in $J$), typically in a randomized $\sim8$\arcsec\ dither
pattern of 27 exposures.  Total exposure times and 3$\sigma$ image
depths in each field are given in Table~\ref{tab:irobs}.  

Only those images with seeing as good as or better than our optical
images (typically $\sim1.2$\arcsec; 0.85\arcsec\ in the Q1700 field)
were incorporated into the final mosaics.
The images from each dither sequence were reduced, registered and
stacked using IDL scripts customized for WIRC data by K. Bundy
(private communication).  The resulting images were then combined and
registered to our optical images using IRAF tasks.  Flux calibration was
performed with reference to the 15--30 2MASS stars in each field.  We have
also used a 45 min image of Q2346-BX404 and
Q2346-BX405 from the near-IR camera NIRC
\citep{nirc} on the Keck I telescope taken on 1 July 2004 (UT),
courtesy of D. Kaplan and S. Kulkarni; this 
image was reduced similarly, and flux-calibrated by matching the
magnitudes of objects in the field to their magnitudes in the
calibrated Q2346 WIRC image.

Photometry was performed as described by \citet{sse+05}.  Briefly,
galaxies were detected (i.e., found to consist of connecting pixels
with flux at least three times the sky $\sigma$ and area greater than
the seeing disk) in the ${\cal R}$-band catalogs as described in
detail by \citet{sas+03}, with total ${\cal R}$ magnitudes determined
by increasing the isophotal detection aperture by a factor of two in
area. ${\cal R}-K$ and ${\cal R}-J$ colors were determined by applying
the ${\cal R}$-band detection isophotes to the $J$ and $K$ images; the
$J$ and $K$ magnitudes were then determined from the total ${\cal R}$
magnitudes and the ${\cal R}-K$ and ${\cal R}-J$ colors.  We also
applied the detection algorithm to the $K$-band images, and then
applied the resulting isophotes to the $J$-band images.  This
procedure resulted in two sets of $J$ and $K$ magnitudes which
typically differed by $\sim0.3$ mag or less, comparable to the
photometric uncertainties derived from the simulations described
below; these differences arise because the ${\cal R}$-band total
aperture may not include all of the $K$-band light, or vice versa.
For the final $J$ and $K$ photometry, we adopted the more significant
of the two magnitudes for each galaxy.  The additional detection in
$K$ represent a change in method from that used by \citet{sse+05};
this change, as well as a slight change in the photometric zeropoints
from new flux calibrations, means that some magnitudes given here may
be slightly different than those previously published
\citep{sep+04,sse+05}.  We have also trimmed noisy, less well-exposed edge
regions of the $K$-band images somewhat more than previous versions
for cleaner detections in $K$, with the result that some objects with
previously published magnitudes are no longer in the $K$-band sample.
Photometric uncertainties were determined by adding a large number of
fake galaxies of known magnitudes to the images, and detecting them in
both $\cal R$ and $K$ bands to mimic our photometry of actual
galaxies.  This process is described in detail by \citet{sse+05}.  As
noted in \S\ref{sec:dynmass}, the photometric apertures used here are
approximately the same size as those used to determine dynamical
masses from the \Ha\ spectroscopy.

\subsection{Mid-IR Imaging}
Two of our fields have been imaged by the Infrared Array Camera (IRAC)
on the Spitzer Space Telescope.  Images of the Q1700 field were
obtained in October 2003 during the ``In-Orbit Checkout'', and have
been previously discussed in detail \citep{bhf+04,sse+05}.  We also
make use of the fully reduced IRAC mosaics of the GOODS-N field, which
were made public in the first data release of the GOODS Legacy project
(M. Dickinson, PI).  These images are described in more detail by
\citet{res+05}.  For both fields we use the photometric procedure
described by \citet{sse+05}.
 
\section{Model SEDs and Stellar Masses}
\label{sec:smass}

\subsection{Modeling procedure}
We determine best-fit model SEDs and stellar population parameters for
the 93 galaxies for which we have $K$-band magnitudes.  Most of these
(87) also have $J$-band magnitudes, and 35 (in the GOODS-N and Q1700
fields) have been observed with the IRAC camera on the Spitzer Space
Telescope.  For modeling purposes we correct the $K$ magnitudes for
\Ha\ emission; the typical correction is 0.1 mag, but for 4/93 objects
(Q2343-BX418, Q2343-BM133, Q1623-BX455, Q1623-BX502), it is
$\gtrsim0.4$ mag.  We use a modeling procedure identical to that
described in detail by \citet{sse+05}, and review the method briefly
here.

Photometry is fit with the solar metallicity \citet{bc03} models; as
shown by \citet{esp+06}, solar metallicity is a reasonable
approximation for the massive galaxies in the sample, and more typical
galaxies have metallicities only slightly less than solar.  Although
we use the models which employ a \citet{s55} IMF, it is well known
that the steep faint-end slope of this IMF below 1 \msun\ overpredicts
the mass-to-light ratio ($M/L$) and stellar mass by a factor of
$\sim2$ (e.g. \citealt{bmkw03,r05}).  For more accurate comparisons of
stellar and dynamical masses we have therefore converted all stellar
masses and star formation rates to the \citet{c03} IMF by dividing by
1.8 (a change in the faint end of the IMF affects only the inferred
total stellar mass and star formation rate, because low mass stars do
not contribute significantly to the flux in the observed passbands).
Note that previous model SEDs of some of the same objects discussed
here used the Salpeter IMF \citep{sep+04,sse+05}.

The \citet{cab+00} starburst attenuation law is used to account for
dust extinction.  We employ a variety of simple star formation
histories of the form ${\rm SFR} \propto e^{(-t_{\rm sf}/\tau)}$, with
$\tau=10$, 20, 50, 100, 200, 500, 1000, 2000 and 5000 Myr, as well as
$\tau=\infty$ (i.e.\ constant star formation, CSF).  We also briefly
consider more complex two component models, as described below; but as
it is generally difficult to constrain the star formation histories
even with simple models, we use only the single component models for
the main analysis.  For each galaxy we consider a grid of models with
ages ranging from 1 Myr to the age of the Universe at the redshift of
the galaxy, with extinctions ranging from $E(B-V)=0.0$ to
$E(B-V)=0.7$, and with each of the values of $\tau$ listed above.  We
compare the $U_nG{\cal R}JK$+IRAC magnitudes of each model (shifted to
the redshift of the galaxy) with the observed photometry, and
determine the normalization which minimizes the value of $\chi^2$ with
respect to the observed photometry.  Thus the reddening and age are
determined by the model, and the star formation rate and total stellar
mass by the normalization.  In most cases all or most values of $\tau$
give acceptable values of $\chi^2$, while the constant star formation
models give the best agreement with star formation rates from other
indicators (see \citealt{ess+06}).  For these reasons we adopt the CSF
model unless it is a significantly poorer fit than the $\tau$ models.
CSF models are used for 74 of the 93 objects modeled, and $\tau$
models for the remaining 19; as discussed by \citet{sse+05}, the most
massive galaxies, with $M_{\star}\gtrsim 10^{11}$ \msun, are usually
significantly better fit by declining models with $\tau=1$ or 2 Gyr.
\citet{ess+06} discuss this issue further.  The adopted parameters and
values of $\tau$ may be found in Table~\ref{tab:sedfit}.

\begin{figure}[htbp]
\plotone{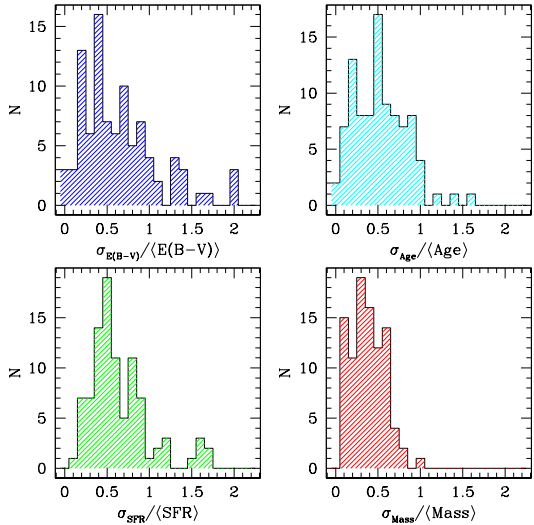}
\caption{The distribution of uncertainties of each parameter fitted by
  our SED modeling code, from the Monte Carlo simulations described in
  the text.  We show histograms of the ratio of the standard deviation
  to the mean of each parameter for all the objects modeled,
  determined iteratively with 3$\sigma$ rejection.  From left to right
  and top to bottom, we show $E(B-V)$, age, star formation rate, and
  stellar mass.  Stellar masses are the best-constrained parameter,
  followed by the age of the star formation episode.}
\label{fig:mchists}
\end{figure}

Parameters derived from this type of modeling are subject to
substantial degeneracies and systematic uncertainties, primarily
because of the difficulty in constraining the star formation history.
These issues have been discussed in detail elsewhere
\citep{pdf01,ssa+01,sse+05}.  Here we simply note that age, reddening,
and the value of $\tau$ suffer most from these degeneracies, and that
the stellar mass is less sensitive to the assumed star formation
history and thus more tightly constrained.  We examine the effects of
the assumed star formation history, and of photometric errors, through
a series of Monte Carlo simulations which we use to determine
uncertainties on the fitted parameters.  The simulations are conducted
as described by \citet{sse+05}; the observed colors of the galaxy are
perturbed by an amount randomly drawn from a Gaussian distribution
with width determined by the photometric errors, and the best-fit SED
is determined as usual for the perturbed colors, including the star
formation history $\tau$ as a free parameter.  We conduct 10,000
trials per object to estimate the uncertainties in each fitted
parameter, and iteratively determine the mean and standard deviation
of each parameter for each object, using 3$\sigma$ rejection to
suppress the effects of outliers.  The resulting mean fractional
uncertainties are $\langle \sigma_x/\langle x\rangle \rangle= 0.7$,
0.5, 0.6 and 0.4 in $E(B-V)$, age, SFR and stellar mass respectively.
The distributions of $\sigma_x/\langle x\rangle$ are shown in
Figure~\ref{fig:mchists}.  The addition of the mid-IR IRAC data
significantly improves our determination of stellar masses; for those
objects with IRAC data (all those in the Q1700 and HDF fields) we find
$\sigma_{M_{\star}}/M_{\star}\sim0.2$, while those without the IRAC
data have $\sigma_{M_{\star}}/M_{\star}\sim0.5$.  The IRAC data also
reduce the uncertainties in the other parameters, though not by as
large an amount.  \citet{sse+05} show that while the IRAC data
significantly reduce the uncertainties in stellar mass, the lack of
these data does not substantially change or bias the inferred mass
itself, especially when the $K$-band magnitude can be corrected for
\Ha\ emission.

\begin{figure*}[htbp]
\epsscale{0.7}
\plotone{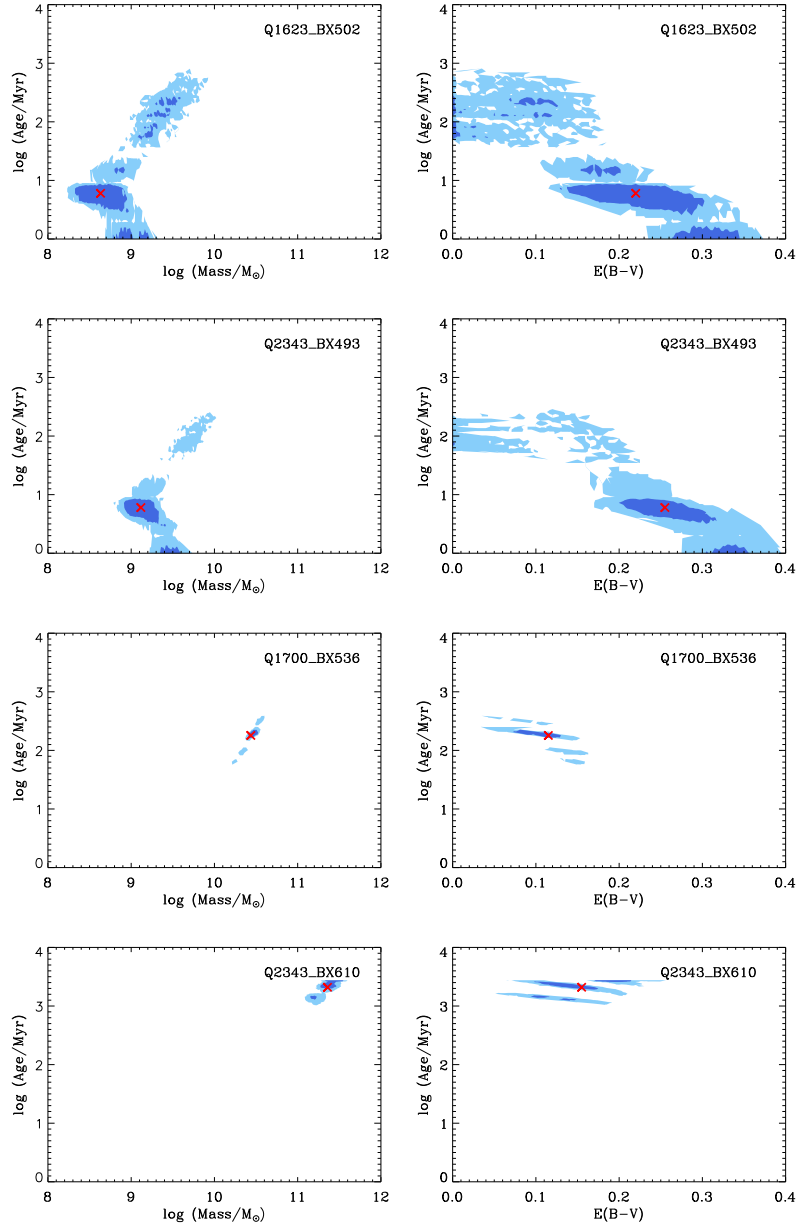}
\caption{Sample confidence intervals for the fitted parameters, for
  four galaxies in the sample.  Contours of stellar mass
  vs.\ age are shown on the left, and $E(B-V)$ vs.\ age on the right.  The dark
  and light blue regions represent 68\% and 95\% confidence intervals
  respectively, and the red $\times$ marks the adopted best fit.  The
  top two objects are characteristic of the young, low stellar mass
  galaxies in the sample; the third row shows a typical galaxy, with
  fit very well-constrained by the addition of mid-IR IRAC data; and
  the bottom panel shows one of the most massive galaxies in the
  sample, which can only be fit by an old stellar population.  Plots
  are shown with the same axes for ease of comparison.}
\label{fig:confint}
\end{figure*}

Confidence interval plots for $M_{\star}$ vs. $E(B-V)$ and age
vs. $E(B-V)$ for selected objects are shown in
Figure~\ref{fig:confint}, where the dark contours represent 68\%
confidence levels and the light 95\%.  The red $\times$ marks our
adopted best fit.  The objects shown are representative of the range
of properties spanned by the NIRSPEC sample.  The first two objects,
Q1623-BX502 and Q2343-BX493, have low stellar masses, young ages and
relatively high values of $E(B-V)$.  The third galaxy, Q1700-BX536,
has properties close to the average of the sample, and also shows how
much smaller the confidence intervals can become with the addition of
the IRAC data.  The last object, Q2343-BX610, is a massive, old galaxy
which can only be fit with $M_{\star}>10^{11}$ \msun\ and age $t_{\rm
  sf}>1$ Gyr.

\begin{figure}[htbp]
\epsscale{1.0}
\plotone{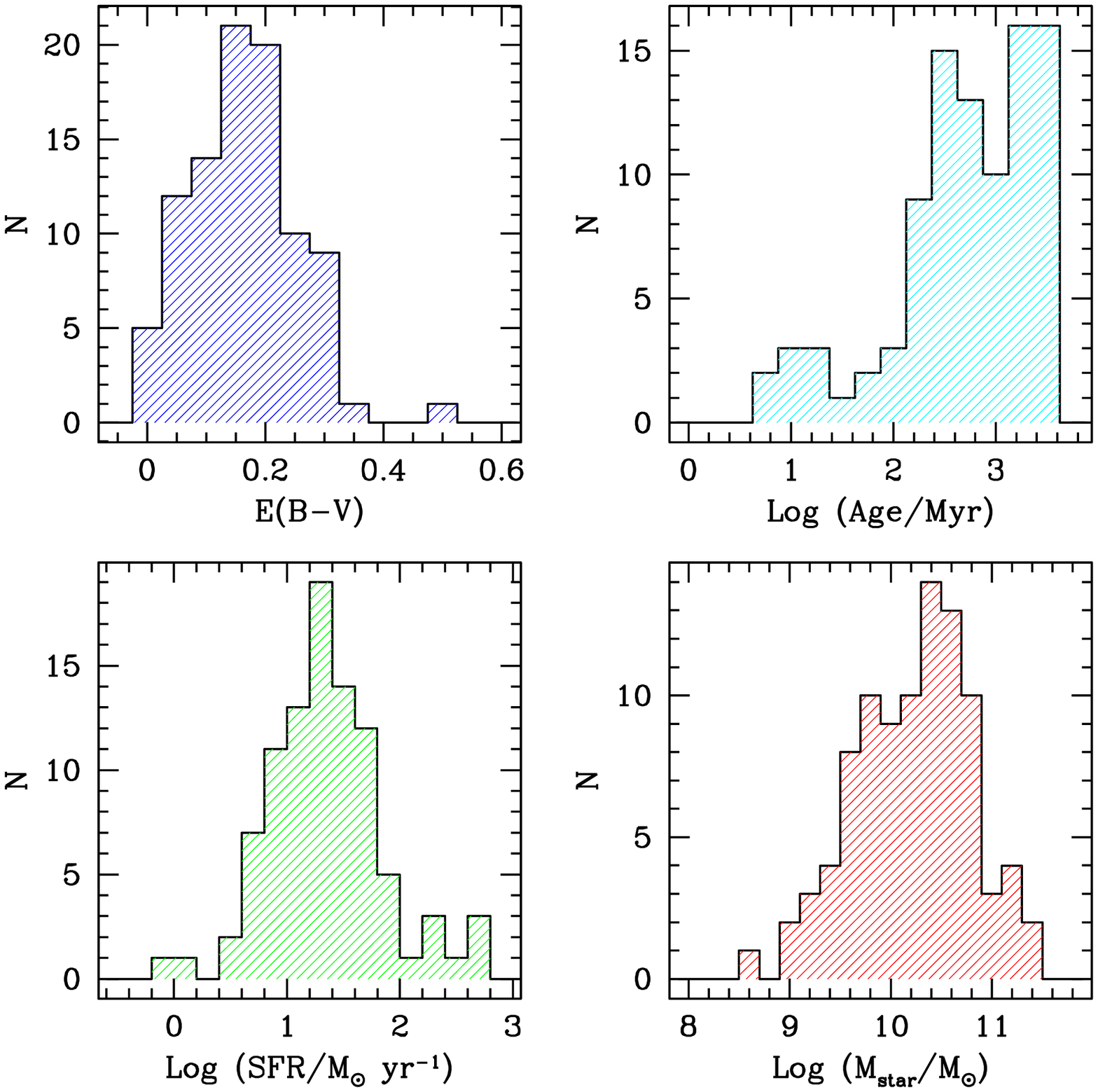}
\caption{Histograms showing the distributions of the results of the
  SED modeling.  From left to right and top to bottom, we show $E(B-V)$,
  age, star formation rate, and stellar mass.  Statistics of the
  distributions are given in the text.}
\label{fig:sedhists}
\end{figure}

\subsection{Model SED results}
The best-fit parameters from the SED modeling are given in
Table~\ref{tab:sedfit}, and their distributions shown in the
histograms in Figure~\ref{fig:sedhists}.  The mean stellar mass is
$3.6\pm0.4 \times 10^{10}$ \msun, and the median is $1.9\pm0.4 \times
10^{10}$ \msun.  The mean age is $1046\pm103$ Myr, and the median age
is $570\pm137$ Myr.  The sample has a mean $E(B-V)$ of $0.16\pm0.01$
and a median of $0.15\pm0.01$.  The mean SFR is $52\pm10$
\msun\ yr$^{-1}$, while the median is $23\pm3$ \msunyr; the difference
between the two reflects the fact that a few objects are best fit with
high SFRs ($>300$ \msunyr).  We discuss the results of the models
further in the following sections and in \citet{ess+06}, where we
compare them with properties determined from the \Ha\ spectra.

\begin{figure}[htbp]
\plotone{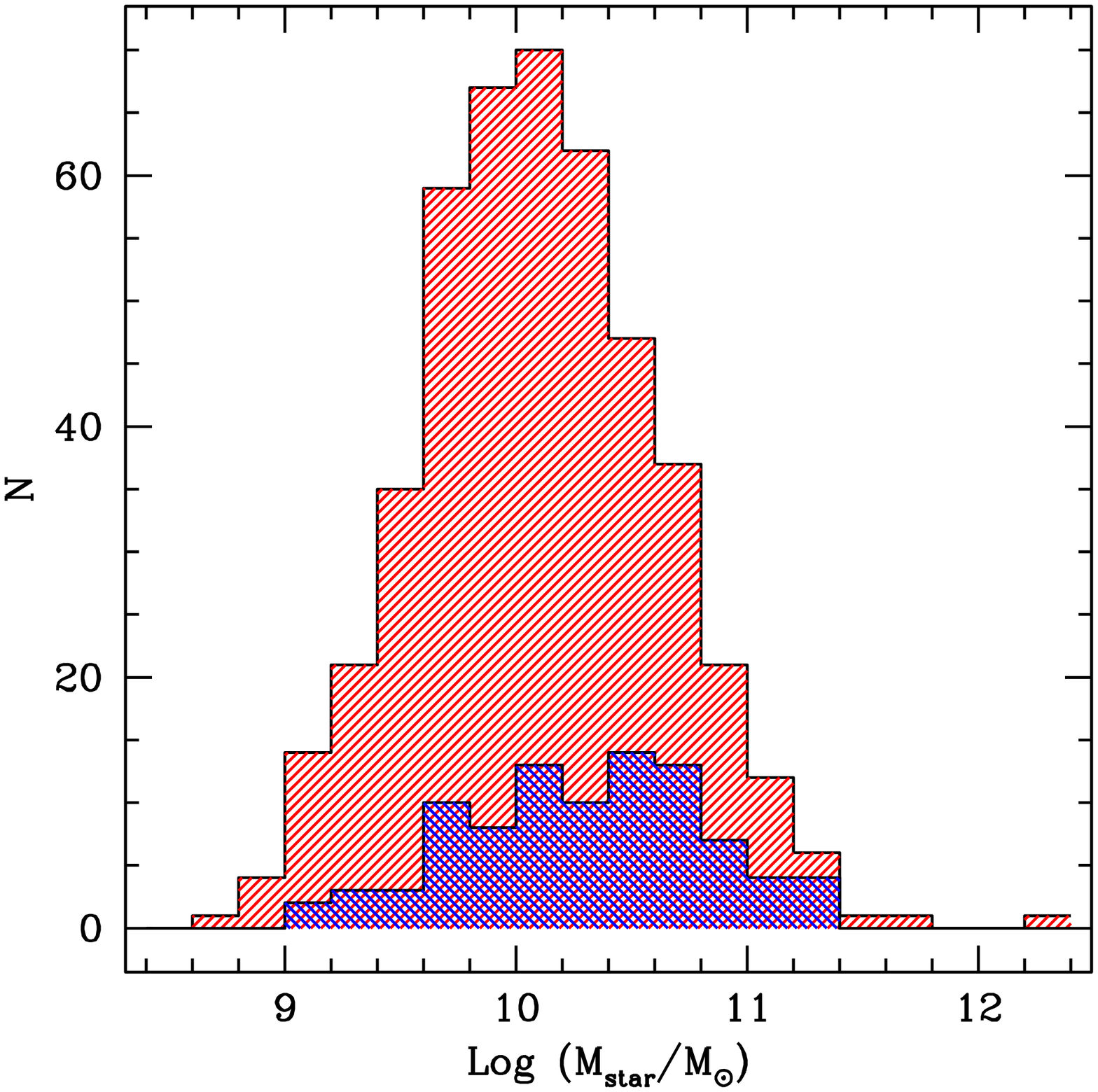}
\caption{A comparison of the stellar masses of the full sample of 461
  UV-selected galaxies (red) and the 93 galaxies for which we have
  model SEDs and \Ha\ spectra (blue).  The mean stellar mass of the
  full sample is $3.0\pm0.5 \times10^{10}$ \msun, while that of the
  \Ha\ subsample is $3.7\pm0.5\times10^{10}$ \msun.  For the purposes
  of this comparison we use constant star formation models for all
  objects, and the $K$ magnitudes of the NIRSPEC sample have not been
  corrected for \Ha\ emission.}
\label{fig:masshist}
\end{figure}

The 93 galaxies whose \Ha\ spectra and SEDs are discussed here are a
subset of a larger sample of 461 objects with model SEDs.  This larger
sample will be discussed in full elsewhere (Erb et al.\ 2006, in
preparation); for the moment we compare the distribution of stellar
masses of the NIRSPEC galaxies with that of the full sample, which is
representative of the UV-selected sample except that it excludes the
$\sim20$\% of objects which are not detected to $K\sim22.5$ (as
discussed by \citet{ess+06}, these are likely to be objects with low
stellar masses and relatively low star formation rates). For the
purposes of comparison with the full sample, we use models for the
NIRSPEC galaxies in which we have not corrected the $K$ magnitude for
\Ha\ emission (since we do not have \Ha\ fluxes for the full sample);
everywhere else in this paper, we use corrected $K$ magnitudes for the
modeling as described above.  We also use CSF models for all galaxies
for this comparison.  The distributions of stellar mass for the two
samples are shown in Figure~\ref{fig:masshist}.  The mean stellar mass
of the subsample for which we have \Ha\ spectra is $3.7\pm0.5 \times
10^{10}$ \msun; this is slightly higher than that of the full
UV-selected sample which has a mean stellar mass of $3.0\pm0.5 \times
10^{10}$ \msun, though the two are consistent within the uncertainties.  A
slightly higher average mass is expected for the NIRSPEC sample, since
some of the galaxies were selected because of their bright $K$
magnitudes or red ${\cal R}-K$ colors.

\subsection{Two component models and maximum stellar masses}
\label{sec:twocomp}
One limitation of this type of modeling is that it is sensitive to
only the current episode of star formation; it is possible for the
light from an older, underlying burst to be completely obscured by
current star formation \citep{pdf01,yde+04,sse+05}.  Although we can
only weakly constrain the star formation histories of individual
galaxies even with simple declining models, more complex two-component
models are a useful tool to assess how much mass could plausibly be
missing.  Two types of two-component models were used to examine this
question.  The first is a maximal mass model, in which we first fit
the $K$-band (and IRAC, when it exists) data with a nearly
instantaneous ($\tau=10$ Myr), maximally old ($z_{\rm form}=1000$)
burst, subtract this model from the observed data, and fit a young
model to the (primarily UV) residuals.  Such models produce stellar
masses $\sim3$ times higher than the single component models on
average, but for galaxies with (single component) $M_{\star} \lesssim
10^{10}$ \msun, the two component total masses are 10--30 times
larger.  Such models are poorer fits to the data, however, and are not
plausible on average because the young components require extreme star
formation rates; the average SFR for the young component of these
models is $\sim900$ \msun\ yr$^{-1}$, $\gtrsim30$ times higher than
the average star formation rate predicted by the \Ha\ and UV emission
(see \citealt{ess+06}).  Nevertheless, we cannot rule out the
possibility that such models may be correct for some small fraction of
the sample.

We also use a more general two-component model, in which we fit a
maximally old burst and a current star formation episode (of any age
and with any of our values of $\tau$) simultaneously, allowing the
total mass to come from any linear combination of the two models.  As
expected given the number of free parameters, such models generally
fit the data as well as or better than the single component models.
They do not, however, usually significantly increase the total mass;
the total two component mass is more than three times larger than the
single component mass in only six cases.  In general, then, the data
do not favor models with large amounts of mass hidden in old bursts.

\section{Velocity Dispersions and Dynamical Masses}
\label{sec:dynmass}
The \Ha\ velocity dispersion $\sigma$ reflects the dynamics of the gas
in the galaxies' potential wells.  Because it requires only a
measurement of the line width, it can be determined for most of the
objects in the sample; it is therefore our most useful kinematic
quantity.  Out of the 114 galaxies with \Ha\ detections, we measure a
velocity dispersion from the \Ha\ line width for 85 objects.  For 11
additional objects we measure an upper limit on the velocity
dispersion because the measured line width is not significantly higher
than the instrumental resolution, as described below.  12 objects have
FWHM less than the instrumental resolution due to noise, and 6 have
been rejected due to significant contamination from night sky lines.
We calculate the velocity dispersion $\sigma=\rm FWHM/2.35$, where
FWHM is the full width at half maximum after subtraction of the
nominal instrumental resolution (15 \AA\ in the K-band) in
quadrature. Uncertainties ($\Delta \sigma_{\rm up}$ and $\Delta
\sigma_{\rm down}$) are determined by calculating $\sigma_{\rm max}$
and $\sigma_{\rm min}$ from the raw line width $W$ and its statistical
uncertainty $\Delta W$; $W+\Delta W$ gives $\sigma_{\rm max}$ and
$W-\Delta W$ gives $\sigma_{\rm min}$.  This results in a larger lower
bound on the error when the line width is close to the instrumental
resolution, and an upper limit on $\sigma$ when $W-\Delta W$ is less
than the instrumental resolution.  In such cases we give one standard
deviation upper limits of $\sigma+\Delta \sigma_{\rm up}$.  As we
discuss further below, we have also measured the spatial extent of the
\Ha\ emission for each object; for 14 objects, the $\rm FWHM_{\Ha}$ is
less than the slit width.  In such cases the effective resolution is
increased by the ratio of the object size to the slit width; we have
made this correction in the calculation of $\sigma$ for these 14
objects.  The correction is usually $<10$\%.

Velocity dispersions are given in column 6 of Table~\ref{tab:kin}.
From the 85 measurements we find $\langle \sigma \rangle = 120\pm9$
\kms\ ($108\pm5$ \kms\ with AGN removed), with a standard deviation of
86 \kms, while counting the upper limits as detections gives $\langle
\sigma \rangle = 112\pm9$ \kms\ ($101\pm5$ \kms\ without AGN) with a
standard deviation of 85 \kms.  The observed line widths could be
caused by random motions, rotation, merging, or, most likely, some
combination of all of these.  Galactic outflow speeds (as measured by
the velocity offsets between \Ha, \lya, and the rest-frame UV
interstellar absorption lines) show no correlation with the line
widths, and the AGN fraction of $\sim4$\% means that broadening by AGN
is usually not significant.  Dynamical masses can be calculated from
the line widths via the relation
\begin{equation}
M_{\rm dyn}=C \sigma^2 r/G
\label{eq:dynmass}
\end{equation}
The factor $C$ depends on the galaxy's mass distribution and velocity
field, and may range from $C\leq1$ to $C\geq5$.  The value of $C$ depends on
the mass density profile, the velocity anisotropy and relative
contributions to $\sigma$ from random motions or rotation, the
assumption of a spherical or disk-like system, and possible
differences in distribution between the total mass and that of tracer
particles used to measure it.  Obviously the definition of the radius
$r$ is also crucial.  Most of these factors are unknown for the
current sample.

Under the assumption that a disk geometry is appropriate for gas-rich
objects (see \S\ref{sec:gasmass}), we begin with the relationship $M_{\rm
  dyn}=v_{\rm true}^2r/G$.  We incorporate an
 average inclination correction $\langle v_{\rm
  true} \rangle = \pi/2\, \langle v_{\rm obs} \rangle$, and 
for $v_{\rm obs}$ use the observed velocity
half-width ($v_{\rm obs}={\rm FWHM}/2= 2.35\sigma/2$).  Combining the
constants and writing the equation in terms of $\sigma$ then gives
$C\simeq3.4$.  For the galaxy size
$r_{\Ha}$ we use half of the spatial extent of the \Ha\ emission,
after deconvolution of the seeing.  The galaxies are spatially
resolved in almost all cases; for those few that are not, we use the
smallest size measured under the same seeing conditions as an upper
limit.  We find a mean and standard deviation $\langle r_{\Ha} \rangle
= 0.7\pm0.3$\arcsec\ ($\sim6$ kpc), approximately the same as the
typical isophotal radius used for photometric measurements.  Because
it is difficult to continuously monitor the seeing while observing
with NIRSPEC, some uncertainty is introduced by our inexact knowledge
of the seeing during each observation. This uncertainty is typically
0.1--0.2\arcsec, but for those objects that are near the resolution
limit it may be much larger.  Because of this issue, the sizes are
uncertain by $\sim30$\%.  Note that most previous calculations of
dynamical mass at high redshift have used $C=5$ (appropriate for a
uniform sphere;
e.g.\ \citealt{pss+01,ess+03,ess+04,sep+04,vff+04,ssc+04}).

Values of $r_{\Ha}$ are given in column 5 of Table~\ref{tab:kin}, and
dynamical masses in column 8.  Using the 85 galaxies with
well-determined $\sigma$, we find a mean dynamical mass $\langle
M_{\rm dyn} \rangle =1.2\pm0.4\times10^{11}$ \msun.  The two largest
dynamical masses in the sample are $M_{\rm dyn}=5.7\times10^{11}$
\msun\ (Q1700-MD174) and $M_{\rm dyn}=3.8\times10^{12}$
\msun\ (Q1700-MD94); both of these objects are AGN, and their broad
\Ha\ lines are therefore not a reflection of the gravitational
potential.  Neglecting these two objects, and the other AGN which have
$M_{\rm dyn}\sim6-9\times10^{10}$ \msun\ (HDF-BMZ1156, Q1623-BX151 and
Q1623-BX663), results in $\langle M_{\rm dyn}\rangle
=6.9\pm0.6\times10^{10}$ \msun, with a standard deviation of
$5.8\times10^{10}$ \msun.  Approximately 25\% of the non-AGN have
dynamical masses $M_{\rm dyn} > 10^{11}$ \msun, with Q1623-BX376
($M_{\rm dyn}=3.7\times10^{11}$ \msun) the largest.

Uncertainties in the dynamical masses are probably dominated by the
constant $C$, which may vary by a factor of a few.  Our line widths
may also occasionally suffer from contamination by night sky lines; we
have removed the most obvious cases, but repeated observations with
varying signal-to-noise ratios (S/N) show that this can occasionally
affect the line profile by up to $\sim30$\%.  Complex spatial and
kinematic structure can also affect the observed velocity dispersion,
as discussed previously \citep{ess+03} and recently emphasized by
\citet{cam05}.  The \Ha\ emission also may not trace the full
potential, especially in low S/N spectra of high redshift objects.
Even in some local starburst galaxies the \Ha\ emission does not fully
sample the rotation curve \citep{lh96}.  Indeed, the mean dynamical
mass we derive here is $\sim15$ times lower than the typical halo mass
as inferred from the galaxies' clustering properties \citep{asp+05}.

\subsection{Comparisons with Stellar Masses}
\label{sec:masscomp}

\begin{figure*}[htbp]
\plotone{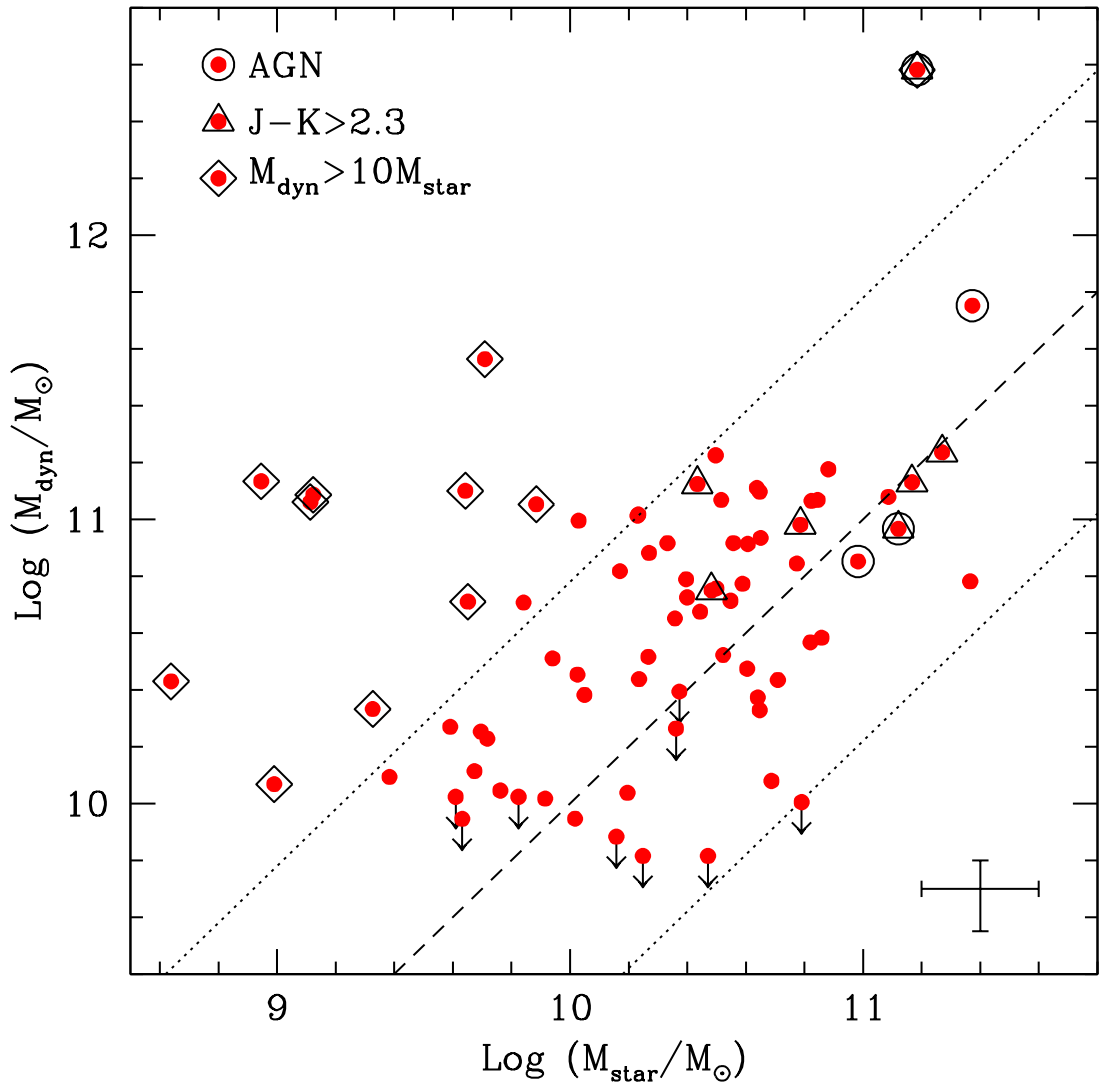}
\caption{Dynamical vs. stellar mass for the 68 galaxies with
  measurements of both quantities.  Excluding AGN, we find $\langle
  M_{\star} \rangle =3.5\pm0.5\times10^{10}$ \msun, while the mean
  dynamical mass is $\langle M_{\rm dyn} \rangle
  =7.1\pm0.7\times10^{10}$ \msun.  The dashed line indicates equal
  masses.  The error bars in the lower right corner show typical
  uncertainties, which for dynamical masses include only measurement errors
  in $r_{\Ha}$ and $\sigma$.  Uncertainties in dynamical masses are
  dominated by the unknown factors in the mass model as discussed in
  the text; incorporating these uncertainties, we estimate approximate
  combined uncertainties of $\sim0.8$ dex, as shown by the dotted lines.
  Galaxies with $M_{\rm dyn}/M_{\star}>10$ are marked with open
  diamonds, and we discuss possible reasons for their discrepancy in
  the text.}
\label{fig:masscomp}
\end{figure*}

Turning next to a comparison of the galaxies' dynamical and stellar
masses, in Figure~\ref{fig:masscomp} we plot dynamical mass
vs.\ stellar mass for the 68 galaxies for which we have determined
both.  The dashed line shows equal masses, while the dotted lines on
either side indicate approximate uncertainties of $\pm0.8$ dex (a
$\sim50$\% uncertainty in both masses, added in quadrature; this
includes systematic uncertainties in the dynamical mass model as well
as the measurement uncertainties shown by the error bars in the lower
right corner).  Excluding AGN, the mean dynamical mass is $\sim2$
times higher than the mean stellar mass.  Significant correlation
between stellar and dynamical mass is observed in the local universe
(e.g.\ \citealt{be00}), but such a correlation will exist only if the
stellar mass makes up a relatively constant fraction of the dynamical
mass over the full range of stellar masses.  If low stellar mass
galaxies have large gas fractions, as is seen in local galaxies
\citep{md97,bd00}, they may not have correspondingly small dynamical
masses.  With these caveats in mind, we perform Spearman and Kendall
$\tau$ correlation tests (neglecting the AGN) and find a probability
$P=0.01$ that the masses are uncorrelated, for a significance of the
correlation of $2.5\sigma$.  This is consistent with our qualitative
expectations from Figure~\ref{fig:masscomp}, where most of the points
lie well within the dotted lines marking a factor of $\sim6$
difference between the masses (though we see a significant number of
outliers with $M_{\rm dyn}\gg M_{\star}$, to be discussed further
below).

Approximately 20\% of the galaxies in the sample have
$M_{\star}>M_{\rm dyn}$, clearly an unphysical situation.  The
discrepancies are usually a factor of 2--3 and always less than a
factor of $\sim6$, which can be easily explained by the factor of a
few uncertainty in both masses.  In any case, because the
\Ha\ linewidths may not sample the full potential it would not be a
surprise to find some objects with $M_{\star}>M_{\rm dyn}$.

It is potentially more interesting to consider the small but
significant fraction of the objects for which $M_{\rm dyn}\gg
M_{\star}$; we focus on the $\sim15$\% of the sample (11/68) for which
$M_{\rm dyn}/M_{\star}>10$ (one of these is the AGN Q1700-MD94, which
has a dynamical mass of $3.8\times10^{12}$ \msun\ due to its very
broad \Ha\ line; this is clearly a problematic estimate of dynamical
mass, and we exclude this object in the following discussion).  These
objects are marked with open diamonds in Figure~\ref{fig:masscomp},
and in many of the subsequent figures as well.  The large differences
between stellar and dynamical masses suggest the following
possibilities: 1) these are galaxies that have recently begun forming
stars, and thus have small stellar masses and large gas fractions; or
2) we significantly underestimate the stellar mass for up to
$\sim15$\% of the sample, presumably because of an undetected old
stellar population.  One should also consider the possibility that
contamination from winds, shock ionization or AGN causes an
overestimate of the dynamical masses of $\sim15$\% of the sample.
There is no evidence for this third possibility, however; except in
the few cases of known AGN, none of the objects show evidence of
ionization from any source other than star formation, either from
their UV spectra or from high \Ntwo/\Ha\ ratios.  The objects with
$M_{\rm dyn}\gg M_{\star}$ have low stellar masses, and as
\citet{esp+06} show in the context of estimating metallicity as a
function of stellar mass, \Ntwo\ is not detected even in a composite
spectrum of 15 such objects.

\begin{figure*}[htbp]
\plotone{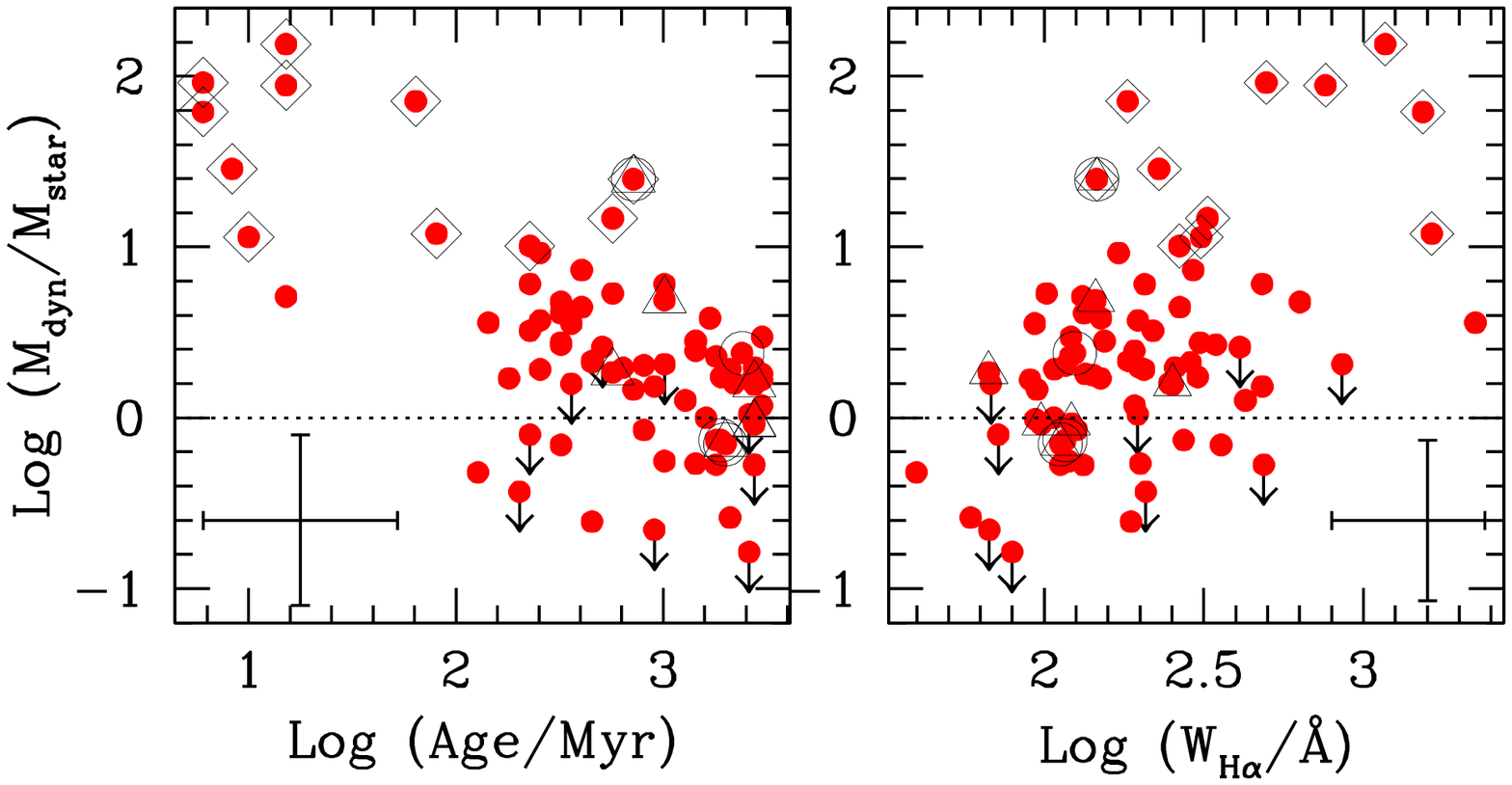}
\caption{Left:  The mass ratio $M_{\rm dyn}/M_{\star}$ vs.\ the
  best-fit age from SED modeling, and right, $M_{\rm dyn}/M_{\star}$
  vs.\ \Ha\ equivalent width $W_{\Ha}$.  Galaxies with high  $M_{\rm
    dyn}/M_{\star}$ have young ages, and also tend to have high
  equivalent widths.  Typical uncertainties are shown by the error
  bars in the corner of each plot.  Symbols are as in Figure~\ref{fig:masscomp}.}
\label{fig:ewage}
\end{figure*}

The best fit ages and \Ha\ equivalent widths (\citet{ess+06} discuss
the equivalent widths more fully ) of the set of galaxies with $M_{\rm
  dyn}/M_{\star}>10$ favor the first possibility.  As shown in
Figure~\ref{fig:ewage}, there is a strong correlation between $M_{\rm
  dyn}/M_{\star}$ and age, and, with somewhat more scatter, between
$M_{\rm dyn}/M_{\star}$ and $W_{\Ha}$ (the correlations have 5 and 4
$\sigma$ significances respectively).  Objects with $M_{\rm dyn} \gg
M_{\star}$ have young best fit ages, and tend to have high
\Ha\ equivalent widths as well, with $200 \lesssim W_{\Ha} \lesssim
1300$ \AA.  Note that these are not correlations of independent
quantities; age and stellar mass are correlated by definition in the
constant star formation models we use for the majority of the sample,
and the equivalent width depends on the $K_s$ magnitude, which is also
closely related to stellar mass.  This, combined with the fact that we
see less dispersion in dynamical mass than stellar mass, accounts for
the correlations in Figure~\ref{fig:ewage}.

It is important to assess whether or not the $M_{\rm dyn} \gg
M_{\star}$ galaxies are really significantly younger than the rest of
the sample, given the well-known degeneracies between age and
extinction in SED modeling and the considerable uncertainties in the
age of a typical object.  Confidence intervals for the age--$E(B-V)$
fits for two of the $M_{\rm dyn} \gg M_{\star}$ galaxies (Q1623-BX502
and Q2343-BX493) are shown in Figure~\ref{fig:confint}.  These are
representative of this set of objects.  Young ages (and
correspondingly high values of $E(B-V)$) are clearly strongly favored,
but there is a tail of acceptable solutions extending to higher ages.
This set of objects is unique in favoring such young ages.  To further
test the significance of the young ages, we have divided the sample
into four quartiles by $M_{\rm dyn}/M_{\star}$ (with 16, 17, 17 and 17
galaxies in the quartiles), and performed K-S tests on the age
distribution of each quartile, making use of the Monte Carlo
simulations described in \S\ref{sec:smass}.  These simulations perturb
the colors of each galaxy according to its photometric errors, and
compute the best-fit model for the perturbed colors.  We conducted
10,000 trials for each galaxy, and generated lists of the best-fit
ages for all trials in each quartile (for 160,000 or 170,000 trial
ages in each quartile).  A two sample K-S test on all pairs of the
four quartiles finds that the probability $P\simeq0$ that the ages in
the quartile with the highest $M_{\rm dyn}/M_{\star}$ are drawn from
the same distribution as the remaining quartiles.  This test suggests
that even though the ages of individual galaxies may not be well
constrained, the younger average age of the subsample with $M_{\rm
  dyn}\gg M_{\star}$ is highly significant.

The best-fit ages from the SED fitting represent the age of the
current star formation episode, however, not necessarily the galaxy as
a whole, which brings us to possibility 2) above: is there a
significant underlying population of old stars in these objects which
causes us to underestimate the stellar mass?  We address this question
through the two-component models described in \S\ref{sec:twocomp},
which fit the SED with the superposition of a young burst and
maximally old population to estimate the maximum stellar mass.  The
maximal mass models, which fit all the flux redward of the $K$-band
with an old burst, and then fit a young model to the UV residuals,
result in stellar masses $\sim$5--30 times larger for those galaxies
with $M_{\rm dyn}/M_{\star}>10$.  If we use the maximal mass models
for these objects rather than the single-component models in the
comparison with dynamical mass, we find $\langle M_{\rm dyn}/M_{\star}
\rangle = 3$, as compared to $\langle M_{\rm dyn}/M_{\star} \rangle =
54$ using the single component models.  However, as described in
\S\ref{sec:twocomp}, these models are poorer fits and result in
implausibly high average SFRs.  A further difficulty is that the mean
age of the young component in these models is 3.5 Myr, approximately
1/20 the typical dynamical timescale for these galaxies.  The more
general two-component modeling, in which the total mass comes from a
varying linear combination of a maximally old burst and an episode of
current star formation, does not favor such extreme models; for the
$M_{\rm dyn} \gg M_{\star}$ galaxies in question, the use of these
more general models increases the average stellar mass by a factor of
three.

More powerful constraints on the ages and masses of the $M_{\rm dyn}
\gg M_{\star}$ objects come from the spectra of the galaxies
themselves.  Using composite spectra of the current sample,
\citet{esp+06} show that there is a strong trend between metallicity
and stellar mass in the \ztwo\ sample, such that the lowest stellar
mass galaxies also have low metallicities.  These objects likely have
large gas fractions, accounting for their relatively low metal content
(we discuss the derivations of the gas fractions further in
\S\ref{sec:gasmass}). These results strongly favor the hypothesis that
the large dynamical masses of these objects are due to a large gas
mass rather than a significant population of old stars.  From both
considerations of dynamical mass and baryonic mass (gas and stars,
\S\ref{sec:gasmass}), it seems clear that objects with a best-fit {\it
  stellar} mass of $M_{\star}\lesssim10^9$ \msun\ are not in fact
significantly less massive than the rest of the sample.

\subsection{Velocity Dispersion and Optical Luminosity}
\label{sec:siglum}
The correlation between the luminosity of elliptical galaxies and
their velocity dispersion $\sigma$ (the Faber--Jackson relation,
$L\propto \sigma^4$; \citealt{fj76}) is well-established in the local
universe.  Many proposals for the origin of this correlation rely on
self-regulating feedback processes; for example, \citet{mqt05} propose
that momentum-driven winds from supernovae and radiation pressure
regulate the luminosity from star formation at $L\propto \sigma^4$.
Other theories (e.g.\ \citealt{bn05,rhc+05} and references
therein) use feedback from accretion onto a supermassive black hole to
produce a correlation between luminosity or stellar mass and velocity
dispersion as a byproduct of the relationship between black hole mass
and velocity dispersion, $M_{\rm BH}\propto \sigma^4$
\citep{fm00,gbb+00,tgb+02}.  In the local universe, these correlations
exist in relatively homogeneous, quiescent galaxies; with the present
data, we can test the relationships between luminosity, stellar mass
and velocity dispersion at a time in which the galaxies are building
up much of their stellar mass.

Previous efforts to find such relations at high redshift have
generally suffered from small sample sizes and resulted in weak or
absent correlations with large amounts of scatter
\citep{pss+01,vff+04,ssc+04}.  We revisit the relation between
$\sigma$ and optical luminosity with a sample of 77 \ztwo\ UV-selected
galaxies, augmented by 21 SCUBA galaxies presented by \citet{ssc+04},
four of the Distant Red Galaxies (DRGs) discussed by \citet{vff+04},
and 16 Lyman-break galaxies at $z\sim3$ (\citealt{pss+01} give
velocity dispersions for nine galaxies observed in $K$ and (usually)
$J$ by \citealt{ssa+01}, and we have subsequently observed seven
more).  Our total sample thus consists of 118 galaxies with a mean
spectroscopic redshift $\langle z \rangle =2.38$ and $\sigma_{\rm
  z}=0.37$.

\begin{figure*}[htbp]
\plotone{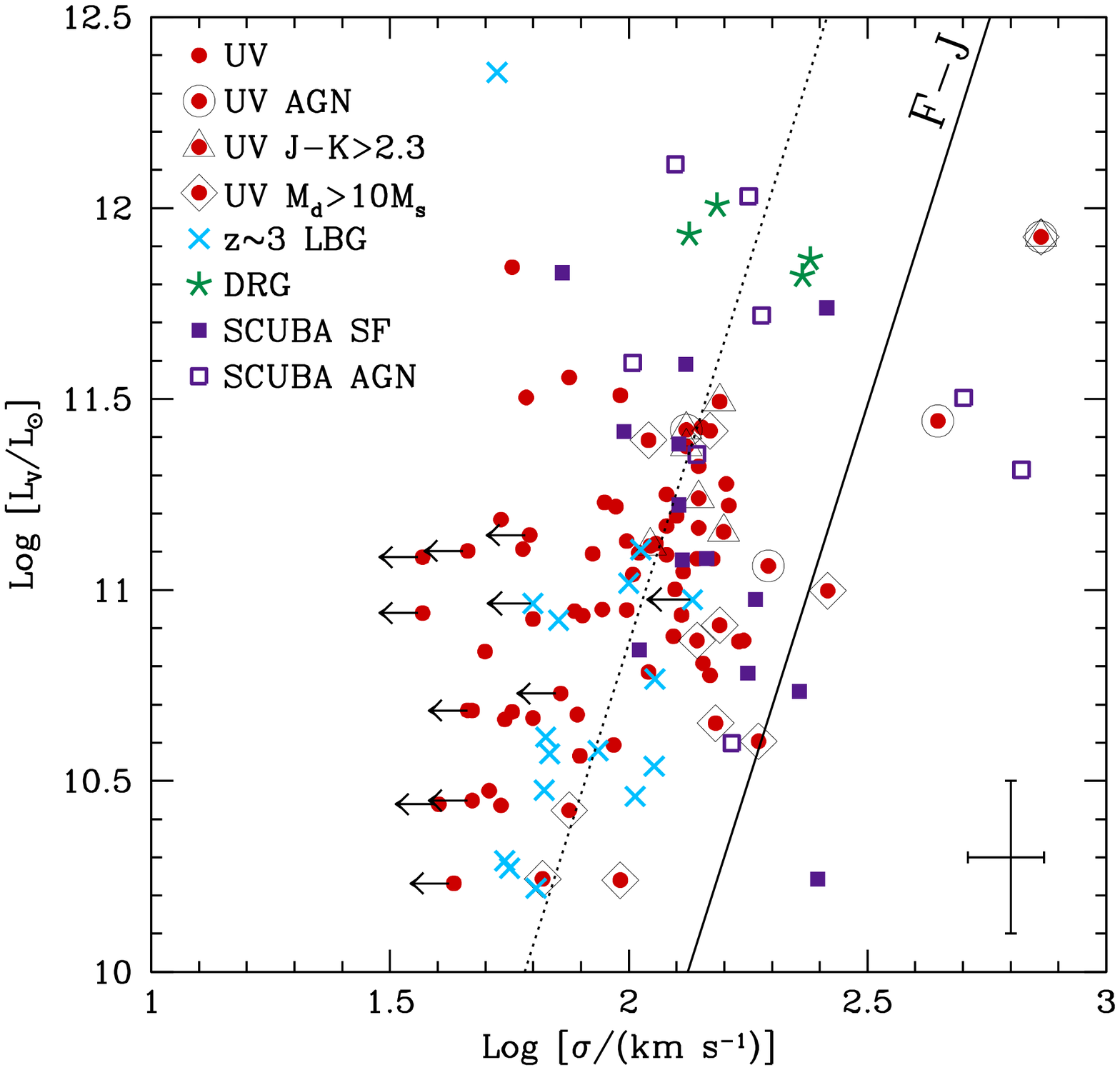}
\caption{Velocity dispersion vs. extinction-corrected
  rest-frame V-band luminosity. Red circles show the current sample,
  blue $\times$s are Lyman break galaxies at \zthree, green stars are
  the DRGs presented by \citet{vff+04}, and the purple squares
  represent the SCUBA galaxies of \citet[][filled squares are those
    classified as star-forming galaxies, and open squares represent
    AGN]{ssc+04}.  The solid line shows the local Faber-Jackson
  relation, and the dotted line shows the local relation shifted upward by
  1.35 dex in order to pass through the median of the data.  Typical
  uncertainties are shown by the error bars in the lower right corner.}
\label{fig:siglum}
\end{figure*}

Figure~\ref{fig:siglum} shows the extinction-corrected rest-frame
V-band luminosity plotted against the velocity dispersion $\sigma$;
the luminosity is determined by multiplying the best-fit SED with the
redshifted $V$ transmission curve, and the extinction correction
employs the \citet{cab+00} extinction law and the best-fit values of
$E(B-V)$ from the SED modeling.  We have used photometry from
\citet{ssa+01} to re-model the SEDs of the $z\sim3$ LBGs, for
consistency with the \ztwo\ sample.  We also use the
extinction-corrected magnitudes $M_V$ from \citet{ssc+04}, and use the
values of $A_V$ determined from SED modeling by \citet{vff+04} to
correct their absolute $V$ magnitudes.  The solid line shows the local
$i$-band\footnote{The slope of the Faber-Jackson relation is the same
  within the uncertainties in all bands, while the zeropoint decreases
  slightly at shorter wavelengths.} Faber-Jackson relation derived
from 9000 Sloan Digital Sky Survey (SDSS) galaxies by \citet{bsa+03}.
Assuming no change in slope, the local relation is offset from the
median of the data by 1.35 dex in luminosity, as shown by the dotted
line.  The data are strongly correlated; the Spearman and Kendall
$\tau$ tests give a probability $P=2\times10^{-5}$ that the data are
uncorrelated, for a significance of $4.1\sigma$.  An attempt to fit
both the slope and the zeropint of the correlation shows that the
slope is not well-constrained, and depends strongly on the fitting
procedure and the relative uncertainties assumed for $L$ and $\sigma$;
however, it is not inconsistent with $\sim4$, the observed slope of
the local Faber-Jackson relation.

\begin{figure}[htbp]
\plotone{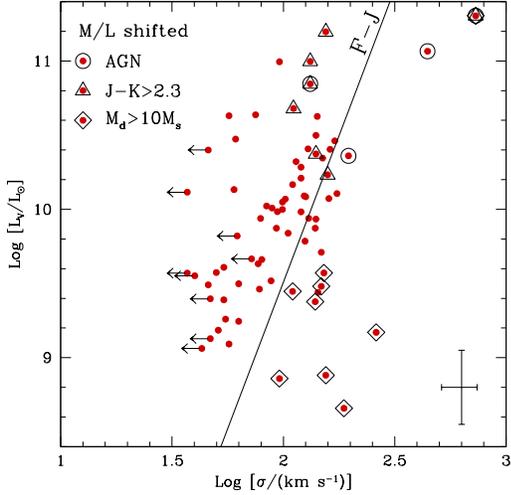}
\caption{The V-band luminosities of the \ztwo\ sample shifted by the
  difference between the observed and local mass-to-light ratios as
  described in the text, plotted vs.\ velocity dispersion $\sigma$.
  The solid line shows the local Faber-Jackson relation.  Typical
  uncertainties are shown by the error bars in the lower right corner.}
\label{fig:mlshift}
\end{figure}

Though there is strong evidence for a correlation between $L$ and
$\sigma$ at $z>2$, the scatter is considerable and too large to be
accounted for by observational errors.  $\sigma$ varies by a factor of
$\sim3$--4 at a given luminosity, and for most values of $\sigma$ we
see at least an order of magnitude variation in $L$.  Some insight
into the sources of the scatter may be obtained by accounting for the
large variation in the mass-to-light ratio $M/L$ observed at \ztwo.
As shown by \citet{sse+05} for a similar sample of UV-selected
galaxies, the rest-frame optical $M/L$ varies by a factor of $\sim80$,
with the highest values of $M/L$ approaching the typical values seen
in local galaxies; we find very similar results for the current
sample.  In Figure~\ref{fig:mlshift} we account for this variation by
decreasing the luminosity of each galaxy in the \ztwo\ sample by the
difference between its value of $M/L$ and the typical $(M/L)_0$ of an
elliptical galaxy in the local universe.  We use $(M/L)_0=3$, as found
by \citet{pss+04} for elliptical galaxies in the SDSS; the average
shift in luminosity is a factor of 10, and the shift ranges from a
factor of 2 to a factor of 120.  The figure approximates what would be
observed at $z\sim0$ if each galaxy were fixed at its current stellar
mass and velocity dispersion and faded to match present-day $M/L$
values.  This prediction is particuarly relevant for the current
sample because the clustering properties of the \ztwo\ galaxies
indicate that they are the likely progenitors of elliptical galaxies
like those seen in the SDSS \citep{asp+05}.  The solid line again
shows the local Faber-Jackson relation, which is a reasonable fit to
many of the points; however, the large remaining scatter and
significant number of outliers are consistent with the expectation
(from the high gas fractions discussed in \S\ref{sec:gasmass} and the
high SFRs found by \citealt{ess+06}) that most of the galaxies are
still actively building up stellar mass.

\begin{figure}[htbp]
\plotone{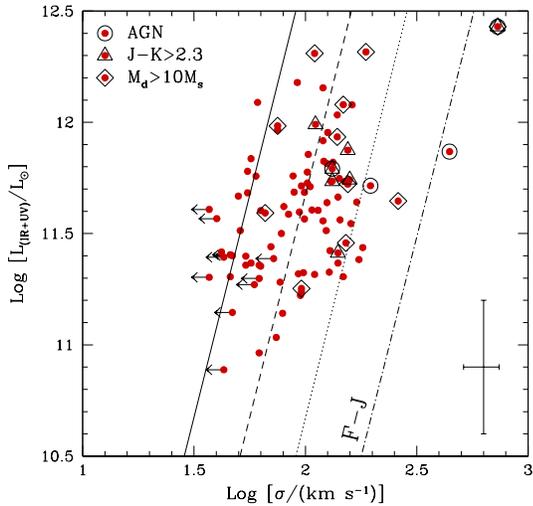}
\caption{ Bolometric luminosity $L_{\rm (IR+UV)}$ calculated from the
  extinction-corrected \Ha\ star formation rates as described in the
  text vs.\ velocity dispersion, compared with the model of
  \citet{mqt05} for gas fractions $f_g=1.0$, 0.1, and 0.01 (solid,
  dashed and dotted lines respectively).  These three lines also
  indicate reasonable uncertainties in the model.  The local
  Faber-Jackson relation is shown by the dot-dash line at
  right. Typical uncertainties are shown by the error bars in the
  lower right corner.}
\label{fig:lbolsig}
\end{figure}

For the current data, the most relevant prediction of the relationship
between $L$ and $\sigma$ at high redshift is that of \citet{mqt05},
who propose a limiting, Eddington-like luminosity for starbursts,
$L_M\simeq (4f_gc/G)\sigma^4$, where $f_g$ refers to the gas fraction.
In this model, momentum-driven winds powered by radiation pressure and
supernovae would expel much of the gas from the galaxy at luminosities
above $L_M$.  At $L_M$, feedback moderates the SFR, and luminosity
does not increase sigificantly beyond $L_M$.  $L_M$ represents the
bolometric luminosity from star formation, so the plot of $L_V$
vs.\ $\sigma$ discussed above is not the most appropriate test of this
prediction.  Instead we estimate the bolometric luminosities $L_{\rm
  (IR+UV)}$ from the extinction-corrected \Ha\ SFRs determined by
\citet{ess+06}.  We invert the prescription of \citet{bpw+05}, who
relate the star formation rate to the bolometric luminosity, to
produce the following relation:
\begin{equation}
L_{(\rm IR+UV)}/\rm L_{\odot}=1.0\times10^{10} \:\: SFR/\msunyr.
\end{equation}
Here $L_{\rm IR}$ represents the total IR luminosity, while $L_{\rm
  UV}$ is the light from unobscured stars.  Before using this
expression we have converted our SFRs to the \citet{k01} IMF used by
\citet{bpw+05}.  We find a mean $\langle L_{\rm (IR+UV)} \rangle =
4.6\times10^{11}$ L$_{\odot}$, slightly higher than the average
bolometric luminosity $\langle L_{\rm (IR+UV)} \rangle =
2.3\times10^{11}$ L$_{\odot}$ estimated for a different sample of
UV-selected \ztwo\ galaxies from 24 \micron\ observations by
\citet{rsf+06}.  The two determinations of $L_{\rm (IR+UV)}$ can be
compared directly for 11 galaxies common to both samples, as shown in
Figure 5 of \citet{rsf+06}; the two measurements agree within the
uncertainties, with $\langle \log L_{\rm bol}^{\Ha} \rangle
=11.46\pm0.27$ and  $\langle \log L_{\rm bol}^{\rm IR+UV} \rangle
=11.59\pm0.20$.  

Figure~\ref{fig:lbolsig} shows the inferred bolometric luminosity
$L_{\rm (IR+UV)}$ plotted against velocity dispersion, along with
lines indicating $L_M$ for gas fractions $f_g=1.0$, 0.1, and 0.01
(from left to right, solid, dashed and dotted lines).  We also show
the local Faber-Jackson relation.  Although the three $L_M$ lines
indicate differences in the gas fraction, they also represent the
uncertainty in $L_M$.  They are not sufficient to discriminate between
gas fractions; the objects with the highest inferred gas fractions
(the $M_{\rm dyn}\gg M_{\star}$ galaxies marked with open diamonds) do
not fall closest to the $f_g=1.0$ line.  Within these uncertainties,
the \citet{mqt05} model appears to be a reasonable description of the
data; the points are strongly correlated, with 4.0$\sigma$
significance and $P=7\times10^{-5}$, and the $L_M$ lines bracket most
of them.

\begin{figure}[htbp]
\plotone{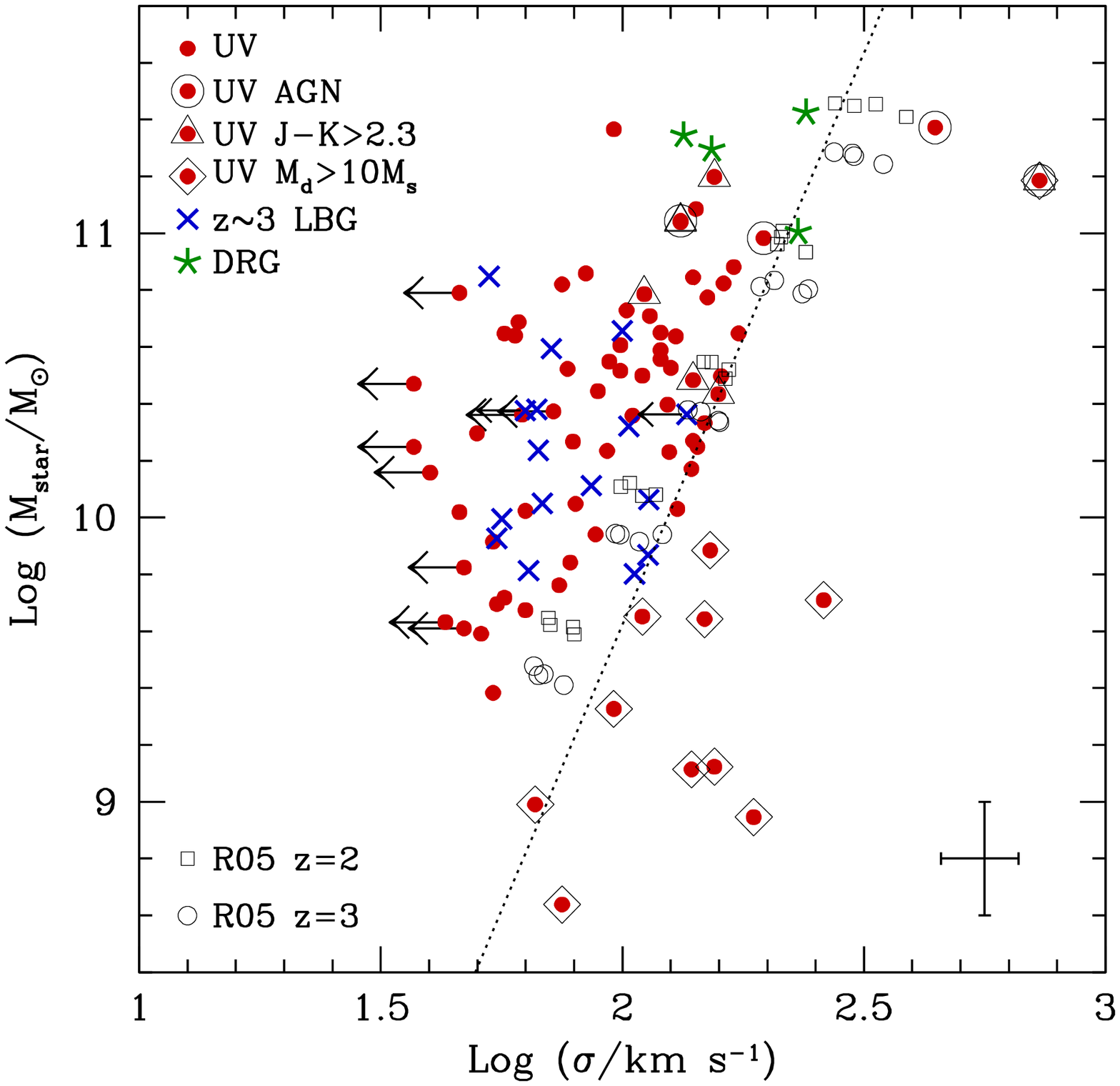}
\caption{Stellar mass vs.\ velocity dispersion.  The dotted line shows
  the local relation between bulge stellar mass and velocity
  dispersion, constructed by combining the $M_{\rm bulge}$--$M_{BH}$
  relation of \citet{mh03} and the $M_{BH}$--$\sigma$ relation of
  \citet{tgb+02}.  Open squares and circles show the results of the
  simulations of \citet{rhc+05}, which test the evolution of the
  $M_{BH}-\sigma$ relation with redshift.  Typical uncertainties are
  shown by the error bars in the lower right corner.}
\label{fig:sigmass}
\end{figure}

The observed or inferred luminosity is a complex superposition of
light from current star formation and from formed stellar mass (and
from dust emission, in the case of bolometric luminosities), and the
star formation rate likely varies on a much shorter timescale than
$\sigma$.  A somewhat less complicated comparison may therefore be
between stellar mass and velocity dispersion, quantities that might be
expected to be related through the $M_{BH}-\sigma$ relation and the
$M_{\rm bulge}-M_{BH}$ relation \citep{mtr+98,mh03} as well as through
the comparison of stellar and dynamical masses discussed above.  We
plot stellar mass $M_{\star}$ vs. $\sigma$ in
Figure~\ref{fig:sigmass}; this is similar to
Figure~\ref{fig:masscomp}, although here we add the $z\sim3$ LBGs and
the DRGs.  The correlation has a significance of 3.6$\sigma$, about
the same as the $\sigma-L$ correlation; the low stellar mass galaxies
with large velocity dispersions discussed above are clear outliers
(marked with open diamonds; they are included in the above correlation
test).  The dotted line is the local relation between bulge stellar
mass and velocity dispersion, constructed by combining the $M_{\rm
  bulge}$--$M_{BH}$ relation of \citet{mh03} and the
$M_{BH}$--$\sigma$ relation of \citet{tgb+02}; it provides a
surprisingly accurate upper envelope in $\sigma$ on the plot, as with
the exception of AGN and the $M_{\rm dyn}\gg M_{\star}$ objects
discussed above (marked with open diamonds), virtually all of the
points lie on or to the left of this line. 

At least some theoretical predictions indicate that the correlation
between $M_{\star}$ and $\sigma$ should be strong at high redshift.
The results of one such study are shown by the open symbols, from the
numerical simulations of \citet[][see their Figure 3]{rhc+05}.  These
simulations test the evolution of the $M_{BH}-\sigma$ relation with
redshift using simulations of merging galaxies, incorporating feedback
from black hole growth such that $\sim0.5$\% of the accreted rest mass
energy heats the gas.  The $M_{BH}-\sigma$ correlation then results
from the regulation of black hole growth by feedback \citep{dsh05}.
The simulation results for $z=2$ (open squares) and $z=3$ (open
circles) describe the upper envelope in observed $\sigma$ reasonably
well, but the data exhibit considerably more scatter.  Given the
parameters of the simulations, this is not a surprise.  The velocity
dispersions in the models (which are the stellar velocity dispersions,
not the gas velocity dispersions we use) are measured well after the
merger and accompanying AGN feedback, when the star formation rate has
dropped to nearly zero \citep{dsh05}, while the data represent
galaxies at a variety of stages of an earlier starburst phase.  The
simulation sample is therefore considerably more homogeneous than the
data set.

A large number of galaxies in our data set have low velocity
dispersions for their stellar mass relative to the local correlation
(or high stellar mass for their velocity dispersion).  In fact this is
the opposite of the trend predicted by the simulations, which indicate
weak evolution with redshift in the $M_{BH}-\sigma$ and
$M_{\star}-\sigma$ relations, in the sense that velocity dispersions
increase at a given stellar mass at higher redshifts (attributed to
the steeper potential wells of high redshift galaxies).  This
difference may reflect the fact that we are probably underestimating
the true velocity dispersion in many cases, since we are only
sensitive to the highest surface brightness star-forming regions,
which may not sample the full potential.  It will be interesting to
see if deeper spectra with larger telescopes result in a higher
average velocity dispersion for this sample, and an accompanying
reduction of the scatter.  This plot also provides a hint of an
evolutionary sequence: if the $M_{\rm dyn}/M_{\star}>10$ objects are
indeed young galaxies with high gas fractions, they may evolve upward
on this plot, significantly increasing their stellar mass with
relatively little change in velocity dispersion until they fall on or
to the left of the dotted line along with the rest of the sample.  It
is also possible that the velocity dispersions of these objects will
{\it decrease} over time, bringing them closer to the local relation,
if they lose a large fraction of their gas to outflows (as suggested
by \citealt{esp+06}).  From the observed correlations of $\sigma$,
$L$, and $M_{\star}$, and their accompanying scatter, we conclude that
the processes which create correlations between these properties in
the local universe are underway at \ztwo.

\subsection{A Comparison with \zthree\ Lyman Break Galaxies}
\label{sec:lbgcomp}
\begin{figure}[htbp]
\plotone{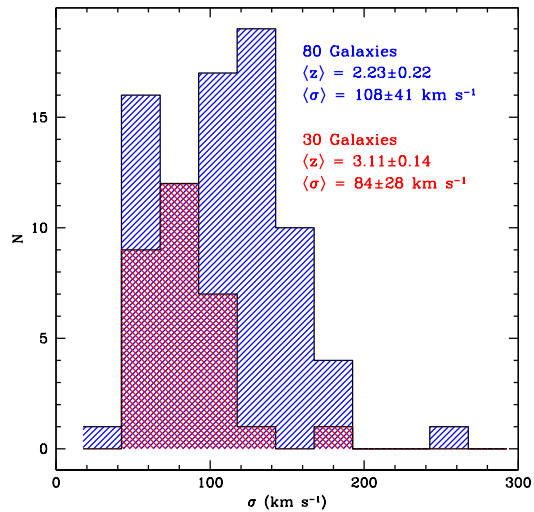}
\caption{The distributions of the velocity dispersions of
  \ztwo\ (larger blue histogram) and \zthree\ (smaller red histogram)
  galaxies.  At \ztwo, $\sigma$ is derived from the width of the
  \Ha\ emission line, while [\ion{O}{3}] is used for the \zthree\ sample.}
\label{fig:sigmahist}
\end{figure}

Close examination of Figure~\ref{fig:siglum} shows that the velocity
dispersions of the \ztwo\ sample and the \zthree\ LBGs have different
distributions.  We compare the two directly in
Figure~\ref{fig:sigmahist}, which shows histograms of the velocity
dispersions of the \ztwo\ galaxies and the LBGs, after excluding AGN.
The mean and the error in the mean of the \ztwo\ sample is $\langle
\sigma \rangle =108 \pm 5$ \kms, while for the \zthree\ LBGs it is
$\langle \sigma \rangle =84 \pm 5$ \kms; the difference between the
two means is of 5$\sigma$ significance.  The \ztwo\ distribution is
significantly broader; objects with low velocity dispersions are
common at \ztwo\ as well as at \zthree, but the lower redshift sample
contains a large number of objects with $\sigma \gtrsim 130$ \kms.
Only one such galaxy is present in the \zthree\ sample.  The
\zthree\ galaxies have an average dynamical mass $\langle M_{\rm dyn}
\rangle = 3.6\pm0.6 \times 10^{10}$ \msun, a factor of two smaller than that
of the \ztwo\ galaxies; the average stellar mass of the LBGs is also a
factor of $\sim2$ lower than that of the \ztwo\ galaxies
\citep{ssa+01}, though this comparison is complicated by sample
selection effects.  As described in detail below, these differences in
velocity dispersion and dynamical mass appear to reflect real
differences between the \ztwo\ and \zthree\ samples.

Possible selection effects are a concern in attempting to understand
the different distributions of $\sigma$ in the two samples.  The
\zthree\ line widths are measured from [\ion{O}{3}]$\lambda$5007,
while \Ha\ is used for the \ztwo\ galaxies.  To see if systematic
differences between line widths measured from the two lines might be
responsible for the effect, we have measured line widths from
[\ion{O}{3}]$\lambda$5007 for 8 of the \ztwo\ galaxies which also have
\Ha\ line widths.  The eight galaxies have \Ha\ velocity dispersions
ranging from 60 to 150 \kms.  There is no systematic difference in the
velocity dispersion from the two lines, with $\langle \sigma_{\Ha}
\rangle = 95$ \kms, and $\langle \sigma_{[\rm OIII]} \rangle = 97$
\kms.  The individual values of $\sigma$ usually agree well, and when
they do not, contamination by sky lines, particularly in the $H$-band
where we measure [\ion{O}{3}], appears to be the source of the
discrepancy. There is therefore no evidence that the use of different
lines is a signifcant issue in the comparison of the two samples.

Another possible factor is the higher redshift of the LBGs.  Surface
brightness is a strong function of redshift, and it is possible that
we are seeing only the brightest, most central regions of the LBGs
compared with flux from a larger region of the \ztwo\ sample.  If this
is true, we might expect to observe less emission from the faint, low
surface brightness portions of the galaxies at \zthree\ than at \ztwo;
in other words, the \zthree\ galaxies may appear smaller, beyond the
expected scaling in size with redshift (e.g.\ \citealt{fdg+04}). We
have measured the spatial extent of 18 of the [\ion{O}{3}] emission
lines, for comparison with the \Ha\ sizes; as in the \ztwo\ sample,
most of the lines are spatially resolved, and $\langle r_{[\rm OIII]}
\rangle = 0.6$\arcsec$\simeq4.5$ kpc.  The mean value of $r_{[\rm
    OIII]}$ is slightly smaller than the mean \Ha\ spatial extent, for
which we find $\langle r_{\Ha} \rangle = 0.7$\arcsec$\simeq6$ kpc.
This difference is almost identical to the mean size difference found
between UV-bright galaxies at $z\sim3$ and $z\sim2.3$ by
\citet{fdg+04}, who find that galaxy radii scale with redshift
approximately as the Hubble parameter $H^{-1}(z)$, in accordance with
theoretical expectations.  In our adopted cosmology, this scaling
predicts that the $z\sim3$ LBGs should have mean radii $\sim70$\%
smaller than the $z\sim2$ sample, entirely consistent with our
observations.  We therefore have reason to believe that the observed
size differences reflect real differences between the two samples,
rather than surface brightness effects.

The selection criteria for observation with NIRSPEC were somewhat
different for the two samples.  The \zthree\ galaxies were chosen
primarily because of their UV brightness, with additional objects
selected because of their proximity to a QSO sightline.  Unlike some
of the observations of \ztwo\ galaxies, objects with bright or red
near-IR magnitudes or colors were not favored.  If such bright or red
objects make up most of those with large line widths, this could be a
plausible explanation for the differences.  An examination of the list
of \ztwo\ galaxies with $\sigma>120$ \kms\ shows that some of them
were indeed selected for their bright or red near-IR properties, but
an approximately equal number were not.  The velocity dispersions of
\ztwo\ galaxies along QSO sightlines, which should be representative
of the sample as a whole, have the same distribution as that of the
full \ztwo\ sample, indicating that the selection of near-IR bright or
red objects is probably not responsible for the presence of objects
with large velocity dispersions in the \ztwo\ sample and their absence
among the LBGs.

Another possibility is that [\ion{O}{3}] is more difficult to detect
in massive galaxies; with increasing metallicity, the ratio of
[\ion{O}{3}]/[\ion{O}{2}] decreases, as does the ratio of the
[\ion{O}{3}] and [\ion{O}{2}] lines to \Hb\ (or \Ha) for metallicities
above $\sim1/3\, Z_{\odot}$.  Thus the relative fluxes of \Ha\ and
[\ion{O}{3}] might be expected to depend strongly on metallicity and
hence stellar mass \citep{thk+04,esp+06}.  Although uncertainties in
flux calibration make the ratios uncertain, in our sample of
\ztwo\ galaxies with both \Ha\ and [\ion{O}{3}] measurements
$F_{\Ha}/F_{\rm [OIII]}$ ranges from $\sim0.5$ to $\sim5$ as stellar
mass increases from $\sim10^9$ to $\sim10^{11}$ \msun, and
[\ion{O}{3}] is not detected for one of the most massive galaxies (we
do detect \Hb\ for this object).  We see that [\ion{O}{3}] is weak in
massive, relatively metal-rich galaxies, and this could contribute to
the absence of objects with large line widths in the \zthree\ NIRSPEC
samples\footnote{In principle we would still recognize
  \zthree\ objects with large velocity dispersions via their broad
  \Hb\ lines, but in many cases \Hb\ is too weak for a reliable
  measurement}.  Not all of the objects with large velocity
dispersions in the \ztwo\ sample have large stellar masses, however;
$\sim60$\% of the objects with $M_{\rm dyn} \gg M_{\star}$ have
$\sigma>130$ \kms, and these low stellar mass objects also have low
metallicities which should result in strong [\ion{O}{3}] lines.  The
absence of such galaxies from the \zthree\ sample (there may be one;
we have no near-IR photometry and no stellar mass estimate for the
\zthree\ galaxy with the largest velocity dispersion) is unlikely to
be explained by selection effects, and probably indicates that low
stellar mass galaxies with large line widths are rarer at
\zthree\ than at \ztwo.

There are reasons to expect physical differences between the
\ztwo\ and \zthree\ samples.  We have already noted that galaxies at
\zthree\ are seen to be more compact than those at \ztwo, in agreement
with expectations from hierarchical galaxy formation theory.  An
analysis of the correlation lengths of the LBGs and the \ztwo\ sample
indicates that the \ztwo\ galaxies reside in halos $\sim3$ times more
massive than those hosting the LBGs \citep{asp+05}.  The smaller
dynamical masses of the \zthree\ galaxies are in good agreement with
this result.  We conclude that the lower average velocity dispersion of the
LBGs is not fully explained by selection effects, and that the
differences in $\sigma$ between the \ztwo\ and \zthree\ samples are
likely to indicate real differences in the typical masses of galaxies in
the two samples.

\begin{figure*}[htbp]
\plotone{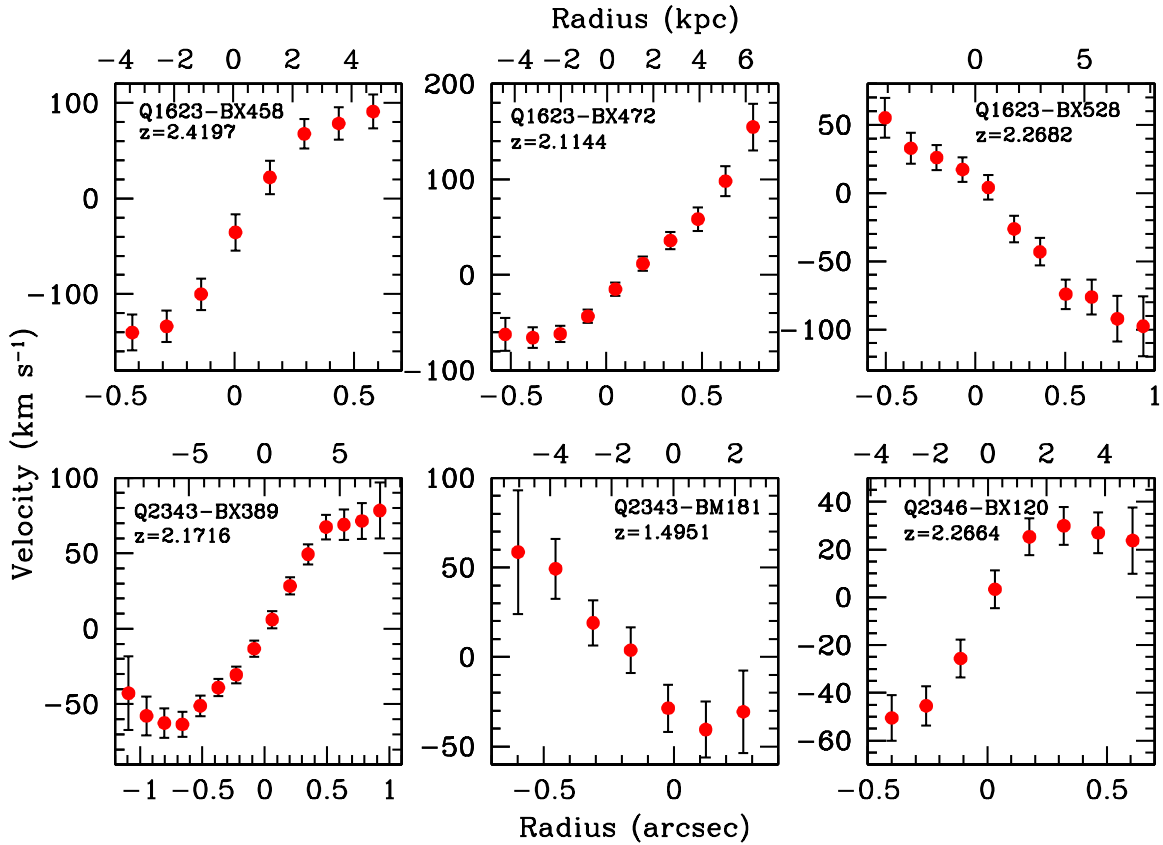}
\caption{Observed velocity as a function of slit position for objects
  with spatially resolved and tilted \Ha\ emission lines.  The seeing
  for these observations was $\sim0.5$\arcsec, so the points shown are
  highly correlated, with approximately four points per resolution
  element.  We plot one point per pixel to show the observed velocity
  field clearly, but the blurring of the seeing means that these
  diagrams do not represent the true velocity structure of the
  galaxies.}
\label{fig:velpos}
\end{figure*}

\section{Spatially Resolved Kinematics}
\label{sec:tilt}
In our initial study of 16 \Ha\ spectra of \ztwo\ galaxies, nearly
40\% of the sample showed spatially resolved and tilted emission lines
\citep{ess+03}.  In the current, enlarged sample of 114 objects the
fraction is much smaller, 14/114 or 12\%; this is probably both
because of exceptionally good conditions during our first observing
run (we observed 6 of the 14 objects during this run) and simply
because of small number statistics.  Velocity shear may be caused by
rotation, merging, or some combination of the two, and whether or not
it is detected in a nebular emission line depends on the size, surface
brightness and velocity structure of the object, its inclination, and
the alignment of the slit with respect to the major axis.  The seeing
during the observations plays a crucial role as well, as
\citet{ess+04} show with repeated observations of the same object.
Inclinations and major axes are unknown for nearly all of our sample,
as is the primary cause of the velocity shear, and therefore this
fraction of 12\% should be considered a lower limit to the fraction of
rotating or merging objects in the \ztwo\ sample.  The true fraction
cannot be determined given the limitations imposed by the seeing.

In order for the tilt of an emission line to be considered
significant, we required that the observed spatial extent be at least
1.5 times larger than the seeing disk, and that the peak-to-peak
amplitude of the shear be at least four times larger than the typical
velocity uncertainty.  Five of the 14 objects with spatially resolved
velocity shear have not been previously discussed (six are described
by \citealt{ess+03}, two by \citealt{ess+04}, and one by
\citealt{sep+04}).  We plot the observed velocities with respect to
systemic as a function of slit position for these five objects in
Figure~\ref{fig:velpos}; we also include Q1623-BX528, the object
discussed by \citealt{sep+04}, because such a diagram was not
presented in that paper.  These diagrams are constructed by stepping
along the slit pixel by pixel, summing the flux of each pixel and its
neighbor on either side to increase the S/N, and measuring the
centroid in velocity at each position.  We emphasize that these are
not rotation curves of the variety plotted for local galaxies.  The
points are highly correlated because of the seeing
($\sim$0.5\arcsec\ for the observations presented here, thus there are
$\sim4$ points per resolution element; we plot one point per pixel to
give a clear picture of the {\it observed} velocity field), and
recovery of the true velocity structure requires a model for the
structure of the object because of the degeneracies created by the
seeing (see \citealt{lse06} for a demonstration).  Although the
physical scale corresponding to the angular scale is given at the top
of each panel for reference, the implied mapping between physical
radius and velocity is not a true representation of the structure of
the object.

\begin{figure}[htbp]
\plotone{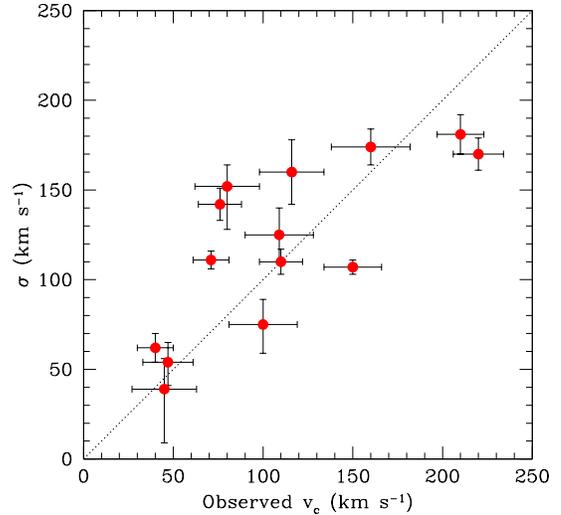}
\caption{A comparison of the velocity dispersion $\sigma$ and the
  observed velocity shear $v_c=(v_{\rm max}-v_{\rm min})/2$ for the 14
  objects with tilted lines.  The dotted line shows equal values.  We
  find $\langle \sigma/v_c \rangle = 1.2$.}
\label{fig:vcsig}
\end{figure}

Although the seeing prevents us from determining the true velocity
field of the objects, we can at least determine a lower limit on the
amplitude of the velocity shear from the uncorrelated endpoints of
each curve.  We calculate the observed $v_c=(v_{\rm max}-v_{\rm
  min})/2$ for each object, where $v_{\rm max}$ and $v_{\rm min}$ are
with respect to the systemic redshift.  This estimate is almost
certainly less than the true velocity shear because the inclinations
of the galaxies are unknown, in most cases we have made no attempt to
align the slit with the major axis, and the seeing effectively reduces
the velocities at the endpoints by mixing the light from the edges of
the galaxy with emission from higher surface brightness, lower
velocity regions toward the center.  Even under ideal conditions, the
\Ha\ emission may not trace the true circular velocity; \citet{lh96}
find that in local starburst galaxies the regions of high \Ha\ surface
brightness sample only the inner, solid-body portion of the rotation
curve.  The observed values of $v_c$ for the 14 objects with velocity
shear are given in Table~\ref{tab:kin}, and in the left panel of
Figure~\ref{fig:vcsig} they are plotted against the velocity
dispersion $\sigma$.  The dotted line marks equal values; we see that
$v_c \sim \sigma$ for most objects, though with considerable scatter.

The relationship between $\sigma$ and the terminal velocity $V_c$ (we
use $v_c$ to refer to our observed velocity shear, and $V_c$ to
indicate the true circular velocity of a disk) has been quantified in
samples of more local galaxies.  \citet{rgci97} find
$\sigma\sim0.6\,V_c$, while \citet{pcd+05} find
$V_c = 1.32\, \sigma +46$ for elliptical and high surface brightness
disk galaxies, for $\sigma \gtrsim 50$ \kms\ and with velocities in
\kms.  It is clear that $\sigma$ and $v_c$ of the galaxies in our sample do
not follow these distributions; instead we find a mean $\langle
\sigma/v_c \rangle \sim 1.2$ (while it is not possible for $V_c$ to be
less than $\sigma$, in the limit as the galaxy becomes spatially
unresolved the {\it observed} $v_c\rightarrow 0$, and
$\sigma/v_c\rightarrow\infty$).  Because our measurements of $\sigma$
are less affected by the seeing, they are undoubtably more reliable
than our measurements of $v_c$.  If we assume for the moment that our
galaxies are rotating disks, we can use one of the local relations to
predict $V_c$ for our sample.  If the \ztwo\ galaxies obey the
relation of \citet{pcd+05}, our
measurements of $v_c$ underestimate the true circular velocities by an
average factor of $\sim2$, and the full \Ha\ sample has an average
$\langle V_c \rangle \sim 190$ \kms.  We have already shown that a
change in the seeing from $\sim0.5$\arcsec\ to $\sim0.9$\arcsec\ can
reduce the observed $v_c$ by a factor of $\sim2$ while changing
$\sigma$ by less than 10\% \citep{ess+04}, so it is not unreasonable
to assume that a change in resolution from $\sim0.1$\arcsec\ or better
to $\sim0.5$\arcsec\ might have a similar effect.  Deep observations
of these objects at high angular resolution will be required to obtain
a true measure of $V_c$.

Next we ask whether or not the galaxies which display velocity shear
are different from the rest of the galaxies in any significant way.
Figure~\ref{fig:vcsig} shows that galaxies with shear span nearly the
full range in velocity dispersion; the mean value of $\sigma$ for the
galaxies with shear is 119 \kms, while the mean $\sigma$ of galaxies
without shear is 120 \kms.  There is mild evidence that galaxies with
shear tend to be older; the median age of the galaxies with shear is
1434 Myr, compared to 509 Myr for the galaxies without shear.  The
galaxies with shear also have slightly higher stellar masses, with a
median of $4.4\times10^{10}$ \msun\ compared to $1.8\times10^{10}$ for
those without shear.  A larger sample of galaxies with shear, and more
uniform observing conditions, are required to determine whether or not
these differences are significant.  Assuming for the moment that they
are, they may suggest that the rotation of mature, dynamically relaxed
galaxies is a more important contribution to our observed shear than
merging, which should not have a preference for older, more massive
galaxies.

Nothing in the data is inconsistent with ordered rotation as the
primary cause of the observed shear.  With this in mind we draw
attention to the position-velocity diagram of Q2343-BX389, shown in
the lower left panel of Figure~\ref{fig:velpos}, which appears to turn
over or flatten at both ends in the manner of locally observed
rotation curves.  This is one of several objects from the
\ztwo\ sample recently observed with SINFONI, the near-IR integral
field spectrograph on the VLT \citep{fgl+06}.  The SINFONI
position-velocity diagram also shows evidence for flattening, but at
higher velocities; the velocity difference is probably due to the
$\sim50^\circ$ misalignment of the NIRSPEC slit with the kinematic
major axis and the deeper SINFONI integration which reveals flux at
larger radii.  Several but not all of the galaxies observed with
SINFONI show signatures of rotating disks, and the role of merging in
\ztwo\ galaxies still remains unclear.  Given its expected importance
at high redshift it is likely that it makes some contribution to our
observed velocities.  Improved disentangling of these effects must
await high angular resolution spectroscopy with the aid of adaptive
optics; such instruments will be able to map the velocity fields of
high redshift galaxies at a resolution impossible with our current
data \citep{lse06}.

\section{Gas and Baryonic Masses}
\label{sec:gasmass}
We have used the stellar and dynamical masses determined above to
infer that the galaxies in our sample have significant gas fractions.
It would be highly desirable to test this directly, for example with
CO measurements, but such data would require extremely long
integration times with current technology, making the assembly of a
large sample of gas masses impossible.  We can, however, exploit
the correlation between star formation and gas density to obtain an
estimate of the gas masses.  In star-forming galaxies in the local
universe, the surface densities of star formation and gas are observed
to follow a \citet{s59} law, $\Sigma_{\rm SFR}=A \Sigma_{\rm gas}^N$,
over more than six orders of magnitude in $\Sigma_{\rm SFR}$
\citep{k98schmidt}.  This empirical relation is usually explained by a
model in which the SFR scales with density-dependent gravitational
instabilities in the gas.  The correlation has not yet been tested at
high redshift because of the lack of measurements of gas masses,
although the one well-studied example, the lensed $z=2.7$ LBG
MS1512-cB58, appears to be consistent with the local Schmidt law
\citep{btg+04}.  Assuming that the galaxies in our sample obey such a
law, we can use the SFRs and the galaxy sizes $r_{\Ha}$ measured from
the spatial extent of the \Ha\ emission to compute their star
formation densities, and thus estimate their gas densities and masses.
We use the global Schmidt law of \citet{k98schmidt}:
\begin{equation}
\Sigma_{\rm SFR}=2.5\times 10^{-4}\left( \frac{\Sigma_{\rm gas}}{1\
  \msun\ {\rm pc}^{-2}}\right)^{1.4} {\rm M_{\odot}\; yr^{-1}\; kpc}^{-2}
\end{equation}
in combination with the conversion from \Ha\ luminosity to SFR from
the same paper,
\begin{equation} {\rm SFR\; (M_{\odot}\;
    yr^{-1})=\frac{L(H\alpha)}{1.26\times10^{41}\; erg\; s^{-1}}}
\end{equation}
to create an IMF-independent relation between our observed
\Ha\ luminosity per unit area and the gas surface density
\begin{equation}
\Sigma_{\rm gas}=1.6\times10^{-27}\left( \frac{\Sigma_{\rm
    H\alpha}}{{\rm erg\; s^{-1}\; kpc}^{-2}}\right)^{0.71} \; {\rm
  M_{\odot}\; pc^{-2}}.
\end{equation}

The radii used by \citet{k98schmidt} to compute surface densities
approximately coincide with the edge of the galaxies' \Ha-emitting
disks; for our surface densities, we take an area equal to the square
of the FWHM of the (continuum-subtracted when necessary) \Ha\ emission
($r_{\Ha}$ in Table~\ref{tab:kin}).  For $L(\Ha)$ we use the
extinction-corrected \Ha\ luminosity, incorporating the factor of 2
aperture correction discussed by \citet{ess+06}.  We then take $M_{\rm
  gas}= \Sigma_{\rm gas}r_{\Ha}^2$ as an estimate of the cold gas mass
associated with star formation, and calculate the gas fraction $\mu
\equiv M_{\rm gas}/(M_{\rm gas}+M_{\star})$.  The removal of the IMF
dependence from the Schmidt law (which assumes a Salpeter IMF in the
conversion from \Ha\ luminosity to SFR) facilitates comparison between
gas, dynamical and stellar masses, as we discuss below.  It is likely
to be difficult to measure the objects' sizes with accuracy, given
their often clumpy UV continuum morphologies as seen in high
resolution images (e.g.\ \citealt{ess+03,ess+04}); the \Ha\ morphology
is likely to be complicated as well.  As noted in \S\ref{sec:dynmass},
the object sizes have typical uncertainties of $\sim30$\%, and the
corrected \Ha\ fluxes of the objects may be uncertain by up to a
factor of $\sim2$; therefore the gas masses of individual objects are
uncertain by a factor of $\sim3$, and the typical fractional
uncertainty in $\mu$ is $\sim$50\%.  Systematic uncertainties due to
the scatter in the Schmidt law itself (0.3 dex) are an additional
source of error.  We focus here on overall trends, which depend on
large numbers of objects and so are better determined than the
parameters of individual galaxies; for example, the error in the mean
gas fraction of a subsample of $\sim15$ objects is 15\% or less.

\begin{figure*}[htbp]
\plotone{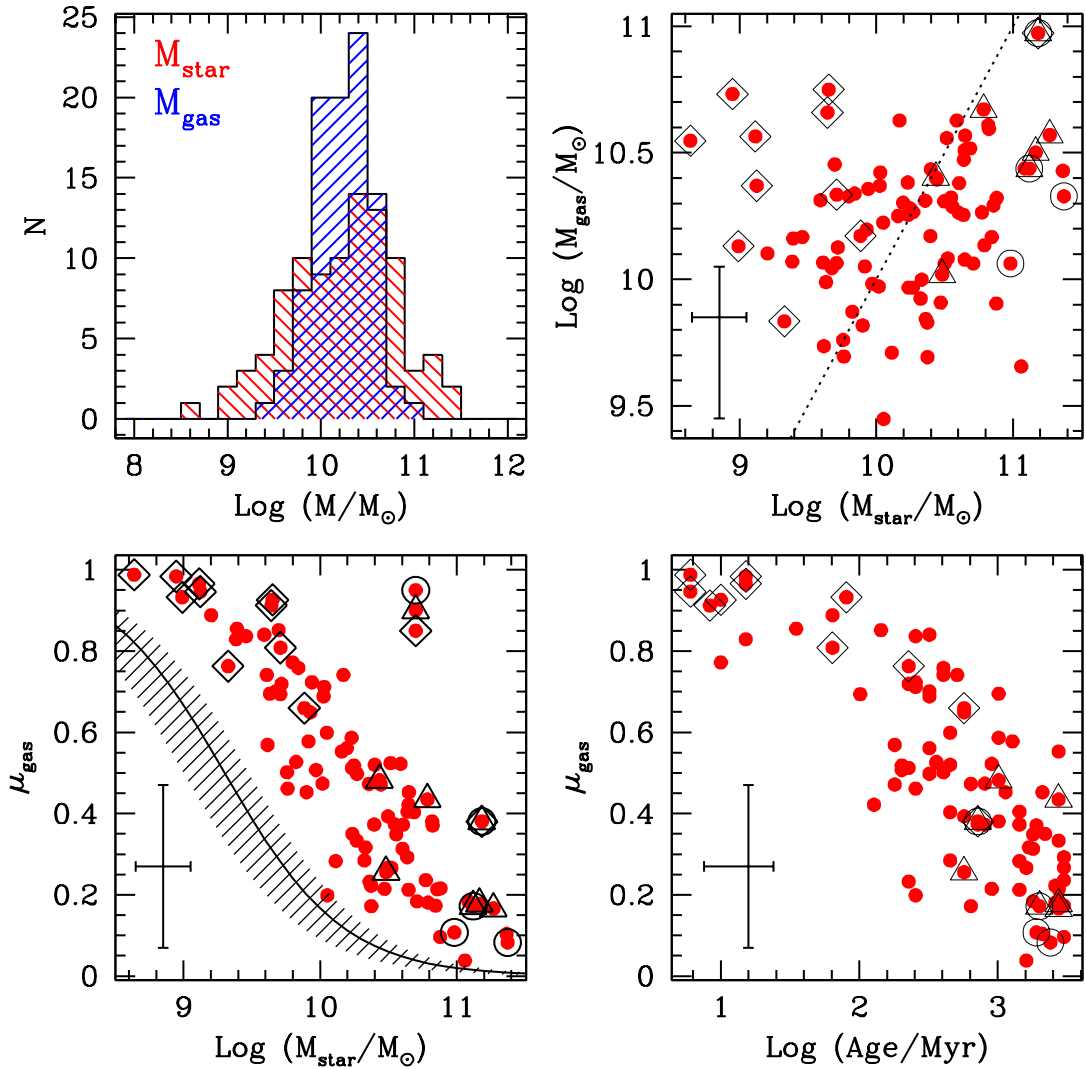}
\caption{The relation of the gas mass inferred from the Schmidt law to
  the stellar population parameters.  Upper left: Histograms of the
  gas (blue) and stellar (red) masses, showing that the range in
  stellar mass is significantly broader than the range in gas mass.
  Upper right: Gas mass vs.\ stellar mass.  Gas masses are relatively
  constant across the sample, but when the galaxies with $M_{\rm
    dyn}/M_{\star}>10$ are not considered we see an increase in gas
  mass with stellar mass.  Lower panels: The gas fraction strongly
  decreases with increasing stellar mass (left) and age (right).  The
  solid line and accompanying shaded region in the lower left panel
  show the minimum detectable gas fraction set by our \Ha\ flux
  detection limit, for the mean (solid line) and range (shaded region)
  of sizes observed.  Symbols are as in Figure~\ref{fig:masscomp}, and
  the error bars in the corner of each plot show typical
  uncertainties.}
\label{fig:gasmass}
\end{figure*}

The mean inferred gas mass is $\langle M_{\rm gas} \rangle =
2.1\pm0.1\times 10^{10}$ \msun, slightly lower than the mean stellar
mass of $\langle M_{\star} \rangle = 3.6\pm0.4\times 10^{10}$ \msun.
The distributions of the two masses are shown in the histograms in the
upper left panel of Figure~\ref{fig:gasmass}, where the narrower blue
histogram shows the gas masses and the broader red histogram shows the
stellar masses.  The range of inferred gas masses is clearly smaller
than the range of stellar masses; this is because the dispersions in
star formation rate and size are smaller than the more than two orders
of magnitude variation we see in stellar mass.  The two masses are
plotted against each other in the upper right panel of
Figure~\ref{fig:gasmass}, where it is apparent that the $M_{\rm
  dyn}/M_{\star}>10$ objects (open diamonds) have large inferred gas
masses, and are exceptions to a general trend of increasing gas mass
with increasing stellar mass.  The absence of points in the lower left
corner of this plot is probably a selection effect, as low mass
galaxies with low star formation rates are unlikely to be detected in
our $K$-band images (they may fall below our ${\cal R}<25.5$ magnitude
limit for selection as well), and we are also less likely to detect
\Ha\ emission in $K$-faint objects.  We next plot the gas fraction
$\mu$ vs.\ stellar mass in the lower left panel of
Figure~\ref{fig:gasmass} and $\mu$ vs.\ age in the lower right.  The
trends of decreasing gas fraction with increasing stellar mass and age
are strong, supporting our hypothesis that the $M_{\rm dyn}\gg
M_{\star}$ objects are young starbursts with high gas fractions (low
mass galaxies with low gas fractions may exist, but are probably too
faint to be detected by our survey).  Local galaxies with low stellar
masses are also observed to have higher gas fractions
\citep{md97,bd00}.

Other studies of galaxies at high and low redshift may provide
additional insight into these results.  \citet{rsf+06} have recently
used 24 \micron\ data from the Spitzer Space Telescope to infer the
bolometric luminosities of both optically-selected galaxies similar to
those considered here and near-IR selected objects \citep{bzk,flr+03}.
They assess the relationship between gas fraction and stellar mass,
finding a similar strong trend which extends to massive, nearly
passively evolving galaxies with low gas fractions that are not
selected by our UV criteria.  They also consider the dust-to-gas ratio
of local and high redshift galaxies, finding that at a given
bolometric luminosity galaxies at $z\sim2$ are $\sim10$ times less
obscured by dust than local galaxies.  Thus we infer that FIR emission
is likely to be weaker in high redshift gas-rich galaxies than in
their low redshift counterparts with similar bolometric luminosities.
Galaxies that are detected at 850 \micron\ have so far provided the
best candidates for direct detection of gas in high redshift galaxies;
\citet{gbs+05} use CO measurements to estimate the gas masses of five
submillimeter galaxies at $z=1$--3.5, finding a median molecular gas
mass of $3.0\times10^{10}$ \msun\ within a typical radius of $\sim2$
kpc.  This is only slightly higher than the typical gas masses we
infer from the Schmidt law, implying that the largest gas masses we
infer may be testable with current technology.  Though the gas masses
of the submm galaxies appear to be similar to those we find here, the
factor of $\sim3$ smaller sizes of the submm galaxies imply a gas
surface density an order of magnitude higher than the average value of
our sample.  Such comparisons must be viewed with caution, of course,
given the very different methods of inferring the gas content and
sizes of the two samples; it is possible that a more direct
measurement of the molecular gas distribution of the current sample
would reveal smaller emitting regions and higher gas surface
densities.

\begin{figure*}[htbp]
\plotone{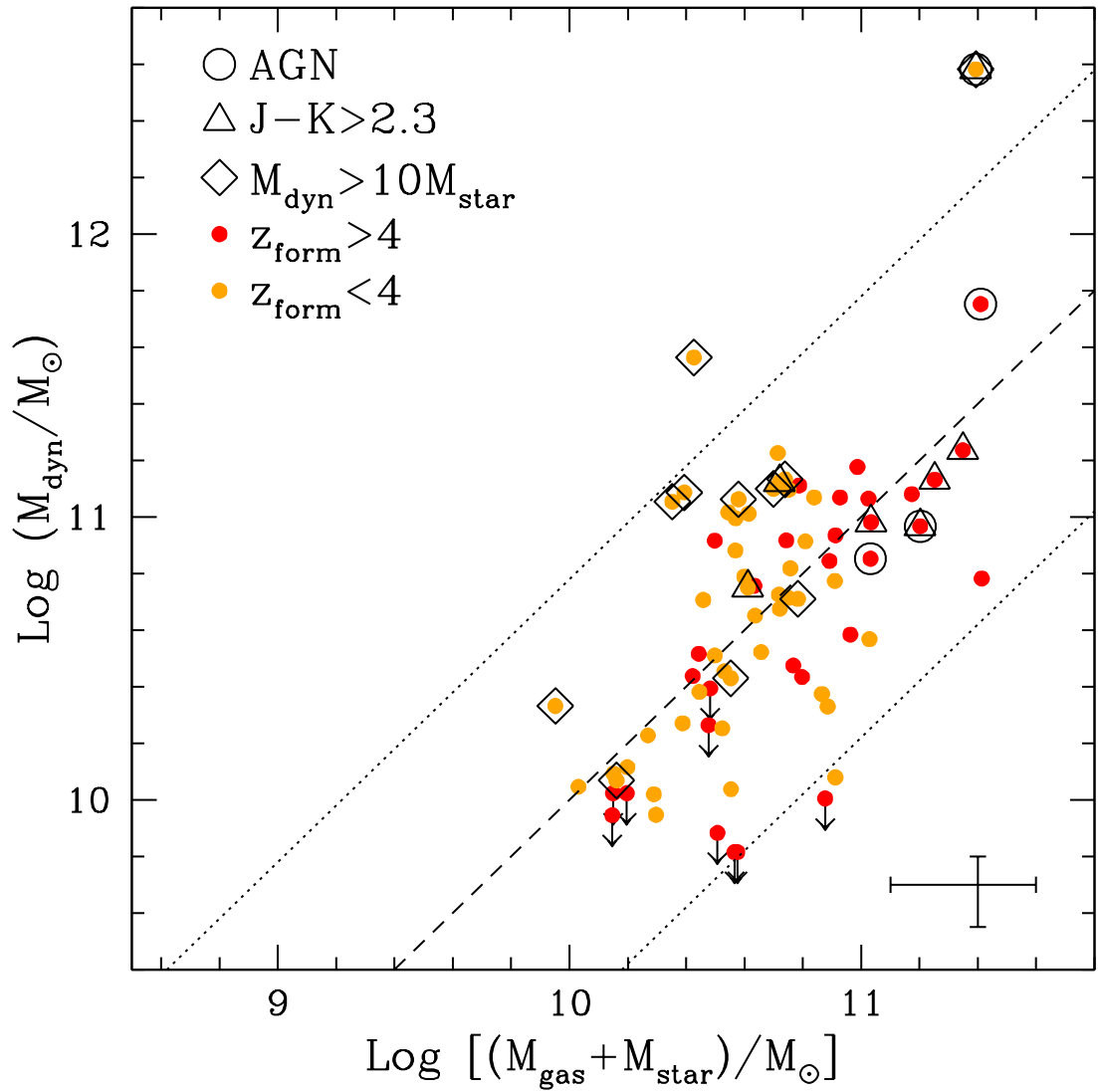}
\caption{The combined gas and stellar mass $M_{\rm gas}+M_{\star}$
  vs.\ dynamical mass $M_{\rm dyn}$; compare
  Figure~\ref{fig:masscomp}. The dashed line shows equal masses, and
  the dotted lines show a factor of $\sim6$ difference between the
  masses.  The correlation is significant at the 4$\sigma$ level, and
  the masses of 85\% of the objects agree to within a factor of three.
  Neglecting AGN, the average dynamical mass is 1.2 times larger than
  the average baryonic mass. Symbols are as in
  Figure~\ref{fig:masscomp}, with the addition that galaxies with
  inferred formation redshifts (discussed in \S\ref{sec:disc}) $z_{\rm
    form}>4$ are shown in red and those with $z_{\rm form}<4$ are
  shown in orange.  Typical uncertainties are shown by the error bars
  in the lower left corner, which for dynamical masses include only
  measurement errors in $r_{\Ha}$ and $\sigma$.  Uncertainties in
  dynamical masses are dominated by the unknown factors in the mass
  model as discussed in the text; the dotted lines indicate
  approximate uncertainties in the mass comparison including this
  additional source of error.}
\label{fig:stargassum}
\end{figure*}

We can now compare the estimated baryonic masses $M_{\rm bar}=M_{\rm
  gas}+M_{\star}$ with the dynamical masses in
Figure~\ref{fig:stargassum}, to see if the addition of the inferred
gas mass improves the agreement between stellar and dynamical mass.
We use the same axes as Figure~\ref{fig:masscomp} to facilitate
comparison, and again mark the line of equal masses (dashed line) and
a factor of $\sim6$ difference between the two masses (dotted lines).
The agreement between the masses is greatly improved with the addition
of the gas; there is a significant correlation (4$\sigma$, with
probability $P=8\times10^{-5}$ that the data are uncorrelated), and
the masses are within a factor of 3 for 85\% of the sample (objects
with limits on $\sigma$ are not included, though most of these are
consistent with the rest of the data, and those that are inconsistent
are highly uncertain).  After excluding AGN (open circles), we find
$\langle M_{\rm gas}+M_{\star} \rangle = 5.8\pm0.6 \times 10^{10}$ \msun;
the mean dynamical mass $\langle M_{\rm dyn} \rangle = 7.1\pm0.7 \times
10^{10}$ \msun\ is 1.2 times higher.  There are only two objects for
which $M_{\rm dyn}/M_{\rm bar}>10$; these are the AGN Q1700-MD94 (upper
right), and the galaxy Q1623-BX376, which has an anomalously large
velocity dispersion that may be influenced by its complicated spatial
structure \citep{ess+03}.

Although the estimates of gas, stellar and dynamical masses considered
here carry significant uncertainties, we emphasize that they are
obtained from independent quantities (\Ha\ luminosity and size,
multi-wavelength photometry, and line width respectively).  Together
they provide a remarkably coherent scenario in which relatively
massive ($M_{\rm gas}+M_{\star}\gtrsim 10^{10}$ \msun, and $M_{\rm
  halo}\sim10^{12}$ \msun; \citealt{asp+05}) star-forming galaxies
have high gas fractions which decrease with age.  The observed
variation of metallicity across the sample also supports this model,
as we describe separately \citep{esp+06}.  In this scenario, the
stellar mass of a galaxy depends strongly on its evolutionary state,
as well as on its total mass.  The lowest stellar mass galaxies in the
sample have low stellar masses because they are young and have
converted or expelled only a small fraction of their gas; such objects
have stellar masses $\gtrsim100$ times lower than the most massive
galaxies in the sample, while their inferred baryonic and dynamical
masses are only $\lesssim30$ times lower than those of the most
massive galaxies.

\section{Conclusions and Discussion}
\label{sec:disc}

The goals of this paper have been to examine the kinematic properties of
star-forming galaxies at \ztwo\ as revealed by their \Ha\ spectra; to
compare their dynamical masses with stellar masses determined from
population synthesis modeling; to look for trends between kinematic
and stellar population properties; and to estimate the stellar and gas
content of the galaxies.  Our main conclusions are as follows:

1.  The sample has a mean \Ha\ velocity disperson $\langle \sigma
\rangle = 101\pm5$ \kms, excluding AGN.  From $\sigma$ and the spatial
extent of the \Ha\ emission we estimate dynamical masses, finding
$\langle M_{\rm dyn} \rangle = 7.1\pm0.7\times10^{10}$ \msun, again
excluding AGN.  Stellar and dynamical masses agree to within a factor
of $\sim6$ for most objects, consistent with observational and
systematic uncertainties.  The masses are correlated with 2.5$\sigma$
significance.  However, 15\% of the galaxies have $M_{\rm
  dyn}>10\times M_{\star}$; these objects have low stellar masses,
young ages and tend to have high \Ha\ equivalent widths, suggesting
that they are young galaxies with large gas fractions.

2.  We combine our \Ha\ results with LBGs at \zthree\ and other
samples from the literature, and find that rest-frame optical
luminosity (corrected for extinction) and velocity dispersion are
correlated with 4$\sigma$ significance.  The large scatter
prevents a robust determination of the slope, but it is not
inconsistent with the local relation $L\propto\sigma^4$.  The high
redshift galaxies have much higher luminosities at a given velocity
dispersion than would be predicted by the local Faber-Jackson
relation.  The observed correlation and its accompanying large scatter
suggest that the processes which produce the $L$--$\sigma$ correlation
in the local universe are underway at \ztwo--3.

3. A comparison of the \Ha\ sample with a similarly rest-frame
UV-selected sample of Lyman break galaxies at \zthree\ shows that the
average velocity dispersion is $\sim30$\% larger at \ztwo\ than at
\zthree, and the average dynamical mass of the \ztwo\ galaxies is
$\sim2$ times larger than that of the \zthree\ LBGs.  This difference
probably reflects real differences in the average masses of the
galaxies at \ztwo\ and \zthree.

4. Fourteen of the 114 galaxies with \Ha\ spectra, or 12\%, have
spatially resolved and tilted emission lines.  On average, the
observed amplitude of the velocity shear $v_c$ is approximately equal
to the velocity dispersion $\sigma$.  If the galaxies are rotating
disks and follow the local relations between the true circular
velocity $V_c$ and $\sigma$, our observations underestimate $V_c$ by
an average factor of $\sim2$, and the galaxies have $\langle V_c
\rangle \sim 190$ \kms.  However, other factors including
galaxy-galaxy mergers, slit position angle, and the seeing FWHM are
significant sources of uncertainty in $v_c$.

5.  Using the empirical correlation between star formation rate per
unit area and gas surface density, we estimate gas masses and gas
fractions for the galaxies in the sample.  The mean gas fraction is
$\sim50$\%, and the gas fraction strongly decreases with increasing
stellar mass and age.  The combined mass in gas and stars is
significantly better correlated with the dynamical mass than is the
stellar mass alone.  The sample spans a much wider range in stellar
mass than dynamical mass (or inferred baryonic mass), indicating that
there is considerable dispersion in the fraction of gas that has been
converted to stars by \ztwo. 

These differing ranges in stellar and baryonic mass imply that
galaxies of similar total masses began forming stars at different
times, and that at a given redshift galaxies with a wide variety of
total masses have just begun forming stars.  For $\sim70$\% of the
galaxies in the \Ha\ sample, the best-fit ages imply formation
redshifts $2\lesssim z_{\rm form} \lesssim 4$; most of these galaxies
have baryonic masses between $\sim10^{10}$ and $\sim10^{11}$ \msun.
The remaining $\sim30$\% have $z_{\rm form}>4$ (including the 15\% of
the sample which have best-fit ages equal to the age of the universe);
these galaxies have $10^{10.4} \lesssim M_{\rm bar} \lesssim
10^{11.4}$ \msun.
$M_{\rm bar}$ and $z_{\rm form}$ are strongly correlated across the
full sample, with 5.4$\sigma$ significance, while those with
$2\lesssim z_{\rm form} \lesssim 4$ are correlated with 2.2$\sigma$
significance and there is no correlation between $M_{\rm bar}$ and
$z_{\rm form}$ within the $z_{\rm form}>4$ group.  The two groups are
shown in Figure~\ref{fig:stargassum}, where galaxies with $z_{\rm
  form}>4$ are shown by red points and galaxies with $z_{\rm form}<4$
are plotted in orange.  Galaxies with $M_{\rm bar}\lesssim10^{11}$ \msun\
fall in both groups.  These results are not unexpected, since
total mass is not the only factor which determines when a galaxy
begins to form stars; large scale environment and interactions with
other galaxies are also significant.  Galaxies in dense environments
will cross the critical density threshold for collapse earlier, and
should therefore begin forming stars at earlier times.  One of the
fields in the present survey contains a significant overdensity of
galaxies at $z=2.3$, and \citet{sas+05} show that galaxies within the
overdensity are significantly older than average.  It is also expected
that mergers will trigger significant star formation events; the
merger-based numerical simulations of \citet[][and references
  therein]{sdh05} are successful in reproducing many properties of
galaxies.  We cannot yet identify the likely triggers for star
formation for galaxies on an individual basis, however; detailed
morphological analysis and an improved quantification of the galaxies'
environments will be useful in this regard.

Because of uncertainties in the sizes and velocity distributions of the
\ztwo\ galaxies, the dynamical masses presented here have much room
for improvement.  These improvements will require deep observations of
a large sample of objects, and in many cases high angular resolution
will be needed in order to discriminate between plausible models of
the velocity field.  Near-IR multi-object and integral field
spectrographs, especially with adaptive optics, will make such
observations feasible, and we anticipate that the galaxies presented
here will be a useful sample from which to select targets for such
observations.  These kinematic measurements will provide otherwise
unobtainable insights about the prevalence of mergers at high redshift
and the growth of massive disk galaxies.  We also anticipate studies
at longer wavelengths which will improve our understanding of the
galaxies' gas masses.  All of these results will lead to a more
refined picture of the assembly of mass at high redshift.

\acknowledgements We thank Andrew Blain, Jonathan Bird, David Kaplan
and Shri Kulkarni for obtaining near-IR images of some of our targets;
the staffs of the Keck and Palomar observatories for their assistance
with the observations; and the anonymous referee for a thoroughly
helpful report that signficantly improved the paper. CCS, DKE and NAR
have been supported by grant AST03-07263 from the US National Science
Foundation and by the David and Lucile Packard Foundation.  AES
acknowledges support from the Miller Institute for Basic Research in
Science, and KLA from the Carnegie Institution of Washington.
Finally, we wish to extend special thanks to those of Hawaiian
ancestry on whose sacred mountain we are privileged to be guests.
Without their generous hospitality, most of the observations presented
herein would not have been possible.


\LongTables
\begin{deluxetable}{l l l l l c c c c c}
\tablewidth{0pt}
\tabletypesize{\footnotesize}
\tablecaption{Galaxies Observed\label{tab:obs}}
\tablehead{
\colhead{Object} & 
\colhead{R.A.} & 
\colhead{Dec.} & 
\colhead{$z_{\Ha}$} & 
\colhead{${\cal R}$\tablenotemark{a}} & 
\colhead{$G-{\cal R}$\tablenotemark{a}} & 
\colhead{$U_n-G$\tablenotemark{a}} &
\colhead{$K_s$\tablenotemark{b}} &
\colhead{$J-K_s$\tablenotemark{b}} &
\colhead{Exposure time (s)}
}
\startdata
CDFb-BN88\tablenotemark{c} & 00:53:52.87 & 12:23:51.25 & 2.2615 & 23.14 & 0.29 & 0.68 & ... & ... & $12\times 720$\\
HDF-BX1055 & 12:35:59.59 & 62:13:07.50 & 2.4899 & 24.09 & 0.24 & 0.81 & ... & ... & $2\times 900$\\
HDF-BX1084 & 12:36:13.57 & 62:12:21.48 & 2.4403 & 23.24 & 0.26 & 0.72 & ... & ... & $5\times 900$\\
HDF-BX1085 & 12:36:13.33 & 62:12:16.31 & 2.2407 & 24.50 & 0.33 & 0.87 & ... & ... & $5\times 900$\\
HDF-BX1086 & 12:36:13.41 & 62:12:18.84 & 2.4435 & 24.64 & 0.41 & 1.09 & ... & ... & $5\times 900$\\
HDF-BX1277 & 12:37:18.59 & 62:09:55.54 & 2.2713 & 23.87 & 0.14 & 0.61 & 21.26 & 0.73 & $3\times 900$\\
HDF-BX1303 & 12:37:11.20 & 62:11:18.67 & 2.3003 & 24.72 & 0.11 & 0.81 & 21.03 & 2.31 & $2\times 900$\\
HDF-BX1311 & 12:36:30.54 & 62:16:26.12 & 2.4843 & 23.29 & 0.21 & 0.81 & 20.48 & 1.56 & $4\times 900$\\
HDF-BX1322 & 12:37:06.54 & 62:12:24.94 & 2.4443 & 23.72 & 0.31 & 0.57 & 20.95 & 2.16 & $6\times 900$\\
HDF-BX1332 & 12:37:17.13 & 62:11:39.95 & 2.2136 & 23.64 & 0.32 & 0.92 & 20.68 & 1.77 & $3\times 900$\\
HDF-BX1368 & 12:36:48.24 & 62:15:56.24 & 2.4407 & 23.79 & 0.30 & 0.96 & 20.63 & 1.81 & $4\times 900$\\
HDF-BX1376 & 12:36:52.96 & 62:15:45.55 & 2.4294 & 24.48 & 0.01 & 0.70 & 22.13 & 1.02 & $4\times 900$\\
HDF-BX1388 & 12:36:44.84 & 62:17:15.84 & 2.0317 & 24.55 & 0.27 & 0.99 & 19.95 & 1.78 & $2\times 900$\\
HDF-BX1397 & 12:37:04.12 & 62:15:09.84 & 2.1328 & 24.12 & 0.14 & 0.76 & 20.87 & 1.08 & $3\times 900$\\
HDF-BX1409 & 12:36:47.41 & 62:17:28.70 & 2.2452 & 24.66 & 0.49 & 1.17 & 20.07 & 2.31 & $2\times 900$\\
HDF-BX1439 & 12:36:53.66 & 62:17:24.27 & 2.1865 & 23.90 & 0.26 & 0.79 & 19.72 & 2.16 & $4\times 900$\\
HDF-BX1479 & 12:37:15.42 & 62:16:03.88 & 2.3745 & 24.39 & 0.16 & 0.79 & 21.30 & 1.57 & $5\times 900$\\
HDF-BX1564 & 12:37:23.47 & 62:17:20.02 & 2.2225 & 23.28 & 0.27 & 1.01 & 19.62 & 1.69 & $2\times 900$\\
HDF-BX1567 & 12:37:23.17 & 62:17:23.89 & 2.2256 & 23.50 & 0.18 & 1.05 & 20.18 & 1.23 & $2\times 900$\\
HDF-BX305 & 12:36:37.13 & 62:16:28.36 & 2.4839 & 24.28 & 0.79 & 1.30 & 20.14 & 2.59 & $4\times 900$\\
HDF-BMZ1156 & 12:37:04.34 & 62:14:46.28 & 2.2151 & 24.62 & -0.01 & -0.21 & 20.33 & 1.71 & $4\times 900$\\
Q0201-B13\tablenotemark{c} & 02:03:49.25 & 11:36:10.58 & 2.1663 & 23.34 & 0.02 & 0.69 & ... & ... & $16\times 720$\\
Q1307-BM1163 & 13:08:18.04 & 29:23:19.34 & 1.4105 & 21.66 & 0.20 & 0.35 & ... & ... & $2\times 900$\\
Q1623-BX151 & 16:25:29.61 & 26:53:45.01 & 2.4393 & 24.60 & 0.14 & 0.88 & ... & ... & $2\times 900$\\
Q1623-BX214 & 16:25:33.67 & 26:53:53.52 & 2.4700 & 24.06 & 0.39 & 1.12 & ... & ... & $4\times 900$\\
Q1623-BX215 & 16:25:33.80 & 26:53:50.66 & 2.1814 & 24.45 & 0.26 & 0.53 & ... & ... & $4\times 900$\\
Q1623-BX252 & 16:25:36.96 & 26:45:54.86 & 2.3367 & 25.06 & 0.07 & 0.55 & ... & ... & $3\times 900$\\
Q1623-BX274 & 16:25:38.20 & 26:45:57.14 & 2.4100 & 23.23 & 0.25 & 0.89 & ... & ... & $3\times 900$\\
Q1623-BX344 & 16:25:43.93 & 26:43:41.98 & 2.4224 & 24.42 & 0.39 & 1.25 & ... & ... & $2\times 900$\\
Q1623-BX366 & 16:25:45.09 & 26:43:46.95 & 2.4204 & 23.84 & 0.41 & 1.03 & ... & ... & $2\times 900$\\
Q1623-BX376 & 16:25:45.59 & 26:46:49.26 & 2.4085 & 23.31 & 0.24 & 0.75 & 20.84 & 1.61 & $4\times 900$\\
Q1623-BX428 & 16:25:48.41 & 26:47:40.20 & 2.0538 & 23.95 & 0.13 & 1.06 & 20.72 & 1.02 & $4\times 900$\\
Q1623-BX429 & 16:25:48.65 & 26:45:14.47 & 2.0160 & 23.63 & 0.12 & 0.56 & 20.94 & 1.46 & $2\times 900$\\
Q1623-BX432 & 16:25:48.73 & 26:46:47.28 & 2.1817 & 24.58 & 0.10 & 0.53 & 21.48 & 1.76 & $4\times 900$\\
Q1623-BX447 & 16:25:50.37 & 26:47:14.28 & 2.1481 & 24.48 & 0.17 & 1.14 & 20.55 & 1.66 & $3\times 900$\\
Q1623-BX449 & 16:25:50.53 & 26:46:59.97 & 2.4188 & 24.86 & 0.20 & 0.61 & 21.35 & 1.88 & $3\times 900$\\
Q1623-BX452 & 16:25:51.00 & 26:44:20.00 & 2.0595 & 24.73 & 0.20 & 0.90 & 20.56 & 2.05 & $3\times 900$\\
Q1623-BX453 & 16:25:50.84 & 26:49:31.40 & 2.1816 & 23.38 & 0.48 & 0.99 & 19.76 & 1.65 & $3\times 900$\\
Q1623-BX455 & 16:25:51.66 & 26:46:54.88 & 2.4074 & 24.80 & 0.35 & 0.87 & 21.56 & 1.83 & $2\times 900$\\
Q1623-BX458 & 16:25:51.58 & 26:46:21.39 & 2.4194 & 23.41 & 0.28 & 0.85 & 20.52 & 1.23 & $4\times 900$\\
Q1623-BX472 & 16:25:52.87 & 26:46:39.63 & 2.1142 & 24.58 & 0.16 & 0.91 & 20.80 & 1.91 & $4\times 900$\\
Q1623-BX502 & 16:25:54.38 & 26:44:09.25 & 2.1558 & 24.35 & 0.22 & 0.50 & 22.04 & 0.99 & $3\times 900$\\
Q1623-BX511 & 16:25:56.11 & 26:44:44.57 & 2.2421 & 25.37 & 0.42 & 1.05 & 21.78 & ... & $4\times 900$\\
Q1623-BX513 & 16:25:55.86 & 26:46:50.30 & 2.2473 & 23.25 & 0.26 & 0.68 & 20.21 & 1.83 & $2\times 900$\\
Q1623-BX516 & 16:25:56.27 & 26:44:08.19 & 2.4236 & 23.94 & 0.30 & 0.82 & 20.41 & 2.29 & $3\times 900$\\
Q1623-BX522 & 16:25:55.76 & 26:44:53.28 & 2.4757 & 24.50 & 0.31 & 1.19 & 20.75 & 2.03 & $4\times 900$\\
Q1623-BX528 & 16:25:56.44 & 26:50:15.44 & 2.2682 & 23.56 & 0.25 & 0.71 & 19.75 & 1.79 & $4\times 900$\\
Q1623-BX543 & 16:25:57.70 & 26:50:08.59 & 2.5211 & 23.11 & 0.44 & 0.96 & 20.54 & 1.31 & $4\times 900$\\
Q1623-BX586 & 16:26:01.52 & 26:45:41.58 & 2.1045 & 24.58 & 0.32 & 0.87 & 20.84 & 1.79 & $4\times 900$\\
Q1623-BX599 & 16:26:02.54 & 26:45:31.90 & 2.3304 & 23.44 & 0.22 & 0.80 & 19.93 & 2.09 & $4\times 900$\\
Q1623-BX663 & 16:26:04.58 & 26:48:00.20 & 2.4333 & 24.14 & 0.24 & 1.02 & 19.92 & 2.59 & $3\times 900$\\
Q1623-MD107 & 16:25:53.87 & 26:45:15.46 & 2.5373 & 25.35 & 0.12 & 1.43 & 22.43 & 1.72 & $4\times 900$\\
Q1623-MD66 & 16:25:40.39 & 26:50:08.88 & 2.1075 & 23.95 & 0.37 & 1.40 & 20.15 & 1.74 & $3\times 900$\\
Q1700-BX490 & 17:01:14.83 & 64:09:51.69 & 2.3960 & 22.88 & 0.36 & 0.92 & 19.99 & 1.54 & $3\times 900$\\
Q1700-BX505 & 17:00:48.22 & 64:10:05.86 & 2.3089 & 25.17 & 0.45 & 1.28 & 20.85 & 2.17 & $4\times 900$\\
Q1700-BX523 & 17:00:41.71 & 64:10:14.88 & 2.4756 & 24.51 & 0.46 & 1.28 & 20.93 & 1.86 & $4\times 900$\\
Q1700-BX530 & 17:00:36.86 & 64:10:17.38 & 1.9429 & 23.05 & 0.21 & 0.69 & 19.92 & 1.23 & $4\times 900$\\
Q1700-BX536 & 17:01:08.94 & 64:10:24.95 & 1.9780 & 23.00 & 0.21 & 0.79 & 19.71 & 1.31 & $4\times 900$\\
Q1700-BX561 & 17:01:04.18 & 64:10:43.83 & 2.4332 & 24.65 & 0.19 & 1.04 & 19.87 & 2.43 & $2\times 900$\\
Q1700-BX581 & 17:01:02.73 & 64:10:51.30 & 2.4022 & 23.87 & 0.28 & 0.62 & 20.79 & 1.94 & $2\times 900$\\
Q1700-BX681 & 17:01:33.76 & 64:12:04.28 & 1.7396 & 22.04 & 0.19 & 0.40 & 19.18 & 1.27 & $4\times 900$\\
Q1700-BX691 & 17:01:06.00 & 64:12:10.27 & 2.1895 & 25.33 & 0.22 & 0.66 & 20.68 & 1.82 & $4\times 900$\\
Q1700-BX717 & 17:00:56.99 & 64:12:23.76 & 2.4353 & 24.78 & 0.20 & 0.61 & 21.89 & 1.49 & $4\times 900$\\
Q1700-BX759 & 17:00:59.55 & 64:12:55.45 & 2.4213 & 24.43 & 0.36 & 1.29 & 21.23 & 1.35 & $2\times 900$\\
Q1700-BX794 & 17:00:47.30 & 64:13:18.70 & 2.2473 & 23.60 & 0.35 & 0.58 & 20.53 & 1.37 & $3\times 900$\\
Q1700-BX917 & 17:01:16.11 & 64:14:19.80 & 2.3069 & 24.43 & 0.28 & 0.95 & 20.03 & 1.95 & $3\times 900$\\
Q1700-MD103 & 17:01:00.21 & 64:11:55.58 & 2.3148 & 24.23 & 0.46 & 1.49 & 19.94 & 1.96 & $900+600$\\
Q1700-MD109 & 17:01:04.48 & 64:12:09.29 & 2.2942 & 25.46 & 0.26 & 1.44 & 21.77 & 1.75 & $4\times 900$\\
Q1700-MD154 & 17:01:38.39 & 64:14:57.37 & 2.6291 & 23.23 & 0.73 & 1.91 & 19.68 & 2.08 & $3\times 900$\\
Q1700-MD174 & 17:00:54.54 & 64:16:24.76 & 2.3423 & 24.56 & 0.32 & 1.50 & 19.90 & ... & $4\times 900$\\
Q1700-MD69 & 17:00:47.62 & 64:09:44.78 & 2.2883 & 24.85 & 0.37 & 1.50 & 20.05 & 2.60 & $4\times 900$\\
Q1700-MD94 & 17:00:42.02 & 64:11:24.22 & 2.3362 & 24.72 & 0.94 & 2.06 & 19.65 & 2.46 & $3\times 900$\\
Q2343-BM133 & 23:46:16.18 & 12:48:09.31 & 1.4774 & 22.59 & 0.00 & 0.19 & 20.50 & 0.66 & $3\times 900$\\
Q2343-BM181 & 23:46:27.03 & 12:49:19.65 & 1.4951 & 24.77 & 0.12 & 0.29 & ... & ... & $4\times 900$\\
Q2343-BX163 & 23:46:04.78 & 12:45:37.78 & 2.1213 & 24.07 & -0.01 & 0.71 & 21.38 & 0.97 & $4\times 900$\\
Q2343-BX169 & 23:46:05.03 & 12:45:40.77 & 2.2094 & 23.11 & 0.19 & 0.72 & 20.75 & 1.05 & $4\times 900$\\
Q2343-BX182 & 23:46:18.04 & 12:45:51.11 & 2.2879 & 23.74 & 0.14 & 0.56 & 21.60 & 0.92 & $4\times 900$\\
Q2343-BX236 & 23:46:18.71 & 12:46:15.97 & 2.4348 & 24.28 & 0.14 & 0.71 & 21.25 & 1.60 & $3\times 900$\\
Q2343-BX336 & 23:46:29.53 & 12:47:04.76 & 2.5439 & 23.91 & 0.40 & 1.15 & 20.80 & 1.75 & $4\times 900$\\
Q2343-BX341 & 23:46:23.24 & 12:47:07.97 & 2.5749 & 24.21 & 0.38 & 0.89 & 21.40 & 2.07 & $3\times 900$\\
Q2343-BX378 & 23:46:33.90 & 12:47:26.20 & 2.0441 & 24.80 & 0.26 & 0.55 & 21.90 & 1.30 & $4\times 900$\\
Q2343-BX389 & 23:46:28.90 & 12:47:33.55 & 2.1716 & 24.85 & 0.28 & 1.26 & 20.18 & 2.74 & $3\times 900$\\
Q2343-BX390 & 23:46:24.72 & 12:47:33.80 & 2.2313 & 24.36 & 0.24 & 0.79 & 21.29 & 1.73 & $4\times 900$\\
Q2343-BX391 & 23:46:28.07 & 12:47:31.82 & 2.1740 & 24.51 & 0.25 & 1.01 & 21.93 & 1.35 & $3\times 900$\\
Q2343-BX418 & 23:46:18.57 & 12:47:47.38 & 2.3052 & 23.99 & -0.05 & 0.37 & 21.88 & 1.76 & $5\times 900$\\
Q2343-BX429 & 23:46:25.25 & 12:47:51.20 & 2.1751 & 25.12 & 0.30 & 0.85 & 21.88 & 1.97 & $4\times 900$\\
Q2343-BX435 & 23:46:26.36 & 12:47:55.06 & 2.1119 & 24.23 & 0.38 & 1.03 & 20.38 & 1.75 & $4\times 900$\\
Q2343-BX436 & 23:46:09.06 & 12:47:56.00 & 2.3277 & 23.07 & 0.12 & 0.47 & 21.04 & 0.73 & $4\times 900$\\
Q2343-BX442 & 23:46:19.36 & 12:47:59.69 & 2.1760 & 24.48 & 0.40 & 1.14 & 19.85 & 2.36 & $5\times 900$\\
Q2343-BX461 & 23:46:32.96 & 12:48:08.15 & 2.5662 & 24.40 & 0.44 & 0.90 & 21.67 & 1.99 & $4\times 900$\\
Q2343-BX474 & 23:46:32.88 & 12:48:14.08 & 2.2257 & 24.42 & 0.31 & 1.15 & 20.56 & 1.73 & $4\times 900$\\
Q2343-BX480 & 23:46:21.90 & 12:48:15.61 & 2.2313 & 23.77 & 0.29 & 1.03 & 20.44 & 1.92 & $4\times 900$\\
Q2343-BX493 & 23:46:14.46 & 12:48:21.64 & 2.3396 & 23.63 & 0.28 & 0.78 & 21.65 & 0.75 & $2\times 900$\\
Q2343-BX513 & 23:46:11.13 & 12:48:32.14 & 2.1092\tablenotemark{d} & 23.93 & 0.20 & 0.41 & 20.10 & 1.87 & $4\times 900$\\
Q2343-BX529 & 23:46:09.72 & 12:48:40.33 & 2.1129 & 24.42 & 0.20 & 0.91 & 21.41 & 1.52 & $2\times 900$\\
Q2343-BX537 & 23:46:25.55 & 12:48:44.54 & 2.3396 & 24.44 & 0.23 & 0.72 & 21.43 & 1.65 & $4\times 900$\\
Q2343-BX587 & 23:46:29.18 & 12:49:03.34 & 2.2430 & 23.47 & 0.32 & 1.12 & 20.12 & 1.82 & $3\times 900$\\
Q2343-BX599 & 23:46:13.85 & 12:49:11.31 & 2.0116 & 23.50 & 0.10 & 0.81 & 20.40 & 1.19 & $4\times 900$\\
Q2343-BX601 & 23:46:20.40 & 12:49:12.91 & 2.3769 & 23.48 & 0.22 & 0.75 & 20.55 & 1.50 & $4\times 900$\\
Q2343-BX610 & 23:46:09.43 & 12:49:19.21 & 2.2094 & 23.58 & 0.34 & 0.75 & 19.21 & 2.24 & $4\times 900$\\
Q2343-BX660 & 23:46:29.43 & 12:49:45.54 & 2.1735 & 24.36 & -0.09 & 0.45 & 20.98 & 2.26 & $2\times 900$\\
Q2343-MD59 & 23:46:26.90 & 12:47:39.87 & 2.0116 & 24.99 & 0.20 & 1.47 & 20.14 & 2.59 & $4\times 900$\\
Q2343-MD62 & 23:46:27.23 & 12:47:43.48 & 2.1752 & 25.29 & 0.21 & 1.23 & 21.45 & 2.23 & $4\times 900$\\
Q2343-MD80 & 23:46:10.79 & 12:48:33.24 & 2.0138 & 24.81 & 0.09 & 1.33 & 21.38 & 1.15 & $4\times 900$\\
Q2346-BX120 & 23:48:26.30 & 00:20:33.16 & 2.2664 & 25.08 & 0.02 & 0.91 & ... & ... & $4\times 900$\\
Q2346-BX220 & 23:48:46.10 & 00:22:20.95 & 1.9677 & 23.57 & 0.29 & 0.90 & 20.82 & ... & $4\times 900$\\
Q2346-BX244 & 23:48:09.61 & 00:22:36.18 & 1.6465 & 24.54 & 0.41 & 1.16 & ... & ... & $4\times 900$\\
Q2346-BX404 & 23:48:21.40 & 00:24:43.07 & 2.0282 & 23.39 & 0.18 & 0.41 & 20.05 & ... & $5\times 900$\\
Q2346-BX405 & 23:48:21.22 & 00:24:45.46 & 2.0300 & 23.36 & 0.08 & 0.60 & 20.27 & ... & $5\times 900$\\
Q2346-BX416 & 23:48:18.21 & 00:24:55.30 & 2.2404 & 23.49 & 0.40 & 0.75 & 20.30 & ... & $3\times 900$\\
Q2346-BX482 & 23:48:12.97 & 00:25:46.34 & 2.2569 & 23.32 & 0.22 & 0.90 & ... & ... & $4\times 900$\\
SSA22a-MD41\tablenotemark{c} & 22:17:39.97 & 00:17:11.04 & 2.1713 & 23.31 & 0.19 & 1.31 & ... & ... & $15\times 720$\\
West-BM115 & 14:17:37.57 & 52:27:05.42 & 1.6065 & 23.41 & 0.28 & 0.36 & ... & ... & $10\times 900$\\
West-BX600 & 14:17:15.55 & 52:36:15.64 & 2.1607 & 23.94 & 0.10 & 0.46 & ... & ... & $5\times 900$\\

\enddata

\tablenotetext{a}{$U_n$, $G$ and ${\cal R}$ magnitudes are AB.}
\tablenotetext{b}{$J$ and $K_s$ magnitudes are Vega.}
\tablenotetext{c}{Observed with the ISAAC spectrograph on the VLT;
  previously discussed by \citet{ess+03}.}
\tablenotetext{d}{Q2343-BX513 was observed a second time with a
  different position angle, and yielded an \Ha\ redshift $z_{\Ha}=2.1079.$}
\end{deluxetable}

\begin{deluxetable}{l c c l}
\tablewidth{0pt}
\tabletypesize{\footnotesize}
\tablecaption{Near-IR Imaging\label{tab:irobs}}
\tablehead{
\colhead{Field} & 
\colhead{Band} & 
\colhead{Exposure time} & 
\colhead{Depth\tablenotemark{a}}\\  
\colhead{} & 
\colhead{} & 
\colhead{(hrs)} &
\colhead{}
}
\startdata
GOODS-N & $J$ & 13.0 & 24.1 \\
        & $K_s$ & 10.3 & 22.6 \\
Q1623   & $J$ & 9.8 & 23.8 \\
        & $K_s$ & 11.2 & 22.3 \\
Q1700   & $J$ & 10.7 & 24.0 \\
        & $K_s$ & 11.0 & 22.2 \\
Q2343   & $J$ & 10.7 & 24.0 \\
        & $K_s$ & 12.1 & 22.3 \\
Q2346   & $K_s$ & 2.6 & 21.2 \\
\enddata

\tablenotetext{a}{Approximate 3$\sigma$ image depth, in Vega magnitudes.}
\end{deluxetable}

\begin{deluxetable}{l l l l c c}
\tablewidth{0pt}
\tabletypesize{\footnotesize}
\tablecaption{Results of SED Fitting\label{tab:sedfit}}
\tablehead{
\colhead{Object} & 
\colhead{$\tau$} & 
\colhead{$E(B-V)$\tablenotemark{a}} & 
\colhead{Age\tablenotemark{b}} & 
\colhead{SFR\tablenotemark{c}} & 
\colhead{$M_{\star}$\tablenotemark{d}} \\
\colhead{} &
\colhead{(Myr)} &
\colhead{} &
\colhead{(Myr)} &
\colhead{(\msunyr)} &
\colhead{($10^{10}$ \msun)} 
}
\startdata
HDF-BMZ1156 & 500 & 0.000 & 1900 & 4 & 9.6\\
HDF-BX1277 & const & 0.095 & 321 & 15 & 0.5\\
HDF-BX1303 & const & 0.100 & 1139 & 7 & 0.8\\
HDF-BX1311 & const & 0.105 & 255 & 34 & 0.9\\
HDF-BX1322 & const & 0.085 & 360 & 19 & 0.7\\
HDF-BX1332 & const & 0.290 & 15 & 159 & 0.2\\
HDF-BX1368 & const & 0.160 & 404 & 37 & 1.5\\
HDF-BX1376 & const & 0.070 & 227 & 9 & 0.2\\
HDF-BX1388 & const & 0.265 & 3000 & 23 & 7.0\\
HDF-BX1397 & const & 0.150 & 1015 & 17 & 1.7\\
HDF-BX1409 & const & 0.290 & 1015 & 27 & 2.7\\
HDF-BX1439 & const & 0.175 & 2100 & 21 & 4.5\\
HDF-BX1479 & const & 0.110 & 806 & 13 & 1.0\\
HDF-BX1564 & 100 & 0.065 & 360 & 9 & 3.3\\
HDF-BX1567 & 50 & 0.050 & 227 & 5 & 2.3\\
HDF-BX305 & const & 0.285 & 571 & 53 & 3.0\\
Q1623-BX376 & const & 0.175 & 64 & 80 & 0.5\\
Q1623-BX428 & 50 & 0.000 & 255 & 1 & 1.1\\
Q1623-BX429 & const & 0.120 & 227 & 23 & 0.5\\
Q1623-BX432 & const & 0.060 & 1278 & 6 & 0.8\\
Q1623-BX447 & 500 & 0.050 & 1434 & 5 & 4.4\\
Q1623-BX449 & const & 0.110 & 2600 & 9 & 2.4\\
Q1623-BX452 & const & 0.195 & 3000 & 14 & 4.3\\
Q1623-BX453 & const & 0.275 & 454 & 107 & 4.9\\
Q1623-BX455 & const & 0.265 & 15 & 58 & 0.09\\
Q1623-BX458 & const & 0.165 & 571 & 55 & 3.1\\
Q1623-BX472 & const & 0.130 & 3000 & 11 & 3.2\\
Q1623-BX502 & const & 0.220 & 6 & 72 & 0.04\\
Q1623-BX511 & const & 0.235 & 571 & 13 & 0.8\\
Q1623-BX513 & const & 0.145 & 454 & 46 & 2.1\\
Q1623-BX516 & const & 0.145 & 1800 & 28 & 5.1\\
Q1623-BX522 & const & 0.180 & 2600 & 24 & 6.2\\
Q1623-BX528 & const & 0.175 & 2750 & 44 & 12.2\\
Q1623-BX543 & const & 0.305 & 8 & 528 & 0.4\\
Q1623-BX586 & const & 0.195 & 1434 & 17 & 2.5\\
Q1623-BX599 & const & 0.125 & 1900 & 35 & 6.7\\
Q1623-BX663 & 1000 & 0.135 & 2000 & 21 & 13.2\\
Q1623-MD107 & const & 0.060 & 1015 & 4 & 0.4\\
Q1623-MD66 & const & 0.235 & 905 & 43 & 3.9\\
Q1700-BX490 & const & 0.285 & 10 & 448 & 0.4\\
Q1700-BX505 & const & 0.270 & 1800 & 20 & 3.6\\
Q1700-BX523 & const & 0.260 & 255 & 42 & 1.1\\
Q1700-BX530 & 50 & 0.045 & 203 & 6 & 1.8\\
Q1700-BX536 & 50 & 0.115 & 180 & 15 & 2.8\\
Q1700-BX561 & 500 & 0.130 & 1609 & 10 & 11.5\\
Q1700-BX581 & const & 0.215 & 35 & 70 & 0.2\\
Q1700-BX681 & const & 0.315 & 10 & 628 & 0.6\\
Q1700-BX691 & 1000 & 0.125 & 2750 & 5 & 7.6\\
Q1700-BX717 & const & 0.090 & 509 & 8 & 0.4\\
Q1700-BX759 & const & 0.230 & 640 & 37 & 2.4\\
Q1700-BX794 & const & 0.130 & 454 & 25 & 1.1\\
Q1700-BX917 & 200 & 0.040 & 806 & 4 & 4.0\\
Q1700-MD103 & const & 0.305 & 1015 & 65 & 6.6\\
Q1700-MD109 & const & 0.175 & 2200 & 8 & 1.7\\
Q1700-MD154 & const & 0.335 & 128 & 347 & 4.4\\
Q1700-MD174 & 1000 & 0.195 & 2400 & 24 & 23.6\\
Q1700-MD69 & 2000 & 0.275 & 2750 & 31 & 18.6\\
Q1700-MD94 & const & 0.500 & 719 & 213 & 15.3\\
Q2343-BM133 & const & 0.115 & 143 & 35 & 0.5\\
Q2343-BX163 & const & 0.050 & 1434 & 9 & 1.3\\
Q2343-BX169 & const & 0.125 & 203 & 46 & 0.9\\
Q2343-BX182 & const & 0.100 & 180 & 23 & 0.4\\
Q2343-BX236 & const & 0.085 & 1680 & 13 & 2.1\\
Q2343-BX336 & const & 0.210 & 321 & 58 & 1.9\\
Q2343-BX341 & const & 0.210 & 102 & 50 & 0.5\\
Q2343-BX378 & const & 0.165 & 255 & 11 & 0.3\\
Q2343-BX389 & const & 0.250 & 2750 & 22 & 6.1\\
Q2343-BX390 & const & 0.150 & 404 & 17 & 0.7\\
Q2343-BX391 & const & 0.195 & 64 & 25 & 0.2\\
Q2343-BX418 & const & 0.035 & 81 & 12 & 0.1\\
Q2343-BX429 & const & 0.185 & 321 & 12 & 0.4\\
Q2343-BX435 & const & 0.225 & 1434 & 30 & 4.4\\
Q2343-BX436 & const & 0.070 & 321 & 33 & 1.1\\
Q2343-BX442 & 2000 & 0.225 & 2750 & 25 & 14.7\\
Q2343-BX461 & const & 0.250 & 15 & 86 & 0.1\\
Q2343-BX474 & const & 0.215 & 2750 & 26 & 7.2\\
Q2343-BX480 & const & 0.165 & 905 & 33 & 3.0\\
Q2343-BX493 & const & 0.255 & 6 & 220 & 0.1\\
Q2343-BX513 & const & 0.135 & 3000 & 20 & 5.9\\
Q2343-BX529 & const & 0.145 & 404 & 14 & 0.6\\
Q2343-BX537 & const & 0.130 & 571 & 15 & 0.8\\
Q2343-BX587 & const & 0.180 & 719 & 49 & 3.5\\
Q2343-BX599 & const & 0.100 & 1609 & 21 & 3.3\\
Q2343-BX601 & const & 0.125 & 640 & 36 & 2.3\\
Q2343-BX610 & 1000 & 0.155 & 2100 & 32 & 23.2\\
Q2343-BX660 & const & 0.010 & 2750 & 5 & 1.4\\
Q2343-MD59 & 2000 & 0.200 & 3000 & 11 & 7.6\\
Q2343-MD62 & const & 0.150 & 2750 & 7 & 1.9\\
Q2343-MD80 & 50 & 0.020 & 255 & 1 & 0.6\\
Q2346-BX220 & 50 & 0.055 & 227 & 4 & 1.7\\
Q2346-BX404 & const & 0.095 & 1800 & 22 & 4.0\\
Q2346-BX405 & 100 & 0.010 & 321 & 7 & 1.6\\
Q2346-BX416 & const & 0.195 & 454 & 55 & 2.5\\

\enddata
\tablenotetext{a}{Typical uncertainty $\langle \sigma_{E(B-V)}/E(B-V)
  \rangle = 0.7$.}
\tablenotetext{b}{Typical uncertainty $\langle \sigma_{\rm Age}/\rm Age
  \rangle = 0.5$.}
\tablenotetext{c}{Typical uncertainty $\langle \sigma_{\rm SFR}/\rm SFR
  \rangle = 0.6$.}
\tablenotetext{d}{Typical uncertainty $\langle \sigma_{M_{\star}}/M_{\star}
  \rangle = 0.4$.}
\end{deluxetable}

\begin{deluxetable}{l l l l c c c c c c}
\tablewidth{0pt}
\tabletypesize{\footnotesize}
\tablecaption{Kinematics\label{tab:kin}}
\tablehead{
\colhead{Object} & 
\colhead{$z_{\Ha}$} & 
\colhead{$z_{\rm abs}$\tablenotemark{a}} & 
\colhead{$z_{\lya}$\tablenotemark{b}} & 
\colhead{$F_{\Ha}$\tablenotemark{c}} & 
\colhead{$r_{\Ha}$\tablenotemark{d}} &
\colhead{$\sigma$\tablenotemark{e}} &
\colhead{$v_c$\tablenotemark{f}} &
\colhead{$M_{\rm dyn}$\tablenotemark{g}} &
\colhead{$M_{\star}$\tablenotemark{h}} \\
\colhead{} &
\colhead{} &
\colhead{} &
\colhead{} &
\colhead{} &
\colhead{(kpc)} &
\colhead{(\kms)} &
\colhead{(\kms)} &
\colhead{($10^{10}$ \msun)} &
\colhead{($10^{10}$ \msun)} 
}
\startdata
CDFb-BN88 & 2.2615 & ... & ... & $2.6 \pm 0.2$ & 7.4 & $96^{+20}_{-20}$ & ... & 5.3 & ...\\
HDF-BX1055 & 2.4899 & 2.4865 & 2.4959 & $2.6 \pm 0.6$ & 3.8 & $<59$ & ... & $<1.0$ & ...\\
HDF-BX1084 & 2.4403 & 2.4392 & ... & $7.3 \pm 0.2$ & 4.8 & $102^{+6}_{-6}$ & ... & 5.1 & ...\\
HDF-BX1085 & 2.2407 & 2.2381 & ... & $1.1 \pm 0.1$ & 2.4 & $...$ & ... & ... & ...\\
HDF-BX1086 & 2.4435 & ... & ... & $1.8 \pm 0.2$ & $<1.9$ & $95^{+20}_{-20}$ & ... & $<1.6$ & ...\\
HDF-BX1277 & 2.2713 & 2.2686 & ... & $5.3 \pm 0.2$ & 5.0 & $63^{+9}_{-10}$ & ... & 1.3 & 0.5\\
HDF-BX1303 & 2.3003 & 2.3024 & 2.3051 & $2.6 \pm 0.5$ & 3.4 & $...$ & ... & ... & 0.8\\
HDF-BX1311 & 2.4843 & 2.4804 & 2.4890 & $8.0 \pm 0.4$ & 5.5 & $88^{+9}_{-9}$ & ... & 3.2 & 0.9\\
HDF-BX1322 & 2.4443 & 2.4401 & 2.4491 & $2.0 \pm 0.2$ & 4.6 & $<47$ & ... & $<1.1$ & 0.7\\
HDF-BX1332 & 2.2136 & 2.2113 & ... & $4.4 \pm 0.3$ & 5.5 & $54^{+16}_{-20}$ & $47\pm14$ & 1.2 & 0.2\\
HDF-BX1368 & 2.4407 & 2.4380 & 2.4455 & $8.8 \pm 0.4$ & 7.4 & $139^{+9}_{-9}$ & ... & 6.6 & 1.5\\
HDF-BX1376 & 2.4294 & 2.4266 & 2.4338 & $2.2 \pm 0.2$ & 2.9 & $96^{+22}_{-24}$ & ... & 2.2 & 0.2\\
HDF-BX1388 & 2.0317 & 2.0305 & ... & $5.8 \pm 0.5$ & 7.9 & $140^{+21}_{-21}$ & ... & 11.7 & 7.0\\
HDF-BX1397 & 2.1328 & 2.1322 & ... & $5.3 \pm 0.5$ & 8.6 & $125^{+22}_{-24}$ & $109\pm19$ & 10.3 & 1.7\\
HDF-BX1409 & 2.2452 & 2.2433 & ... & $8.5 \pm 0.6$ & 7.0 & $158^{+18}_{-18}$ & ... & 13.3 & 2.7\\
HDF-BX1439 & 2.1865 & 2.1854 & 2.1913 & $8.8 \pm 0.3$ & 8.2 & $120^{+8}_{-8}$ & ... & 8.6 & 4.5\\
HDF-BX1479 & 2.3745 & 2.3726 & 2.3823 & $2.5 \pm 0.2$ & 4.3 & $46^{+18}_{-21}$ & ... & 0.9 & 1.0\\
HDF-BX1564 & 2.2225 & 2.2219 & ... & $8.6 \pm 0.7$ & 15.4 & $99^{+16}_{-18}$ & ... & 11.7 & 3.3\\
HDF-BX1567 & 2.2256 & 2.2257 & ... & $4.0 \pm 0.6$ & 2.6 & $<62$ & ... & $<1.8$ & 2.3\\
HDF-BX305 & 2.4839 & 2.4825 & ... & $4.2 \pm 0.4$ & 3.8 & $140^{+22}_{-24}$ & ... & 5.6 & 3.0\\
HDF-BMZ1156\tablenotemark{i} & 2.2151 & ... & ... & $5.4 \pm 0.4$ & $<1.9$ & $196^{+24}_{-26}$ & ... & $<7.1$ & 9.6\\
Q0201-B13 & 2.1663 & ... & ... & $2.4 \pm 0.1$ & 7.4 & $62^{+10}_{-10}$ & ... & 2.2 & ...\\
Q1307-BM1163 & 1.4105 & 1.4080 & ... & $28.7 \pm 1.2$ & 6.7 & $125^{+8}_{-8}$ & ... & 8.2 & ...\\
Q1623-BX151\tablenotemark{i} & 2.4393 & ... & ... & $3.5 \pm 0.6$ & 8.4 & $98^{+36}_{-40}$ & ... & 6.3 & ...\\
Q1623-BX214 & 2.4700 & 2.4674 & ... & $5.3 \pm 0.4$ & 4.1 & $55^{+14}_{-15}$ & ... & 1.0 & ...\\
Q1623-BX215 & 2.1814 & 2.1819 & ... & $4.8 \pm 0.3$ & 5.0 & $70^{+15}_{-16}$ & ... & 2.0 & ...\\
Q1623-BX252 & 2.3367 & ... & ... & $1.1 \pm 0.3$ & $<1.9$ & $...$ & ... & ... & ...\\
Q1623-BX274 & 2.4100 & 2.4081 & 2.4130 & $9.5 \pm 0.3$ & 4.8 & $121^{+9}_{-9}$ & ... & 5.5 & ...\\
Q1623-BX344 & 2.4224 & ... & ... & $17.1 \pm 0.8$ & 4.1 & $92^{+9}_{-10}$ & ... & 2.7 & ...\\
Q1623-BX366 & 2.4204 & 2.4169 & ... & $7.8 \pm 1.5$ & 8.4 & $103^{+39}_{-44}$ & ... & 7.0 & ...\\
Q1623-BX376 & 2.4085 & 2.4061 & 2.4153 & $5.3 \pm 0.7$ & 7.0 & $261^{+54}_{-54}$ & ... & 36.6 & 0.5\\
Q1623-BX428 & 2.0538 & 2.0514 & 2.0594 & $2.7 \pm 0.5$ & 1.4 & $...$ & ... & ... & 1.1\\
Q1623-BX429 & 2.0160 & 2.0142 & ... & $5.1 \pm 0.5$ & 6.7 & $57^{+22}_{-27}$ & ... & 1.7 & 0.5\\
Q1623-BX432 & 2.1817 & ... & ... & $5.4 \pm 0.3$ & 4.6 & $54^{+15}_{-16}$ & ... & 1.0 & 0.8\\
Q1623-BX447 & 2.1481 & 2.1478 & ... & $5.6 \pm 0.3$ & 5.3 & $174^{+15}_{-15}$ & $160\pm22$ & 12.5 & 4.4\\
Q1623-BX449 & 2.4188 & ... & ... & $3.5 \pm 0.9$ & $<1.9$ & $<72$ & ... & $<2.5$ & 2.4\\
Q1623-BX452 & 2.0595 & 2.0595 & ... & $4.4 \pm 0.4$ & 9.6 & $129^{+20}_{-20}$ & ... & 12.9 & 4.3\\
Q1623-BX453 & 2.1816 & 2.1724 & 2.1838 & $13.8 \pm 0.2$ & 4.1 & $61^{+4}_{-4}$ & ... & 1.2 & 4.9\\
Q1623-BX455 & 2.4074 & 2.4066 & ... & $18.8 \pm 1.1$ & 5.0 & $187^{+15}_{-15}$ & ... & 13.6 & 0.09\\
Q1623-BX458 & 2.4194 & 2.4174 & ... & $4.3 \pm 0.5$ & 8.4 & $160^{+27}_{-27}$ & $116\pm18$ & 16.8 & 3.1\\
Q1623-BX472 & 2.1142 & 2.1144 & ... & $3.9 \pm 0.2$ & 6.0 & $110^{+10}_{-10}$ & $110\pm12$ & 5.7 & 3.2\\
Q1623-BX502 & 2.1558 & 2.1549 & 2.1600 & $13.2 \pm 0.4$ & 6.2 & $75^{+8}_{-8}$ & ... & 2.7 & 0.04\\
Q1623-BX511 & 2.2421 & ... & ... & $3.4 \pm 0.3$ & 6.2 & $152^{+34}_{-36}$ & $80\pm18$ & 11.3 & 0.8\\
Q1623-BX513 & 2.2473 & 2.2469 & 2.2525 & $3.3 \pm 0.3$ & 3.4 & $...$ & ... & ... & 2.1\\
Q1623-BX516 & 2.4236 & 2.4217 & ... & $5.2 \pm 1.0$ & 2.6 & $114^{+15}_{-16}$ & ... & 2.7 & 5.1\\
Q1623-BX522 & 2.4757 & 2.4742 & ... & $2.8 \pm 0.3$ & 6.2 & $<46$ & ... & $<1.0$ & 6.2\\
Q1623-BX528 & 2.2682 & 2.2683 & ... & $7.7 \pm 0.4$ & 7.9 & $142^{+14}_{-14}$ & $76\pm12$ & 12.0 & 12.2\\
Q1623-BX543 & 2.5211 & 2.5196 & ... & $8.6 \pm 0.7$ & 7.4 & $148^{+24}_{-26}$ & ... & 12.6 & 0.4\\
Q1623-BX586 & 2.1045 & ... & ... & $5.1 \pm 0.4$ & 5.3 & $124^{+18}_{-20}$ & ... & 6.2 & 2.5\\
Q1623-BX599 & 2.3304 & 2.3289 & 2.3402 & $18.1 \pm 0.4$ & 5.8 & $162^{+8}_{-8}$ & ... & 11.6 & 6.7\\
Q1623-BX663\tablenotemark{i} & 2.4333 & 2.4296 & 2.4353 & $8.2 \pm 0.3$ & 7.0 & $132^{+12}_{-14}$ & ... & 9.3 & 13.2\\
Q1623-MD107 & 2.5373 & ... & ... & $3.7 \pm 0.4$ & 3.6 & $<43$ & ... & $<0.9$ & 0.4\\
Q1623-MD66 & 2.1075 & 2.1057 & ... & $19.7 \pm 0.3$ & 5.3 & $120^{+4}_{-4}$ & ... & 5.9 & 3.9\\
Q1700-BX490 & 2.3960 & 2.3969 & 2.4043 & $17.7 \pm 0.6$ & 5.5 & $110^{+9}_{-9}$ & ... & 5.1 & 0.4\\
Q1700-BX505 & 2.3089 & ... & ... & $3.6 \pm 0.3$ & 7.4 & $120^{+20}_{-21}$ & ... & 8.3 & 3.6\\
Q1700-BX523 & 2.4756 & ... & ... & $4.7 \pm 0.5$ & 7.7 & $130^{+26}_{-26}$ & ... & 9.9 & 1.1\\
Q1700-BX530 & 1.9429 & 1.9411 & ... & $12.2 \pm 0.7$ & 6.2 & $<37$ & ... & $<0.6$ & 1.8\\
Q1700-BX536 & 1.9780 & ... & ... & $11.3 \pm 0.7$ & 7.9 & $89^{+14}_{-15}$ & ... & 4.7 & 2.8\\
Q1700-BX561 & 2.4332 & 2.4277 & ... & $1.9 \pm 0.6$ & $<1.9$ & $...$ & ... & ... & 11.5\\
Q1700-BX581 & 2.4022 & 2.3984 & ... & $4.0 \pm 0.7$ & 4.3 & $...$ & ... & ... & 0.2\\
Q1700-BX681 & 1.7396 & 1.7398 & 1.7467 & $6.3 \pm 0.2$ & 8.6 & $...$ & ... & ... & 0.6\\
Q1700-BX691 & 2.1895 & ... & ... & $7.7 \pm 0.3$ & 6.7 & $170^{+14}_{-14}$ & $220\pm14$ & 15.0 & 7.6\\
Q1700-BX717 & 2.4353 & ... & 2.4376 & $3.8 \pm 0.4$ & 4.8 & $<47$ & ... & $<1.1$ & 0.4\\
Q1700-BX759 & 2.4213 & ... & ... & $1.3 \pm 0.5$ & 2.4 & $...$ & ... & ... & 2.4\\
Q1700-BX794 & 2.2473 & ... & ... & $6.8 \pm 0.4$ & 4.8 & $80^{+14}_{-14}$ & ... & 2.4 & 1.1\\
Q1700-BX917 & 2.3069 & 2.3027 & ... & $7.4 \pm 0.5$ & 10.6 & $99^{+12}_{-14}$ & ... & 8.2 & 4.0\\
Q1700-MD69 & 2.2883 & 2.288 & ... & $8.2 \pm 0.9$ & 9.4 & $155^{+15}_{-16}$ & ... & 17.2 & 18.6\\
Q1700-MD94\tablenotemark{i} & 2.3362 & ... & ... & $2.8 \pm 0.3$ & 9.6 & $730^{+82}_{-82}$ & ... & 382 & 15.3\\
Q1700-MD103 & 2.3148 & ... & ... & $4.1 \pm 0.8$ & 8.4 & $75^{+21}_{-24}$ & $100\pm19$ & 3.7 & 6.6\\
Q1700-MD109 & 2.2942 & ... & ... & $8.9 \pm 0.7$ & 4.1 & $93^{+24}_{-26}$ & ... & 2.7 & 1.7\\
Q1700-MD154 & 2.6291 & ... & ... & $7.5 \pm 0.5$ & 8.2 & $57^{+34}_{-50}$ & ... & 2.1 & 4.4\\
Q1700-MD174\tablenotemark{i} & 2.3423 & ... & ... & $12.9 \pm 1.4$ & 3.6 & $444^{+50}_{-50}$ & ... & 56.5 & 23.6\\
Q2343-BM133 & 1.4774 & 1.4769 & ... & $28.7 \pm 0.8$ & 7.7 & $55^{+6}_{-6}$ & ... & 1.8 & 0.5\\
Q2343-BM181 & 1.4951 & 1.4952 & ... & $3.4 \pm 0.5$ & 7.7 & $39^{+17}_{-30}$ & $45\pm18$ & 1.0 & ...\\
Q2343-BX163 & 2.1213 & ... & ... & $2.2 \pm 0.4$ & 4.1 & $...$ & ... & ... & 1.3\\
Q2343-BX169 & 2.2094 & 2.2105 & 2.2173 & $4.7 \pm 0.3$ & 3.1 & $...$ & ... & ... & 0.9\\
Q2343-BX182 & 2.2879 & 2.2857 & 2.2909 & $2.4 \pm 0.3$ & 2.6 & $...$ & ... & ... & 0.4\\
Q2343-BX236 & 2.4348 & 2.4304 & 2.4372 & $3.1 \pm 0.6$ & 4.8 & $148^{+40}_{-42}$ & ... & 8.2 & 2.1\\
Q2343-BX336 & 2.5439 & 2.5448 & 2.5516 & $4.3 \pm 0.6$ & 5.0 & $140^{+33}_{-34}$ & ... & 7.6 & 1.9\\
Q2343-BX341 & 2.5749 & 2.5715 & ... & $4.0 \pm 0.6$ & 2.4 & $...$ & ... & ... & 0.5\\
Q2343-BX378 & 2.0441 & ... & ... & $4.5 \pm 1.0$ & 7.4 & $...$ & ... & ... & 0.3\\
Q2343-BX389 & 2.1716 & 2.1722 & ... & $12.0 \pm 0.4$ & 10.1 & $111^{+8}_{-8}$ & $71\pm10$ & 9.6 & 6.1\\
Q2343-BX390 & 2.2313 & 2.2290 & ... & $4.9 \pm 0.5$ & 10.8 & $78^{+21}_{-24}$ & ... & 5.1 & 0.7\\
Q2343-BX391 & 2.1740 & 2.1714 & ... & $4.2 \pm 0.2$ & 6.0 & $...$ & ... & ... & 0.2\\
Q2343-BX418 & 2.3052 & 2.3030 & 2.3084 & $8.0 \pm 0.2$ & 3.6 & $66^{+6}_{-6}$ & ... & 1.2 & 0.1\\
Q2343-BX429 & 2.1751 & ... & ... & $4.8 \pm 0.3$ & 9.4 & $51^{+16}_{-20}$ & ... & 1.9 & 0.4\\
Q2343-BX435 & 2.1119 & 2.1088 & 2.1153 & $8.1 \pm 0.4$ & 8.9 & $60^{+12}_{-14}$ & ... & 2.4 & 4.4\\
Q2343-BX436 & 2.3277 & 2.3253 & 2.3315 & $7.2 \pm 0.4$ & 9.1 & $63^{+10}_{-12}$ & ... & 2.8 & 1.1\\
Q2343-BX442 & 2.1760 & ... & ... & $7.1 \pm 0.3$ & 10.6 & $132^{+9}_{-9}$ & ... & 13.5 & 14.7\\
Q2343-BX461 & 2.5662 & 2.5649 & 2.5759 & $7.0 \pm 0.7$ & 7.7 & $139^{+22}_{-24}$ & ... & 11.5 & 0.1\\
Q2343-BX474 & 2.2257 & 2.2263 & ... & $5.0 \pm 0.3$ & 7.0 & $84^{+12}_{-12}$ & ... & 3.8 & 7.2\\
Q2343-BX480 & 2.2313 & 2.2297 & 2.2352 & $3.0 \pm 0.2$ & 3.4 & $<37$ & ... & $<0.7$ & 3.0\\
Q2343-BX493 & 2.3396 & 2.3375 & 2.3447 & $5.3 \pm 0.9$ & 6.5 & $155^{+39}_{-42}$ & ... & 12.2 & 0.1\\
Q2343-BX513\tablenotemark{j} & 2.1092 & 2.1090 & 2.114 & $10.1 \pm 0.4$ & 4.1 & $150^{+9}_{-9}$ & ... & 7.0 & 5.9\\
Q2343-BX529 & 2.1129 & 2.1116 & 2.1190 & $3.5 \pm 0.5$ & $<1.9$ & $...$ & ... & ... & 0.6\\
Q2343-BX537 & 2.3396 & ... & ... & $5.2 \pm 0.3$ & 5.3 & $...$ & ... & ... & 0.8\\
Q2343-BX587 & 2.2430 & 2.2382 & ... & $5.5 \pm 0.4$ & 7.7 & $94^{+14}_{-15}$ & ... & 5.2 & 3.5\\
Q2343-BX599 & 2.0116 & 2.0112 & ... & $4.5 \pm 0.4$ & 7.2 & $77^{+21}_{-22}$ & ... & 3.3 & 3.3\\
Q2343-BX601 & 2.3769 & 2.3745 & 2.3823 & $7.4 \pm 0.4$ & 5.3 & $105^{+12}_{-12}$ & ... & 4.5 & 2.3\\
Q2343-BX610 & 2.2094 & 2.2083 & 2.2129 & $8.1 \pm 0.4$ & 8.4 & $96^{+9}_{-10}$ & ... & 6.1 & 23.2\\
Q2343-BX660 & 2.1735 & 2.1709 & 2.1771 & $9.4 \pm 0.4$ & 6.2 & $<40$ & ... & $<0.8$ & 1.4\\
Q2343-MD59 & 2.0116 & 2.0107 & ... & $2.9 \pm 0.4$ & 5.5 & $...$ & ... & ... & 7.6\\
Q2343-MD62 & 2.1752 & 2.1740 & ... & $2.3 \pm 0.3$ & 6.7 & $79^{+27}_{-30}$ & ... & 3.3 & 1.9\\
Q2343-MD80 & 2.0138 & 2.0116 & ... & $3.2 \pm 0.3$ & 2.6 & $74^{+16}_{-16}$ & ... & 1.1 & 0.6\\
Q2346-BX120 & 2.2664 & ... & ... & $5.3 \pm 0.3$ & 6.2 & $62^{+12}_{-12}$ & $40\pm10$ & 1.9 & ...\\
Q2346-BX220 & 1.9677 & 1.9664 & ... & $10.3 \pm 0.6$ & 6.5 & $143^{+14}_{-14}$ & ... & 10.4 & 1.7\\
Q2346-BX244 & 1.6465 & 1.6462 & 1.6516 & $5.4 \pm 0.8$ & 7.7 & $42^{+27}_{-42}$ & ... & 1.0 & ...\\
Q2346-BX404 & 2.0282 & 2.0270 & 2.0348 & $13.9 \pm 0.3$ & 3.6 & $102^{+3}_{-3}$ & ... & 3.0 & 4.0\\
Q2346-BX405 & 2.0300 & 2.0298 & 2.0358 & $14.0 \pm 0.2$ & 5.8 & $50^{+4}_{-4}$ & ... & 1.1 & 1.6\\
Q2346-BX416 & 2.2404 & 2.2407 & ... & $12.1 \pm 0.7$ & 4.3 & $126^{+12}_{-14}$ & ... & 5.3 & 2.5\\
Q2346-BX482 & 2.2569 & 2.2575 & ... & $11.2 \pm 0.3$ & 9.8 & $133^{+4}_{-4}$ & ... & 13.7 & ...\\
SSA22a-MD41 & 2.1713 & ... & ... & $7.9 \pm 0.1$ & 9.6 & $107^{+6}_{-6}$ & $150\pm16$ & 8.2 & ...\\
West-BM115 & 1.6065 & 1.6060 & ... & $5.9 \pm 0.4$ & 5.5 & $128^{+15}_{-15}$ & ... & 6.9 & ...\\
West-BX600 & 2.1607 & ... & ... & $6.3 \pm 0.4$ & 8.2 & $181^{+16}_{-16}$ & $210\pm13$ & 21.5 & ...\\

\enddata
\tablenotetext{a}{Vacuum redshift of the UV insterstellar absorption
  lines.  We give a value only when the S/N of the spectrum is
  sufficient for a precise measurement.}
\tablenotetext{b}{Vacuum redshift of the \lya\ emission line, when
  present.}
\tablenotetext{c}{Observed flux of \Ha\ emission line, in units of
  $10^{-17}$ erg s$^{-1}$ cm$^{-2}$.}
\tablenotetext{d}{Approximate spatial extent of the \Ha\ emission (FWHM),
  after subtraction of the seeing in quadrature.}
\tablenotetext{e}{Velocity dispersion of the \Ha\ emission line.}
\tablenotetext{f}{For tilted emission lines, the velocity shear $(v_{\rm max}-v_{\rm min})/2$,
  where $v_{\rm max}$ and $v_{\rm min}$ are with respect to the
  systemic redshift.}
\tablenotetext{g}{Dynamical mass $M_{\rm dyn}=3.4\sigma^2r_{\Ha}/G$.}
\tablenotetext{h}{Stellar mass, from SED modeling.}
\tablenotetext{i}{AGN}
\tablenotetext{j}{Q2343-BX513 was observed twice with NIRSPEC, with
  position angles differing by 9\degr.  The first observation yielded
  $z_{\Ha}=2.1079$ and $\sigma=58$ \kms, and the second
  $z_{\Ha}=2.1092$ and $\sigma=150$ \kms.  It also has two
  \lya\ emission redshifts, $z_{\lya}=2.106$ and $z_{\lya}=2.114$.}
\end{deluxetable}

\end{document}